\documentclass[prd,twocolumn,aps,letterpaper,amsmath,amssymb,superscriptaddress,floatfix,longbibliography,nofootinbib]{revtex4-2}
\usepackage{dsfont}
\usepackage{graphicx}
\usepackage[caption=false]{subfig}
\usepackage{enumitem}
\allowdisplaybreaks 
\usepackage[hypertexnames=true]{hyperref}

\usepackage{float}
\makeatletter
\let\newfloat\newfloat@ltx
\let\float@end\float@end@ltx
\makeatother

\usepackage{algorithm}
\usepackage{algpseudocode}
\usepackage[capitalize]{cleveref}
\Crefname{appsec}{appendix}{appendices}
\hypersetup{
    colorlinks=true,       
    linkcolor=blue,          
    citecolor=blue,        
    filecolor=blue,      
    urlcolor=blue           
}

\usepackage{mathtools}
\usepackage{bm}
\usepackage{amsthm}
\usepackage{amssymb}
\usepackage{mathrsfs}

\newtheorem{theorem}{Theorem}
\newtheorem{corollary}{Corollary}
\DeclareMathOperator{\rank}{rank}
\DeclareMathOperator{\vectorization}{vec}
\DeclareMathOperator{\Tr}{Tr}
\DeclareMathOperator{\gap}{gap}
\DeclareMathOperator{\erf}{erf}
\DeclareMathOperator{\diag}{diag}
\DeclareMathOperator{\interior}{int}
\DeclareMathOperator*{\argmin}{arg\,min}
\DeclareMathOperator*{\argmax}{arg\,max}
\DeclareMathOperator*{\minimize}{minimize}
\DeclareMathOperator*{\maximize}{maximize}
\DeclareMathOperator{\primal}{p}
\DeclareMathOperator{\dual}{d}
\newcommand{\identity}{\mathds{1}}
\newcommand{\R}[0]{\ensuremath{\mathbb{R}}}
\newcommand{\C}[0]{\ensuremath{\mathbb{C}}}
\newcommand{\N}[0]{\ensuremath{\mathbb{N}}}
\renewcommand{\H}[0]{\ensuremath{\mathbb{H}}}
\newcommand{\Sym}[0]{\ensuremath{\mathbb{S}}}
\newcommand{\Nop}[0]{p} 
\newcommand{\Nsample}[0]{N}
\newcommand{\Ncrit}[0]{N_{\rm crit}}
\newcommand{\tmin}[0]{t_{0}}
\newcommand{\Nstates}[0]{N_{\rm states}}
\newcommand{\polar}[1]{{#1}^\circ}
\newcommand{\norm}[1]{\lVert #1 \rVert}
\newcommand{\projection}[1]{\bm{\mathcal{P}}_{#1}}
\newcommand{\metric}[0]{g}
\newcommand{\Hk}[1]{H^{[#1]}}
\newcommand{\muflat}{\momseq^{\flat}}
\newcommand{\Hflat}{H^{[d]\flat}}
\newcommand{\lastflat}{\moment{2d-2}^{\flat}}

\newcommand{\expectation}[0]{\mathbb{E}}
\newcommand{\normal}[0]{\mathcal{N}}

\newcommand{\bra}[1]{\ensuremath{\langle #1 |}}   
\newcommand{\ket}[1]{\ensuremath{| #1 \rangle}}   
\newcommand{\avg}[1]{\ensuremath{\langle #1 \rangle}}

\newtheorem{optprob}{Problem}[section]

\newenvironment{optproblem}[1]%
{\begin{optprob}[#1]\begin{equation}\begin{aligned}}%
{\end{aligned}\end{equation}\end{optprob}}

\newcommand{\st}{\operatorname*{subject~to}}

\newcommand{\momseq}{\bm{\mu}}
\newcommand{\moment}[1]{\mu_{#1}}
\newcommand{\dualmomseq}{M}
\newcommand{\centralpath}{\varepsilon}

\newcommand{\Tmeas}{\ensuremath{\mathcal{T}_\text{meas}}}

\newcommand{\kappavar}{\lambda}

\makeatletter
\newcommand{\StateAligned}[1]{%
    \State
    \hangindent=\dimexpr\ALG@tlm-\algorithmicindent\relax
    \hangafter=1\relax
    #1%
}
\makeatother

\begin{document}
\title{
Positivity, Noise, and Cones
}
\begin{abstract}
    As a consequence of causality, Euclidean-time correlation functions in quantum field theory are subject to constraints in the form of positivity conditions on certain Hankel matrices.
    Such conditions define convex conic structures.
    Correlation functions computed numerically using Monte Carlo techniques generally contain statistical fluctuations that violate the positivity conditions.
    This work describes a projection procedure, based on the convex conic geometry induced by the positivity conditions and a statistical covariance matrix, that can be used to restore positivity, providing a unique decomposition for noisy correlators into causal and acausal components.
    In contrast to previous literature, this work explicitly leverages the conic structure and not just the convex nature of the causal space.
    The projection can be computed using primal-dual interior-point methods.
    Constrained projection gives rise to observable-specific frequentist confidence intervals for smeared spectral densities in the causal bootstrap of [\emph{Phys.\ Rev.\ D} 114, 074510 (2026)].
    Analytic results are supported with numerical examples.
\end{abstract}

\author{Ryan Abbott}
\affiliation{Physics Department, Columbia University, New York, NY 10027, USA}
\author{Claire X. Davis}
\affiliation{Department of Physics, Colorado State University, Fort Collins, CO 80523, USA}
\author{Sarah Fields}
\affiliation{Physics Department, Columbia University, New York, NY 10027, USA}
\author{William I. Jay}
\email{william.jay@colostate.edu}
\affiliation{Department of Physics, Colorado State University, Fort Collins, CO 80523, USA}
\author{Patrick Oare}
\affiliation{
Physics Department, Brookhaven National Laboratory, Upton, NY 11973, USA}
\author{Matteo Saccardi}
\email{matteo.saccardi@colostate.edu}
\affiliation{Department of Physics, Colorado State University, Fort Collins, CO 80523, USA}

\date{November 18, 2026}

\maketitle

\section{Introduction \label{sec:intro}}

Causality imposes stringent conditions on correlation functions in quantum field theory (QFT), often in the form of a positivity condition or a constraint on their complex analytic structure.
However, the rigidity of complex analytic functions which makes such conditions so powerful from a formal perspective can sometimes form an obstruction to their application to data, experimental or theoretical, known with finite precision.

Constraints from causality, together with exact Euclidean inputs, give rise to tight rigorous bounds on smeared spectral quantities.
The bounds are calculable either via Nevanlinna--Pick (NP)~\cite{Bergamaschi:2023xzx} interpolation or moment problems~\cite{Abbott:2025snz}.
Ref.~\cite{Fields:2025glg} has initiated a program to include the effects of statistical uncertainties in NP interpolation (see also Refs.~\cite{nogaki2023nevanlinnajl,Nogaki:2023mut,PhysRevLett.126.056402,Huang:2023gpb} for additional field-theoretic applications of NP interpolation theory).
Lawrence has shown that bounds from noisy data are also accessible via convex optimization~\cite{Lawrence:2024hjm,Lawrence:2025nmh}.
Ref.~\cite{Abbott:2026wdw} has shown that bounds are accessible from finite-dimensional semidefinite programs (SDPs) for an important class of problems and that these bounds reproduce the tight bounds of NP interpolation and moment problems, given the same inputs.
Closely related work in Ref.~\cite{Mutzel:2026vyw} provides a different perspective on computational tractability of the SDPs, employing a limiting sequence of finite-dimensional ``relaxations.''

The goal of the present work is to lay out the geometric underpinnings for a new covariance-weighted causal decomposition, which uniquely and unambiguously partitions noisy samples into causal and acausal pieces, together with an algorithmic scheme based on convex optimization to carry out the projection in practice.
The resulting decomposition is given in \cref{eq:conic_decomp}.
The present proposal is in sympathy with previous works in Refs.~\cite{Yu_2024,Ostmeyer:2025igc}, which focus clearly on the positivity of certain Hankel matrices.

In contrast to prior work in the literature on lattice field theory, the present work develops a geometric picture to analyze noisy Euclidean correlators based on the conic, and not just convex, structure of causal constraints; a key novel element is the interplay between the conic geometry and the inner product induced by the covariance matrix.
In particular, we formulate a conic projection for lattice correlators that enforces spectral-positivity constraints respecting the full covariance matrix of any input data, without prescribing a finite-state spectral model.
The projected data nevertheless possess a unique, minimal spectral representation. 

The present work also studies the statistical behavior of the projection in terms of the mean squared error ($L_2$ risk) and effective degrees of freedom. 
A short calculation and dimensional argument give a variational bound on the critical sample size below which acausal fluctuations become unavoidable. 

Constrained projections provide the basis for defining frequentist confidence intervals in the causal bootstrap of Ref.~\cite{Abbott:2026wdw}.
We construct an observable-specific profile-likelihood statistic and calibrate it through a parametric bootstrap procedure.
Tests of the resulting \emph{conic calibration} on realistic mock data demonstrate the desired frequentist coverage.\footnote{
    During the completion of this work, we became aware that Christoph Lehner has independently developed related ideas for statistical interpretation of bounds from the causal bootstrap~\cite{Lehner:2026inprep}.
}

The conic geometry discussed in this work is closely related to the notion of \emph{positive geometry} that has recently been used to constrain the structure of QFTs (see, e.g., Refs.~\cite{Arkani-Hamed:2020blm,Arkani-Hamed:2018ign,Bellazzini:2020cot}).
Here spectral positivity, a consequence of causality, provides strong constraints on the space of causal correlation functions.
In contrast to previous work, however, the noise in the correlation function plays a crucial role, adding a secondary geometric structure.

The current proposal for causal projection is phrased in terms of a convex optimization problem and is intimately connected to the ellipsoidal statistical regularization (cf.\ Chapter~6 of Ref.~\cite{wolkowicz2000handbook}) employed in Refs.~\cite{Lawrence:2024hjm,Lawrence:2025nmh,Abbott:2026wdw,Mutzel:2026vyw}.
Numerical methods based on semidefinite programming are not new to theoretical physics.
The renaissance in the numerical conformal bootstrap~\cite{Poland:2018epd} has been based in part on the realization that established techniques from convex optimization~\cite{wolkowicz2000handbook} could be brought to bear on bootstrap problems~\cite{Simmons-Duffin:2015qma}.
New applications of SDPs continue to appear in theoretical physics, including bootstrap approaches to matrix quantum mechanics~\cite{Lin:2025iir,Lin:2023owt,Lin:2024vvg,Cho:2024kxn,Lin:2025srf}, lattice gauge theories and quantum many-body systems~\cite{Kazakov:2024ool,Scheer:2024eyu,Cho:2024owx,Fawzi:2023fpg}, scattering amplitudes~\cite{He:2024nwd,Guerrieri:2024ckc,deRham:2025vaq}, and sign problems in lattice field theory~\cite{Lawrence:2022vsb,Lawrence:2023sfc}.

The remainder of the work is structured as follows.
\Cref{sec:positivity} reviews well-known consequences of positivity constraints, placing the present work in the context of the existing  literature on lattice field theory and establishing notation.
 \Cref{sec:cones} describes the conic geometry implied by positivity constraints; this geometric perspective is novel to the present work.
\Cref{sec:statistics} considers statistical consequences of the conic geometry when correlation functions are viewed as (approximately) Gaussian random variables.
\Cref{sec:causal_bootstrap} connects conic projection to the causal bootstrap and develops an observable-specific statistical calibration.
\Cref{sec:comparison} gives a detailed comparison of the present work to existing methods and recent proposals in the literature.
\Cref{sec:numerics} provides numerical examples.
\Cref{sec:conclusions} presents conclusions.
Technical results are collected in two appendices.
\section{Review of spectroscopy and constraints from positivity \label{sec:positivity}}

Consider a quantum field theory with inverse temperature $\beta$.
Consider a positive-definite Hermitian matrix of two-point correlation functions at positive Euclidean time $t>0$
\begin{align}
    C_{ab}(t)
    &= \int d^3\bm{x}\, \avg{
        O_a(t,\bm{x}) O^\dagger_b(0,\bm{0})
    }
    \label{eq:two_point_function} \\
    &=
    \int_{0}^{\infty}  d\omega\, \rho_{ab}(\omega)
    \left[
    \frac{e^{-\omega t}  + e^{-\omega(\beta - t)}}
    {1\pm e^{-\beta\omega}}
    \right]
    \label{eq:Laplace_transform_T}
    \\
    &\stackrel{\beta\to\infty}{=}
    \int_{0}^{\infty} d\omega\, \rho_{ab}(\omega) e^{-\omega t},
    \label{eq:Laplace_transform}
\end{align}
where $\left\{O_a : a\in\{1,2,\dots,\Nop\}\right\}$ is a set of suitable local operators and $\rho_{ab}(\omega)$ is the spectral function.
For each positive time, $C_{ab}(t) \in \H_p$, where $\H_p$ is the space of $p\times p$ Hermitian matrices.
Throughout this work, the operators $\{O_a\}$ are taken to transform under an irreducible representation, or irrep, of the relevant global symmetry group so that the operators couple to states with definite quantum numbers.
Causality and reflection positivity imply that the correlation function and spectral function are both positive semidefinite.
It will be assumed, as an additional but standard physical hypothesis (see Refs.~\cite{Luscher:1990ck,Blossier:2009kd}), that linearly independent operators have been chosen so that the correlation function is actually positive in the sense that $C_{ab}(t) \succ 0$ for all $t>0$.
Spectroscopy refers to the problem of determining properties of the spectral function from the correlation function, i.e., inverting the Laplace transform in \cref{eq:Laplace_transform}.

At zero temperature and in a finite spatial volume, the spectral function consists of a discrete but generally infinite sum over finite-volume energy levels with a given set of quantum numbers,\footnote{
    We use the standard relativistic normalization of states in a finite spatial volume,
    \begin{align*}
        \langle n', \bm{p}'|n, \bm{p}\rangle
        = 2 E_n(\bm{p}) L^3 \delta_{nn'} \delta_{\bm{p}\bm{p}'}.
    \end{align*}
}
\begin{align}
    \rho_{ab}(\omega)
    &= \sum_{n=1}^{\infty}
        \frac{\langle 0| O_a | n\rangle \langle n| O_b^\dagger |0 \rangle}{2 E_n}
        \delta(\omega-E_n)\\
    &\equiv \sum_{n=1}^{\infty} Z_{an} Z_{bn}^* \delta(\omega-E_n).
    \label{eq:spectral_function_FV}
\end{align}
The overlaps will often be written as a vector $\bm{Z}_n \in \C^p$.
States with vanishing overlaps ($\bm{Z}_n = \bm{0}$), for example for reasons of symmetry, are excluded from the sum.
Positivity of the spectral function implies that, for each energy level $n$, the matrices of overlap factors
\begin{align}
W_n \equiv \bm{Z}_{n} \bm{Z}_{n}^\dagger \in \H_p
\label{eq:Wn_def}
\end{align}
are positive semidefinite.
When distinct states in the physical Hilbert space share the same energy, these states are said to be degenerate.
Unless otherwise indicated, all energy levels are hereafter assumed to be non-degenerate.
The role of possible degeneracies is discussed briefly in \cref{app:degeneracies}.

Numerical lattice field theory gives access to unbiased statistical estimators for the correlation functions $C_{ab}(t)$ of the lattice quantum field theory.
Unless otherwise noted, the present work uses lattice units $a=1$ so that $t\in\N$.
When viewed as vectors indexed by $t\in\{1,2,\dots, m\}$, correlators will be written in boldface as $\bm{C}\in (\H_{\Nop})^m$.
The choice of the letter $m$ is meant to stand for the number of ``moments,'' a perspective described in \cref{ssec:moments} below.
Clearly, $m \leq N_t$ with $N_t$ the temporal extent of the lattice, and in some cases it may be desirable to restrict the number of moments further.\footnote{
Baryon correlators, for example, are generically asymmetric about the midpoint, with distinct correlator values at each Euclidean time.
For mesonic correlators that happen to be symmetric, one may consider averaging about the midpoint so that $m\leq N_t/2$.}
For scalar-valued correlators, $\bm{C}\in \R^m$ without loss of generality.

The central limit theorem guarantees that, for a sufficiently large number of independent and identically distributed random samples $\hat{\bm{C}}'_1, \hat{\bm{C}}'_2, \dots, \hat{\bm{C}}'_\Nsample$, the sample mean will be Gaussian distributed:
\begin{align}\label{eq:ChatN}
    \hat{\bm{C}}_{\Nsample} = \frac{1}{\Nsample}\sum_{i=1}^{\Nsample} \hat{\bm{C}}'_i \stackrel{\Nsample\to\infty}{\sim} \normal(\bm{C}, \Sigma_{\Nsample}),
\end{align}
where $\Sigma_\Nsample \equiv \Sigma/\Nsample$ is the covariance matrix of the sample mean and where $\Sigma$ is the covariance matrix associated with $C(t)$.
Here and throughout the present work, hatted variables denote instances of a random variable that follow a known distribution, e.g., $\normal(\bm{C}, \Sigma_N)$ in \cref{eq:ChatN}.
All covariance matrices are assumed to be Hermitian and positive definite.
For notational brevity, the subscript on $\Sigma_\Nsample$ will be omitted when the value of $\Nsample$ is immaterial.
As usual, the inverse covariance matrix defines a real inner-product structure via
\begin{align}
    \langle \bm{u}, \bm{v} \rangle_{\Sigma^{-1}} \equiv
    \bm{u}^\dagger \Sigma^{-1} \bm{v}
    =
    \bm{u}_{tab}^\dagger \Sigma^{-1}_{tab,t'a'b'} \bm{v}_{t'a'b'},
    \label{eq:Cov_inner_product}
\end{align}
where $\bm{u}, \bm{v} \in (\H_p)^m$.
The induced norm is $\lVert \bm{a} \rVert_{\Sigma^{-1}}$, with
\begin{align}\label{eq:a2}
    \lVert \bm{a} \rVert^2_{\Sigma^{-1}} \equiv \langle \bm{a}, \bm{a} \rangle_{\Sigma^{-1}} = \bm{a}^\dagger \Sigma^{-1} \bm{a},
    &&
    \bm{a} \in (\H_p)^m .
\end{align}
When the relevant covariance matrix is clear from context and a statistical interpretation is being emphasized, particularly for a difference $\bm{a}-\bm{b}\in (\H_p)^m$, the previous equation will be written
\begin{align}\label{eq:chi2_ab}
    \chi^2(\bm{a}-\bm{b}) = (\bm{a}-\bm{b})^\dagger \Sigma^{-1}(\bm{a}-\bm{b}).
\end{align}

\subsection{The generalized eigenvalue problem \label{sec:GEVP}}

A common and successful approach to spectroscopy phrases the problem using the variational method~\cite{Michael:1982gb} in terms of a generalized eigenvalue problem (GEVP)~\cite{Luscher:1990ck}
\begin{align}
    C(t)\, \bm{u}_n(t,t_0) = \lambda_n(t,t_0)\, C(t_0)\, \bm{u}_n(t,t_0),
    \label{eq:GEVP}
\end{align}
where $C(t) \succ 0$ for all $t>0$ and $t_0>0$ is an arbitrary reference time.
The eigenvectors are $\bm{u}_n \in \C^p$, and the generalized eigenvalues $\lambda_n$ are positive real numbers.
At large $t_0$ and $t>t_0$, they satisfy~\cite{Blossier:2009kd,Luscher:1984is}
\begin{align}
\lambda_n(t,\tmin) = e^{-E_n (t-\tmin)} \left[1 + \mathcal{O}\big(e^{-(E_{p+1}-E_n) (t-\tmin)}\big)\right],
\label{eq:gevp_approx}
\end{align}
where $E_{p+1}-E_n>0$.
As $t\to\infty$, the energies extracted from the generalized eigenvalues of a $\Nop \times \Nop$ matrix of correlation functions converge to the $\Nop$ lowest finite-volume energy levels that overlap with operators in the given variational set.
The eigenvectors are conventionally normalized such that
\begin{align}
    \bm{u}^\dagger_n(t, \tmin) C(\tmin) \bm{u}_m(t, \tmin) = \delta_{nm}.
\end{align}
With this normalization for the eigenvectors, the $n^{\mathrm{th}}$ matrix of spectral weights is accessible for large Euclidean times via
\begin{gather}
    \lambda_n(t, t_0)\, C(\tmin) \bm{u}_n(t, \tmin) \bm{u}^\dagger_n(t, \tmin) C(\tmin)
    \stackrel{t\gg\tmin}{\longrightarrow} e^{-E_n t} W_n,
\end{gather}
with $W_n$ as in \cref{eq:Wn_def}.
For finite $t$, the generalized eigenvalues give variational upper bounds for the energy levels.
The problem can be rephrased equivalently as an ordinary eigenvalue problem of the form
\begin{align}
    L^{-1} C(t) {L^\dagger}^{-1} \bm{v}_n  = \lambda_n \bm{v}_n, && \bm{v}_n = L^\dagger \bm{u}_n,
\end{align}
where $C(t_0) = L L^\dagger$ is the Cholesky decomposition~\cite{Fischer:2020bgv}.

\subsection{Moment problems and Hankel matrices \label{ssec:moments}}

Spectroscopy can be formulated as a classical moment problem~\cite{Abbott:2025snz}.
Correlation functions sampled at equally spaced positive Euclidean times correspond to moments of matrix-valued positive distributions and can be written in the generic form
\begin{align}\label{eq:C_lambda_t}
    C_{ab}(t) = \int_I d\lambda\, \tilde{\rho}_{ab}(\lambda) \lambda^t.
\end{align}
In other words, $C_{ab}(t)$ is the $t^\mathrm{th}$ moment of the positive distribution $\tilde{\rho}_{ab}(\lambda)$.
A correlation function can therefore also be referred to as a moment sequence.
The support $I \subseteq \R$ is determined by the transfer-matrix eigenvalues: the conventional cases are zero-temperature Wilson-like correlators ($I=[0,1]$, with $\lambda=e^{-\omega}$ in lattice units), finite-temperature Wilson correlators ($I=\R^+$), and staggered correlators with ``oscillating'' contributions ($I=\R$)~\cite{Abbott:2025snz}.\footnote{
Certain staggered spectroscopy problems involve opposite-parity contributions with negative spectral weights.
This case, which can be viewed as a signed Hamburger moment problem, is discussed in \cref{sec:signed_hamburger}.}
These cases correspond, respectively, to the Hausdorff, Stieltjes, and Hamburger moment problems.
In the zero-temperature Wilson case, for instance,
$\tilde{\rho}$ is related to the spectral function in \cref{eq:Laplace_transform} by a Jacobian factor:
$\rho_{ab}(\omega) = \lambda(\omega) \tilde{\rho}_{ab}(\lambda(\omega))$.

When it is useful to emphasize the fact that Euclidean-time correlators can be viewed as Hermitian matrix moment sequences, they will be denoted by the Greek letter $\mu$:
\begin{align}
    \moment{t} \equiv C(t) \in \H_\Nop, && \momseq \equiv \bm{C} \in (\H_\Nop)^m,
    \label{eq:moment_defs}
\end{align}
where $m$ counts the number of available moments (i.e., time slices).

Classical moment problems possess an additional notion of positivity, which is phrased in terms of (shifted) block Hankel matrices
\begin{align}
    H^{[d]}(\momseq) &\equiv [\mu_{i+j}]_{i,j=0}^{d-1}
    =
    \begin{bmatrix}
        \moment{0} & \moment{1} & \moment{2} & \dots \\
        \moment{1} & \moment{2} & \moment{3} & \dots \\
        \moment{2} & \moment{3} & \moment{4} & \dots \\
        \vdots & \vdots & \vdots & \ddots \\
    \end{bmatrix}
    \label{eq:hankel_matrix_definition}\\
    H^{(1),[d]}(\momseq) & \equiv [\mu_{i+j+1}]_{i,j=0}^{d-1}
    =
    \begin{bmatrix}
        \moment{1} & \moment{2} & \moment{3} & \dots \\
        \moment{2} & \moment{3} & \moment{4} & \dots \\
        \moment{3} & \moment{4} & \moment{5} & \dots \\
        \vdots & \vdots & \vdots & \ddots \\
    \end{bmatrix}
    \label{eq:shifted_hankel_matrix_definition},
\end{align}
where the blocks are defined in \cref{eq:moment_defs}.
The superscript $[d]$ will be dropped when the dimension $d$ is clear from the context or when discussing the infinite Hankel matrix.
To see why these matrices encode spectral positivity, let $v_i \in \mathbb{C}^p$ and consider a polynomial $p_d(\lambda) = \sum_{i=0}^{d-1} \lambda^i v_i$.
Writing $v = (v_0, \dots, v_{d-1})^T$ and defining $H^{(r),[d]}=[\mu_{i+j+r}]_{i,j=0}^{d-1}$ and $H^{(0),[d]} \equiv H^{[d]}$ then gives
\begin{equation}
\begin{split}
    v^\dagger H^{(r),[d]} v &= \sum_{i,j=0}^{d-1} v_i^\dagger \mu_{i+j+r} v_j\\
    &= \int_I d\lambda\, p_d(\lambda)^\dagger \tilde\rho(\lambda) p_d(\lambda) \lambda^r \,.
\end{split}
\end{equation}
Thus, $H \succeq 0$ follows from positive spectral weight. Support on $\lambda \geq 0$ additionally imposes $H^{(1)} \succeq 0$, while support on $\lambda \leq 1$ implies $H-H^{(1)} \succeq 0$.
The result is that a positive spectral function with suitable support necessarily implies positivity of the relevant Hankel matrices.

Remarkably, for the full, infinite moment sequences, these conditions are also sufficient conditions for existence of a positive matrix-valued spectral function solving the moment problem~\cite{schmudgen2017moment,Hamburger:1920}:
\begin{align}
    \text{Hamburger:}\quad & H \succeq 0
    \label{eq:Hamburger_positivity}
    \\
    \text{Stieltjes:}\quad &
    H, H^{(1)} \succeq 0
    \label{eq:Stieltjes_positivity}
    \\
    \text{Hausdorff:}\quad &
    H^{(1)},
    H - H^{(1)} \succeq 0.
    \label{eq:Hausdorff_positivity}
\end{align}
The conditions in \cref{eq:Hausdorff_positivity}
suffice to guarantee the uniqueness of the solution to the Hausdorff moment problem.
For the Stieltjes and Hamburger moment problems, an additional determinacy condition is required to guarantee uniqueness~\cite{Hamburger:1920,schmudgen2017moment}.
The condition $H\succeq0$ corresponds to the existence of a Hermitian transfer matrix and states with positive norm, $H^{(1)} \succeq 0$ means that the transfer matrix has positive eigenvalues, and $H-H^{(1)} \succeq0$ is the statement that eigenvalues of the transfer matrix are restricted to the unit interval.

A finite temporal lattice gives access to a truncated moment sequence, $(\mu_0, \dots, \mu_{m-1}) \in (\H_p)^m$ with $m=2d-1$, where both $d$ and $m$ are positive integers.
Unsurprisingly, information from truncated moment sequences is no longer sufficient in general to determine spectral functions uniquely.
However, rigorous pointwise bounds on smeared spectral quantities can be computed~\cite{Kovalishina1984}.
Connections between positive semidefinite Hankel matrices and the spectral decomposition of truncated moment sequences are discussed in \cref{sec:causal_boundary,app:spectral_decomp}.

The moment representation is to be applied only at time separations for which the assumed positive spectral decomposition holds.
In lattice QCD applications, a positive spectral representation may only be justified for temporal separations $t\geq \tmin$: $t > 0$ is required to avoid the contribution of contact terms in correlation functions, while for improved actions $\tmin$ is typically some small finite value.
For a zero-temperature Wilson-like spectral decomposition, one may avoid contact terms or short-time artifacts by starting at $\tmin>0$~\cite{Bailas:2020qmv,Bruno:2024fqc,Giusti:2026mcy} and using
\begin{align}
    C_{ab}(t+\tmin) = \int_{0}^{\infty} d\omega\, \left[ \rho_{ab}(\omega) e^{-\omega \tmin} \right] e^{-\omega t}.
\end{align}
The bracketed quantity is positive semidefinite, as this only changes the spectral weights but not their positivity, analogously to the choice of reference time in the GEVP; see \cref{eq:GEVP} above.
Thus the conditions of \cref{eq:Hamburger_positivity,eq:Stieltjes_positivity,eq:Hausdorff_positivity} may be applied to the Hankel matrices constructed starting at $C_{ab}(\tmin)$.
For some improved actions, the transfer matrix may not be Hermitian at finite lattice spacing~\cite{Luscher:1984is}.
In that case, exact Hankel positivity is an additional assumption, and not a consequence of simply discarding short-time data; we neglect such lattice artifacts in the following.

For later convenience, it is useful to record that when the moment sequence has a known spectral function as in \cref{eq:spectral_function_FV}, the Hankel matrix itself has a decomposition of the form
\begin{align}
    H &= \sum_{n=1}^{\infty}
    (\bm{V}_n \bm{V}_n^T) \otimes W_n,
    \label{eq:Hankel_spectral_decomp}
    \\
    \bm{V}_n & = (1, \lambda_n, \lambda_n^2, \dots)^T, \quad W_n \succeq 0,
    \label{eq:def_vandermonde_vectors}
\end{align}
where $W_n$ are the matrices in \cref{eq:Wn_def}; the tensor-product ordering corresponds to time indices labeling the outer blocks of $H^{[d]}$.
The vectors $\bm{V}_n$ are Vandermonde vectors.
If the moments start at $t_0>0$, the weights in this decomposition are correspondingly replaced by $e^{-E_n t_0} W_n$.

Finally, Hankel matrices are notoriously ill-conditioned. For positive-definite scalar $d\times d$ Hankel matrices, Beckermann's lower bound on the condition number for $d \geq 3$ scales exponentially with $d$ as~\cite{beckermann2000condition}
\begin{align}
    \kappa(H) \geq\gamma^{d-1}/(16d),
    &&
    \gamma=e^{4~{\rm Catalan}/\pi}\approx 3.210,
    \label{eq:Beckermann}
\end{align} independent of the particular values of the entries in the Hankel matrix, where ``${\rm Catalan}$'' refers to Catalan's constant.
For moment sequences with finite numerical precision or with small statistical fluctuations, this ill-conditioning obstructs direct application of methods that rely on positivity, including the bounds derived for the exact interpolation problem in Ref.~\cite{Kovalishina1984}.
This motivates relaxing the exact interpolation condition with a covariance-weighted projection onto the appropriate closed positivity cone.

\subsection{Convex optimization}

Convex optimization was introduced by Lawrence in the context of spectral reconstructions in Ref.~\cite{Lawrence:2024hjm}, focusing on smeared spectral observables of the form\footnote{
    For the matrix-valued case, the integrand is implicitly understood in terms of the trace pairing as in Ref.~\cite{Abbott:2026wdw}.
}
\begin{align}\label{eq:rho_kappa}
    \rho[\mathscr{K}] \equiv \int d\omega \,\rho(\omega) \mathscr{K}(\omega).
\end{align}
This notation emphasizes that the spectral function in \cref{eq:spectral_function_FV} is a singular distribution, which acts as a linear functional on test functions like the smearing kernel $\mathscr{K}(\omega)$.
Taking for simplicity the zero-temperature Wilson case by choosing a scalar $\mathscr{K}(\omega)=e^{-\omega t}$ recovers the Euclidean correlator $\rho[\mathscr{K}] = C(t)$.
Consider a noisy moment sequence $\hat{\momseq}$, assumed to be sampled from a distribution with mean $\momseq$ and covariance matrix $\Sigma$.
For a given spectral function $\rho(\omega)$, consider the residual $\delta \hat{\momseq} \in \R^m$ with components
\begin{align}
    \delta \hat{\mu}_{t} \equiv \hat{\moment{}}_{t} - \rho[e^{-\omega t}].
\end{align}
The squared norm of the residual is a random variable $\chi^2(\delta\hat{\momseq})$, cf.~\cref{eq:a2,eq:chi2_ab}.
For noisy scalar-valued data, Ref.~\cite{Lawrence:2024hjm} introduced the following convex optimization problem:

\begin{optproblem}{Primal spectral reconstruction problem for noisy data}
\label{eq:opt-problem-def-inexact}
    \minimize_{\rho } \quad & \rho[\mathscr{K}]  \\
    \st \quad & \rho \geq 0\\
    \mathrm{and} \quad & \chi^2(\delta\hat{\momseq}) \leq \chi_\mathrm{cut}^2.
\end{optproblem}
\noindent
The final line is an ellipsoidal constraint that defines a notion of statistical consistency which relaxes the interpolation condition $\rho[e^{-\omega t}] = \hat\mu_t$, retrieved as $\chi^2_\mathrm{cut} \to 0$ for exact data. Ellipsoidal constraints are a standard approach to handle problems with noisy input data via semidefinite programming~\cite{wolkowicz2000handbook}.
The same constraint was used recently in Refs.~\cite{Abbott:2026wdw,Mutzel:2026vyw}.

The choice of $\chi_\mathrm{cut}^2$ determines the size of the feasible set.
Clearly, a solution will not always exist, e.g., $\chi_\mathrm{cut}^2=0$ for generic noisy input data with $H \not\succeq 0$.
This observation implies the existence of a minimal cutoff $\chi_{\rm min}^2$ for which a solution exists.
This minimum value can be identified explicitly as the solution to the following convex quadratic problem:
\begin{optproblem}{Primal conic projection}
\label{eq:opt-problem-conic-projection}
\minimize_{\rho } \quad & \chi^2(\delta \hat{\momseq})\\
\st \quad & \rho \geq 0.\\
\end{optproblem}
\noindent
This problem will also be cast as an optimization problem over a set of moments whose Hankel matrices satisfy the relevant positivity conditions.
For reasons that will become clear in the next section, this problem will be referred to as conic projection.
A primary goal of the remainder of the paper is to describe the geometric and statistical properties of solutions to \cref{eq:opt-problem-conic-projection} and their implications for smeared spectral reconstructions and spectroscopy.
A key question is how to interpret these solutions correctly in statistical inference problems that arise in practical lattice QCD calculations.
\section{Conic decomposition of acausal moment sequences \label{sec:cones}}

This section presents a novel, unique decomposition of noisy moment sequences.
The decomposition arises from the interplay between the conic geometry implied by Hankel positivity conditions and a suitable inner-product structure, usually but not necessarily defined by a covariance matrix.
\Cref{sec:convex-analysis-defns} begins by reviewing definitions from convex analysis.
\Cref{sec:causal_cone} defines the causal cone, $\mathcal{K}$.
\Cref{sec:causal_boundary} discusses the boundary of the causal cone, $\partial\mathcal{K}$.
\Cref{sec:moreau} uses the inner-product structure to construct the polar cone $\polar{\mathcal{K}}$; the Moreau decomposition furnishes a unique decomposition of a noisy moment sequence into boundary elements on the causal and polar cones.
\Cref{sec:primal-dual} introduces the primal-dual formulation of Moreau decomposition.
\Cref{sec:conic_decomposition} collects the preceding results in a novel conic decomposition formula for noisy moment sequences.

\subsection{Definitions from convex analysis}\label{sec:convex-analysis-defns}

To keep the present work self-contained, this section reviews standard concepts and terminology appearing in convex analysis.
A textbook presentation of cones is given in Ref.~\cite{aliprantis2007cones}, while a recent review of conic regression was given in Ref.~\cite{dimiccoli2016fundamentalsconeregression}.

A subset $C \subseteq \R^n$ is said to be a \emph{cone} if $\bm{x}\in C$ implies $s\bm{x} \in C$ for all $s>0$.
A \emph{convex cone} $C \subset \R^n$ is a cone satisfying $\alpha \bm{x} + \beta \bm{y} \in C$ for all $\bm{x},\bm{y}\in C$ and all $\alpha,\beta>0$.
A cone is said to be \emph{closed} when it is a closed subset of $\R^n$ with respect to the usual topology on $\R^n$.
Suppose that an inner product is also present, i.e., that $(\R^n, \langle \cdot, \cdot\rangle)$ is a mathematical Hilbert space.
Given a cone $C \subset \R^n$,
the \emph{polar cone} is
\begin{align}
    \polar{C} = \{ \bm{y} : \langle \bm{x}, \bm{y}\rangle \leq 0 \text{ for all } \bm{x} \in C\}.
\end{align}
The polar cone is always a convex cone, regardless of the convexity of $C$.
All cones considered below are closed, convex cones.
The definitions extend immediately to block moment sequences $\momseq \in (\H_p)^m$ with a real inner product, since $(\H_p)^m$ is a finite-dimensional real vector space.

\subsection{The causal cone \label{sec:causal_cone}}

The positivity conditions on Hankel matrices in \cref{eq:Hamburger_positivity,eq:Stieltjes_positivity,eq:Hausdorff_positivity}, for example $H\succeq 0$ for the Hamburger moment problem, define convex conic structures, which are the foundation of convex analysis.
In the problem at hand, the relevant conic structures are defined in terms of the linear maps associated with the Hankel matrices in \cref{eq:hankel_matrix_definition,eq:shifted_hankel_matrix_definition}.

To see concretely how these maps arise, consider a block moment sequence $\bm{\mu} \in (\H_p)^m$ with $m=2d-1$, $d\in\N^+$.
The motivation to consider odd values of $m$ is the definition of the Hankel matrix in \cref{eq:hankel_matrix_definition}, which contains an odd number of total moments.
\Cref{eq:hankel_matrix_definition} defines the following linear maps between $(\H_p)^m$ and $\H_{dp}$:
\begin{align}
    H : (\H_p)^m \to \H_{dp}, && \text{with }
        \momseq \mapsto H(\momseq)
    \label{eq:hankel_linear_map}
    \\
    \bm{h}: \H_{dp} \to (\H_p)^m, && \text{with }
    \dualmomseq \mapsto \bm{h}(\dualmomseq).
    \label{eq:hankel_transpose_linear_map}
\end{align}
In physical terms, these maps move between the space where moment sequences live, $(\H_\Nop)^m$, and the space where Hankel matrices live, $\H_{dp}$.
The latter map $\bm{h}$ is the dual (or transpose) map of $H$, defined implicitly by the natural inner-product structure on $\H_{dp}$,
\begin{align}
    \Tr[\dualmomseq^\dagger H(\momseq)] \equiv
    \momseq^\dagger \bm{h}(\dualmomseq).
\end{align}
The linear map $H$ is injective (one-to-one), while $\bm{h}$ is surjective (onto).
In components, the $k^\mathrm{th}$ entry of $\bm{h}(\dualmomseq)$ is the sum of the elements on the $k^\mathrm{th}$ antidiagonal of $\dualmomseq$.
For the scalar-valued case ($p=1$), $H: \R^m \to \Sym^d$, where $\Sym^d$ denotes the space of symmetric $d\times d$ matrices.
Note that the linear maps $H$ and $\bm{h}$ do not by themselves impose or imply any particular positivity conditions;
positivity is instead related to the properties of a given moment sequence $\bm{\mu}$.
Although only $H$ is needed to express the physical positivity constraints in \cref{sec:positivity} above, the dual map $\bm{h}$ will play a key role in enforcing these constraints using Lagrange multipliers below.

\Cref{eq:hankel_linear_map,eq:hankel_transpose_linear_map} employ a notation in which ``primal'' quantities like the moment sequence $\momseq$ are written with lowercase letters, while the corresponding ``dual'' quantities like $\dualmomseq$ are written with uppercase letters.
This notation will be used throughout the rest of the manuscript.
Distinguishing primal and dual variables in this way helps clarify the structure of later convex optimization problems.

The causal cone $\mathcal{K}$ for the Hamburger moment problem in \cref{eq:Hamburger_positivity} is defined by
\begin{align}
   \mathcal{K}
   \equiv \{
    \momseq\in (\H_p)^m : H(\momseq) \succeq 0
   \}.
   \label{eq:def_causal_cone}
\end{align}
Entirely analogous cones can be defined for the Stieltjes and Hausdorff moment problems using \cref{eq:Stieltjes_positivity} and \cref{eq:Hausdorff_positivity}, respectively.
For example, the causal cone for the Stieltjes moment problem is
\begin{align}
\mathcal{K}_{\rm Stieltjes}
   \equiv \{
    \momseq\in (\H_\Nop)^m : H(\momseq) \succeq 0, H^{(1)}(\momseq) \succeq 0
   \},
   \label{eq:def_causal_cone_Stieltjes}
\end{align}
where the shifted Hankel matrix is computed using all the available moments:
\begin{align}
H^{(1)}: (\H_\Nop)^{m-2} &\to \H_{(d-1)\Nop}\\
(\mu_1, \dots, \mu_{m-2}) &\mapsto
    \begin{pmatrix}
    \mu_1     & \mu_2   & \cdots  & \mu_{d-1} \\
    \mu_2     & \ddots  &         & \vdots \\
    \vdots    &         & \ddots  & \mu_{m-3}\\
    \mu_{d-1} & \cdots  & \mu_{m-3} & \mu_{m-2} \\
    \end{pmatrix}.
\end{align}
The map $\bm{h}^{(1)}: \H_{(d-1)\Nop} \to (\H_p)^{m-2}$ is defined similarly.
In physical terms, the causal cone is the space of Euclidean-time data that satisfy the positivity constraints from causality.
Interior moment sequences $\momseq \in \interior \mathcal{K}$ always admit a spectral decomposition (see \cref{thm:flat}(c) in \cref{app:flat}).
As discussed in the next section, a general boundary moment sequence $\momseq \in \partial \mathcal{K}$ may contain both a spectral component and a so-called endpoint excess.

\subsection{The boundary of the causal cone \label{sec:causal_boundary}}

The boundary of the causal cone corresponds to moment sequences with singular, positive semidefinite Hankel matrices.
For block Hankel matrices, the boundary is
\begin{align}
\partial \mathcal{K} =
\{\momseq \in (\H_p)^m : H(\momseq)\succeq0,\det H(\momseq)=0\}.
\end{align}
Points on the boundary of the causal cone are therefore indexed by $\rank(H) \in \N$.
A moment sequence $\momseq \in \partial \mathcal{K}$ will be referred to as a \emph{singular moment sequence}.

The remainder of this subsection examines singular positive Hankel matrices from two different perspectives.
\Cref{sec:boundary_known} analyzes the rank assuming a moment sequence with known spectral content, which highlights the interplay between Euclidean-time dependence and the choice of operator set.
\Cref{sec:boundary_unknown} considers the general case of a moment sequence with unknown spectral content; this perspective will be useful for noisy moment sequences projected to the boundary.

\subsubsection{Boundary points with known spectral content \label{sec:boundary_known}}

Suppose that a moment sequence $\momseq \in (\H_p)^m$ is given with known spectral decomposition.
The rank of a block Hankel matrix depends on the interplay between the Euclidean-time dependence and the choice of operator set.
Being positive semidefinite,
the Hankel matrix in \cref{eq:Hankel_spectral_decomp} can be written as a Gram matrix
\begin{align}
H   &= \sum_{n=1}^{\Nstates}
\bm{V}_n \bm{V}_n^T \otimes
W_n
    = Q Q^\dagger
    \in \H_{d \Nop}.
\end{align}
Suppose that each state $n$ is non-degenerate so that $\rank W_n=1$.
Then the block matrix $Q \in \C^{d\Nop \times \Nstates}$ is given by
\begin{align}
Q   &\equiv
    \begin{pmatrix}
        \bm{V}_1\otimes \bm{Z}_1 &
        \bm{V}_2\otimes \bm{Z}_2 &
        \cdots                    &
        \bm{V}_{\Nstates}\otimes \bm{Z}_{\Nstates}
    \end{pmatrix}.
\end{align}
As for any Gram matrix, $\rank(H)=\rank(Q Q^\dagger)=\rank(Q)$.
Thus
\begin{align}
    \rank(H) = \rank(Q) \leq \min(\Nop d, \Nstates),
\end{align}
where $\Nstates$ is the number of states in the known spectral decomposition.
The same result also holds for degenerate states provided degeneracies are included in the counting for $\Nstates$.
Equality holds when the columns of $Q$ are linearly independent.
Since the Vandermonde vectors $\bm{V}_n$ defined in \cref{eq:def_vandermonde_vectors} are automatically linearly independent for distinct $\lambda_n$ and $\Nstates \leq d$, the available spectral content ultimately depends on a sufficiently expressive operator set that maximizes $\rank(Q)$.
Note that in the scalar case $\Nop=1$, positivity immediately implies equality:
\begin{align}
\rank(H) = \min(d, \Nstates), && (\Nop=1).
\end{align}

\subsubsection{Boundary points with unknown spectral content \label{sec:boundary_unknown}}

The situation is somewhat richer for a generic singular moment sequence $\momseq \in (\H_p)^m$.
Any positive semidefinite matrix can be written in a factorized form as a Gram matrix, i.e., $H(\momseq)=QQ^\dagger$ as in the previous section.
However, additional structure is necessary to identify an underlying spectral decomposition for $\momseq$.

Note that this structure is automatically present for interior points with $H \succ 0$, but the spectral decomposition is not unique.
This fact agrees with practical experience in numerical lattice QCD, where numerical methods based on the GEVP or Krylov-subspace techniques readily deliver a spectral decomposition that interpolates the input Euclidean data (see \cref{eq:block_Lanczos}) and approximates the true spectral function associated with the exact transfer matrix.

In principle, moment sequences on the boundary of the causal cone require additional care.
Recall that the present setting involves an odd number of moments.
The unique minimal spectral decomposition (with the smallest rank of the associated weight matrix) of the first $(m-1)$ moment blocks can be extracted, for example numerically using Krylov-subspace methods (see \cref{cor:main}(a) in \cref{app:definitions}).
As discussed in \cref{app:spectral_decomp}, the final $m^\mathrm{th}$ moment presents a few different possibilities, although the result can be stated simply.
The boundary moment sequence takes the form
\begin{align}
\momseq &= \sum_{n=1}^{\Nstates} \bm{V}_n \otimes W_n  + \momseq_\Delta,\\
\momseq_\Delta &\equiv (0, \dots, 0, \mu_\Delta),
\end{align}
where the sum on the first line is the spectral decomposition from the first $(m-1)$ moments.
The term $\momseq_\Delta$ is a possible \emph{endpoint excess}, which controls whether or not additional moments can be specified in a manner consistent with causality.
Additional details regarding the endpoint excess are given in \cref{app:spectral_decomp}.
The minimal Krylov representation reproduces all projected moments only when $\momseq_\Delta=0$.
Insofar as the present proposal envisions projecting noisy moment sequences and then applying Krylov methods to the first $m-1$ moments, the possible existence of $\momseq_\Delta$ is a question of principle rather than of practice.

\subsection{Moreau decomposition of acausal data\label{sec:moreau}}

A geometric decomposition of statistical fluctuations arises through the interplay between an inner-product space $((\H_p)^m, \langle\cdot, \cdot \rangle_g)$
and the linear structure of the Hankel map, \cref{eq:hankel_linear_map}, used to define the causal cone in \cref{eq:def_causal_cone}.
The inner product is defined generically as
\begin{align}
\langle \bm{u}, \bm{v} \rangle_\metric = \bm{u}^\dagger g \bm{v}.
\end{align}
It will often be convenient, but not necessary, to define the inner product using an inverse covariance matrix as in \cref{eq:Cov_inner_product}.
Even when $\metric=\Sigma^{-1}$ is chosen, the following geometric construction does not necessarily assume Gaussianity.
Of course, all of the familiar warnings apply in the presence of distributions with multiple length scales, including fat-tailed distributions; practical applications of the resulting geometry must assess carefully whether $\Sigma^{-1}$ provides the natural notion of distance in a given problem.
Gaussianity will be assumed in \cref{sec:statistics} below when analyzing properties of random variables under this geometry.

Given this inner product, the polar causal cone is defined via
\begin{align}
    \mathcal{K}^\circ \equiv \{ \momseq' \in (\H_p)^m : \langle \momseq, \momseq' \rangle_\metric \leq 0 \text{ for all } \momseq \in \mathcal{K}\}.
    \label{eq:polar_cone_definition}
\end{align}
Note well that the definition of $\polar{\mathcal{K}}$ includes both the Hankel linear map $H(\cdot)$ and the inner product $\langle \cdot, \cdot \rangle_\metric$.\footnote{
A closely related \emph{dual cone} can be defined with the opposite sign for the inequality.
The dual cone can be understood as the space of positive polynomials~\cite{Nesterov2000}.
}
As expected for a cone, the definition is independent of an overall positive rescaling of the inner product.
When $\metric=\Sigma^{-1}$, this says that the conic geometry is independent of the overall statistical (``$\sqrt{\Nsample}$'') scaling.
Roughly speaking, the polar cone $\polar{\mathcal{K}}$ is the space of acausal fluctuations.

\begin{figure}[t!]
    \centering
    \includegraphics[width=0.99\linewidth]{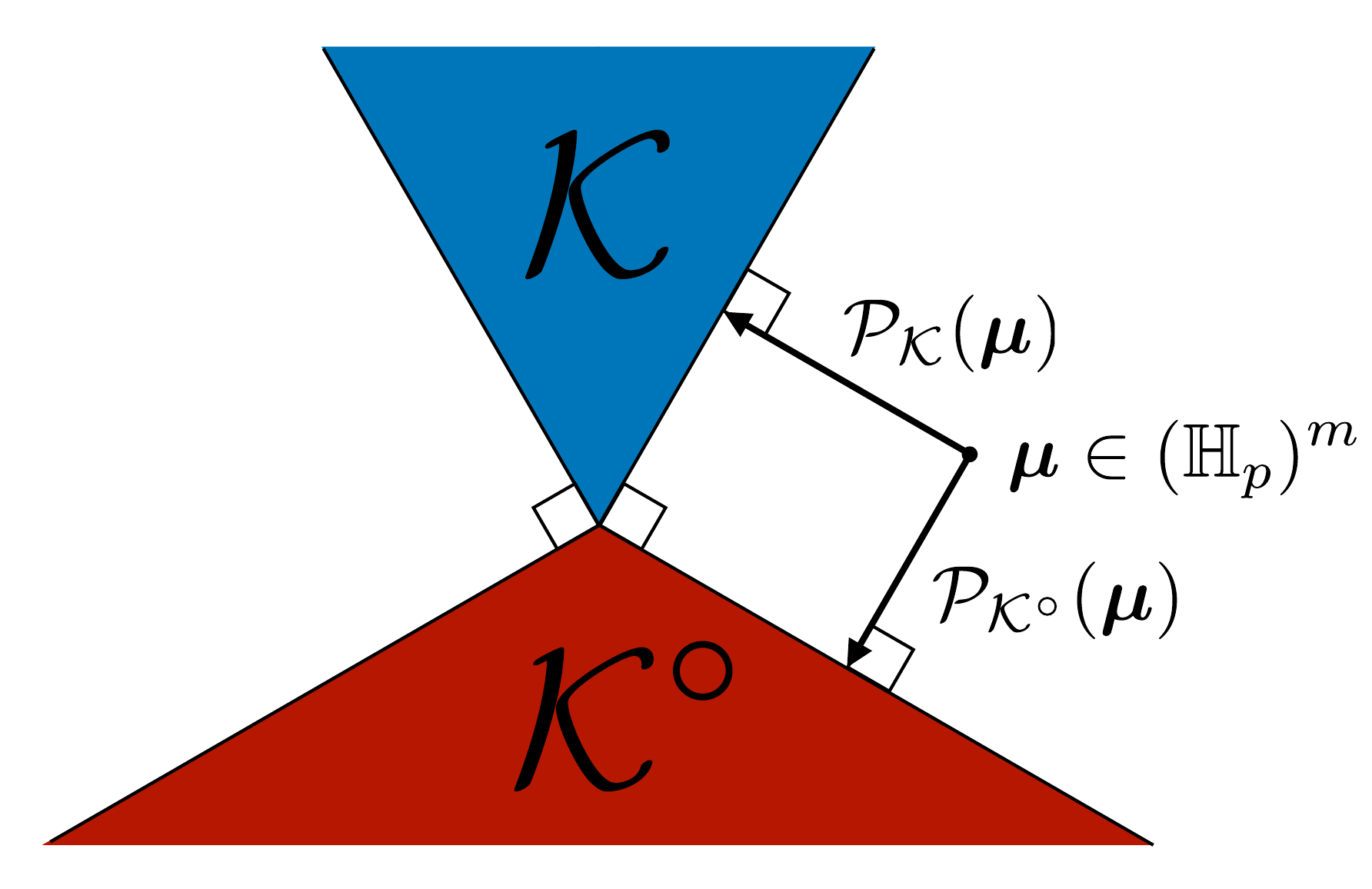}
    \caption{
    Schematic diagram of a generically acausal $\momseq\in(\H_p)^m$ under Moreau projection $\momseq = \projection{\mathcal{K}}(\momseq) + \projection{\polar{\mathcal{K}}}(\momseq)$ onto the boundaries of the causal cone $\mathcal{K}$ and the polar cone $\polar{\mathcal{K}}$, adapted from Fig.~1 of Ref.~\cite{dimiccoli2016fundamentalsconeregression}.
    The causal cone $\mathcal{K}$ is defined by \cref{eq:def_causal_cone}.
    The geometry is specified by the right angles between $\mathcal{K}$ and $\polar{\mathcal{K}}$, which are defined with respect to a chosen inner product.
    In this work, the inner product is induced by the covariance matrix via \cref{eq:Cov_inner_product}.
    \label{fig:moreau}
    }
\end{figure}

As usual, the existence of an inner product provides the means to decompose larger spaces into smaller ones.
For cones, the relevant decomposition is known as the Moreau decomposition.
The Moreau decomposition~\cite{moreau1962decomposition,dimiccoli2016fundamentalsconeregression} of a generic $\momseq \in (\H_p)^m$ is given by
\begin{align}
    \momseq
    &=
    \projection{\mathcal{K}}(\momseq) + \projection{\polar{\mathcal{K}}}(\momseq),
\end{align}
where the projection mappings are defined via
\begin{align}
    \begin{split}
    \projection{\mathcal{K}}(\momseq)
    =&
    \argmin_{\momseq'\in (\H_\Nop)^m} \lVert\momseq -\momseq' \rVert^2_\metric  \,\st \momseq' \in \mathcal{K},
    \end{split}
    \label{eq:projection_K}
    \\
    \begin{split}
    \projection{\polar{\mathcal{K}}}(\momseq)
    =&
    \argmin_{\momseq'\in (\H_\Nop)^m} \lVert\momseq -\momseq' \rVert^2_\metric \, \st \momseq' \in \polar{\mathcal{K}}.
    \end{split}
    \label{eq:projection_Kpolar}
\end{align}
\Cref{fig:moreau} schematically illustrates the geometry of the Moreau projection.
Given the geometry in \cref{fig:moreau}, the Moreau projection is clearly optimal in the sense that $\projection{\mathcal{K}}(\momseq)$ is guaranteed by \cref{eq:projection_K} to be the point most consistent with the starting data for the chosen inner product.

It is useful to record a few further properties enjoyed by the projection mappings.
Although nonlinear, they are idempotent and orthogonal in the sense that
\begin{align}
    \projection{\mathcal{K}} \projection{\polar{\mathcal{K}}}(\momseq) &=
    \projection{\polar{\mathcal{K}}} \projection{\mathcal{K}}(\momseq)
    = 0.
\end{align}
Points outside the causal cone are projected to the boundaries
\begin{align}
    &\projection{\mathcal{K}}(\momseq) \in \partial \mathcal{K}
    &&
    \text{for all }\momseq \not\in \mathcal{K},
    \\
    &\projection{\polar{\mathcal{K}}}(\momseq) \in \partial \polar{\mathcal{K}}
    &&
    \text{for all }\momseq \not\in \polar{\mathcal{K}}.
\end{align}
Causal elements are fixed points satisfying
\begin{align}
    \projection{\mathcal{K}}(\momseq) = \momseq,
    &&
    \text{for all }\momseq \in \mathcal{K}.
\end{align}

Elements of the polar cone can be parameterized in terms of a suitable dual moment sequence $\dualmomseq \in \H_{dp}$.
To see this, let $\momseq' \in \polar{\mathcal{K}} \subset (\H_p)^m$.
By definition of the polar cone, $\langle \momseq', \momseq \rangle_g \leq 0$ for all $\momseq \in \mathcal{K}$.
That is, $\momseq'^\dagger g \momseq \leq 0$.
Since the linear map $\bm{h}: \H_{dp} \to (\H_p)^m$ is surjective, there exists $M \in \H_{dp}$ such that
\begin{align}
\momseq' = -g^{-1} \bm{h}(M) \in \polar{\mathcal{K}}, && M \succeq 0.
\end{align}
The positive semidefinite condition is tied to the definition of the polar cone.
That is, with $\momseq \in \mathcal{K}$ the defining condition reads:
\begin{align}
0
&\leq -\momseq'^\dagger g \momseq
=
\bm{h}^\dagger(M)g^{-1} g \momseq
= \bm{h}^\dagger(M) \momseq\\
&= \Tr[ M^\dagger H(\momseq)].
\end{align}
This condition is satisfied by any $M \succeq 0$.
Conversely, although $M$ is only defined up to elements in the kernel of $\bm{h}$, a positive semidefinite representative always exists.
As the next section shows, it is the optimal Lagrange multiplier that leads to the dual formulation of the problem.
Its existence is related to the fact that a non-negative polynomial in one variable can be written as a sum of squares (cf.\ Theorem~17.1 of Ref.~\cite{Nesterov2000}).

\subsection{Primal-dual formulation of Moreau projection \label{sec:primal-dual}}

This section shows how the projection mappings defined in \cref{eq:projection_K,eq:projection_Kpolar} may be understood in terms of the primal-dual formulation of \cref{eq:projection_K}.
For concreteness, this section considers the Hamburger moment problem, for which the fundamental problem is the following convex quadratic program:
\begin{align}
\begin{split}
    \minimize_{\momseq\in (\H_p)^m} &\quad \primal(\momseq) \equiv\frac{1}{2} \chi^2(\hat{\momseq}-\momseq)\\
    \st &\quad H(\momseq) \succeq 0,
\end{split}
\label{eq:primal_problem}
\end{align}
where $\hat{\momseq} \in (\H_p)^m$ is a fixed acausal input with $H(\hat{\momseq}) \not\succeq 0$.
A conventional factor of $1/2$ is included to increase the similarity to the dual formulation below.

\Cref{eq:primal_problem} will be called the primal problem, and $\primal(\momseq)$ is the primal objective function.
The semidefinite constraint $H(\momseq)\succeq 0$ can be enforced using a Lagrange multiplier $\dualmomseq \in \H_{d\Nop}$, leading to the Lagrange function
\begin{align}
    \mathcal{L}(\momseq, \dualmomseq)
    &=
    \frac{1}{2} \chi^2(\hat{\momseq}- \momseq) - \Tr[\dualmomseq^\dagger H(\momseq)] \\
    &= \frac{1}{2} \chi^2(\hat{\momseq}- \momseq) - \momseq\cdot\bm{h}(\dualmomseq).
\end{align}
Minimizing the Lagrangian over $\momseq$ at fixed $M \succeq 0$ gives $\Sigma^{-1}(\momseq - \hat\momseq) = \bm{h}(M)$ and yields the dual problem:
\begin{align}
\begin{split}
    \maximize_{\dualmomseq \in \H_{d\Nop}}
    \quad
    & \dual(\dualmomseq) = -\frac{1}{2} \bm{h}^T(\dualmomseq) \Sigma \bm{h}(\dualmomseq) - \bm{h}^T(\dualmomseq) \hat{\momseq} \\
    \text{subject to}
    \quad
    & \dualmomseq \succeq 0.
\end{split}
\end{align}
The quantity $\dual(\dualmomseq)$ is called the dual objective function.
In convex optimization, strong duality occurs when the values of the primal and dual objective function agree at the optimum.
Points $(\momseq, \dualmomseq) \in (\H_\Nop)^m \times \H_{d\Nop}$ with $H(\momseq)\succeq 0$ and $\dualmomseq\succeq 0$ are called feasible.
Provided a strictly feasible point exists, Slater's condition guarantees that strong duality holds.

The difference between the primal and dual objective functions is known as the duality gap.
For any feasible $(\momseq, \dualmomseq)$, the duality gap is non-negative, since
\begin{align}
    \gap(\momseq, \dualmomseq)
    &= \primal(\momseq) - \dual(\dualmomseq)\\
    &= \frac{1}{2} \bm{r}^T(\momseq,\dualmomseq) \Sigma \bm{r}(\momseq,\dualmomseq) + \langle \dualmomseq, H(\momseq) \rangle \geq 0,
\end{align}
with the second line the result of a brief calculation using
$\bm{r}(\momseq, \dualmomseq)
\equiv \Sigma^{-1} (\momseq-\hat{\momseq}) - \bm{h}(\dualmomseq)$,
which is a residual measuring the distance from stationarity, since $\partial\mathcal{L}/\partial\momseq=0$ is realized at $\bm{r}=0$.

Primal-dual interior-point algorithms iteratively construct the solution to the primal and dual problems, starting from a feasible point and remaining in the interior of the feasible set at each iteration.
The particular approach to the solution depends on the algorithm.
The central path is defined by the so-called perturbed complementarity condition
\begin{align}
    \dualmomseq H(\momseq) = \centralpath \identity,
    && \centralpath \geq 0.
    \label{eq:perturbed complementarity}
\end{align}
Together with stationarity and primal-dual feasibility, \cref{eq:perturbed complementarity} defines the strictly feasible central path to the global optimum as $\centralpath\to 0$.

Convex quadratic problems can also be reduced to SDPs.
Reduction of \cref{eq:opt-problem-conic-projection} to an SDP follows by introducing a slack variable $\chi^2_\text{cut}$, yielding an equivalent minimization problem\footnote{This reduction holds for both scalar and matrix-valued moment sequences, provided the latter are suitably vectorized.}
\begin{equation}
    \begin{aligned}
        \minimize_{\momseq \in \R^m, \chi^2_{\rm cut}\in \R} \quad &\chi^2_\text{cut}\\
        \st \quad&  \chi^2(\momseq - \hat{\momseq}) \leq \chi^2_\text{cut}, \\
            \quad& H(\momseq) \succeq 0.
    \end{aligned}
\end{equation}
Using the Schur complement~\cite{Abbott:2026wdw,Mutzel:2026vyw}, the condition $\chi^2(\momseq-\hat{\momseq}) \leq \chi^2_\text{cut}$ can be replaced with the matrix positivity condition:
\begin{equation}
    \Gamma[\momseq, \chi^2_\text{cut}] \equiv \begin{pmatrix}
        \Sigma & (\momseq - \hat{\momseq}) \\
        (\momseq - \hat{\momseq})^\dag & \chi^2_\text{cut}
\end{pmatrix} \succeq 0.
\end{equation}
Since the matrix $\Gamma[\momseq, \chi^2_\text{cut}]$ is linear in both $\momseq$ and $\chi^2_\text{cut}$, the following optimization is an SDP:
\begin{equation}
    \begin{aligned}
        \minimize_{\momseq \in \R^m, \chi^2_{\rm cut}\in \R} \quad& \chi^2_\text{cut} \\
        \st \quad& \Gamma[\momseq, \chi^2_\text{cut}] \succeq 0, \\
            \quad& H(\momseq) \succeq 0.
    \end{aligned}
    \label{eq:sdp_reduction}
\end{equation}
\Cref{eq:sdp_reduction} can be solved conveniently with standard software packages including SDPA-GMP~\cite{SDPA-GMP}, a version of SDPA~\cite{Yamashita:2003:SDPA,Yamashita:2010:SDPA} supporting arbitrary-precision inputs.
Arbitrary-precision arithmetic becomes necessary to avoid numerical difficulties associated with the Hankel positivity constraint in light of Beckermann's bound, \cref{eq:Beckermann}.
Standard interior-point methods for solving SDPs have polynomial complexity~\cite{Alizadeh1995,VandenbergheBoyd1996,wolkowicz2000handbook}.

\subsection{Final formula for the conic decomposition\label{sec:conic_decomposition}}

Given the choice of an inner product $\langle\cdot, \cdot \rangle_g$ used to define the projection mappings, a generic $\momseq \in (\H_{p})^m$ with $m = 2d-1$ can be written uniquely in terms of a conic decomposition as
\begin{subequations}
\label{eq:conic_decomp}
\begin{align}
\momseq
    &= \projection{\mathcal{K}}(\momseq) + \projection{\polar{\mathcal{K}}}(\momseq), \\
\projection{\mathcal{K}}(\momseq)
    &= \sum_{n=1}^{\Nstates} \bm{V}_n \otimes W_n + \momseq_\Delta \in \mathcal{K},\\
\projection{\polar{\mathcal{K}}}(\momseq)
    &= - g^{-1} \bm{h}(M) \in \polar{\mathcal{K}}.
\end{align}
\end{subequations}
\noindent
The weight matrices satisfy $W_n \in \H_p$ and $W_n \succeq 0$.
The Vandermonde vector is $\bm{V}_n = (1, \lambda_n, \lambda_n^2, \dots, \lambda_n^{m-1}) \in \R^m$.
The possible endpoint excess $\momseq_\Delta$ is discussed in \cref{app:spectral_decomp}.
Finally, $M\in \H_{dp}$ is a positive semidefinite dual moment sequence.
This decomposition is a central conceptual result and distinguishing feature of the present work.

Each element of the decomposition has a clear interpretation.
The projection onto $\mathcal{K}$ consists of two terms.
The first term is explicitly causal and consists of a spectral sum with a dynamically determined $\Nstates$.
The second term, which plays no practical role in the present discussion, is a possible endpoint excess that controls whether additional moments can be specified consistent with causality.
The polar component of the Moreau decomposition represents acausal noise and is determined by the nearest dual-feasible obstruction to positivity, i.e., by $M \succeq 0$ satisfying the complementarity condition
\begin{align}
    \dualmomseq H(\mathcal{P}_{\mathcal{K}}(\momseq)) = 0.
\end{align}

\section{Statistical aspects of conic projection \label{sec:statistics}}

The conic geometry defined in the previous section furnishes a decomposition of a given moment sequence into causal and acausal components.
The decomposition is unique, up to the choice of a suitable inner-product structure.
The goal of the present section is to develop a class of statistical estimators based on this decomposition.

To begin, suppose that $\momseq$ is a causal moment sequence satisfying $H(\momseq) \succ 0$ and that $\hat{\momseq}'$ is an unbiased statistical estimator satisfying $\expectation[\hat{\momseq}'] = \momseq$.
Given independent and identically distributed samples $\hat{\momseq}'_i$, the central limit theorem implies that the sample mean will be asymptotically normally distributed,
\begin{align}
\hat{\momseq}_{\Nsample} = \frac{1}{\Nsample} \sum_{i=1}^{\Nsample} \hat{\momseq}'_i
\sim \normal(\momseq, \Sigma/\Nsample).
\label{eq:CLT}
\end{align}
To make contact with the previous section, we will identify the inner product with the one induced by the covariance matrix ($g=\Sigma^{-1}$), assumed to be finite and invertible.
As usual, correct interpretation of practical inference problems becomes more complicated if the normality in \cref{eq:CLT} is relaxed, the treatment of which exceeds the scope of the present work.

\Cref{sec:bayesian} begins with a Bayesian description of the problem and situates conic projection within a particular statistical inference problem.
\Cref{sec:critical_fluctuations} estimates the size of critical statistical fluctuations, above which conic projection---or something akin to it---becomes necessary.
\Cref{sec:SURE} analyzes the conic projection using Stein's unbiased risk estimate.

\subsection{Bayesian interpretation of conic projection \label{sec:bayesian}}

Conic projection has an appealing interpretation in terms of a non-parametric Bayesian inference problem.
The connection arises because the projection mapping $\projection{\mathcal{K}}$ in \cref{eq:projection_K} is equivalent to maximizing the following posterior distribution
\begin{align}
    P(\momseq|\hat{\momseq}_N) \propto L(\momseq;\hat{\momseq}_N) \pi(\momseq),
    \label{eq:posterior}
\end{align}
where $\momseq$ is the optimization variable, $\hat{\momseq}_N$ is some generically acausal input data, and the likelihood and prior are given by
\begin{align}
\begin{split}
    L(\momseq;\hat{\momseq}_N)
    & \equiv
    P(\hat{\momseq}_N|\momseq) \propto \exp\left[ - \frac{1}{2} \chi^2(\momseq-\hat{\momseq}_N) \right],\\
    \pi(\momseq) &= \begin{cases}
        1, & H(\momseq) \succeq 0,\\
        0, & H(\momseq) \not\succeq 0.
    \end{cases}
\end{split}
\label{eq:bayesian_formulation}
\end{align}
The constant prior is improper (in the sense of not being a normalizable probability distribution) on the unbounded cone, but the Gaussian likelihood yields a finite, nonzero posterior normalization; it is not a flat prior on the underlying spectral functions.
The prior in \cref{eq:bayesian_formulation} is appropriate for the Hamburger moment problem; the positivity constraints for the other moment problems are handled analogously.
Note that Gaussianity has been invoked in writing down the likelihood function.

The inference is non-parametric in the sense that no particular finite-state spectral model is required to define or compute the projection mapping, which corresponds to the maximum \emph{a posteriori} (MAP) estimator in Bayesian analysis,
\begin{align}
\hat{\momseq}_{\rm MAP} = \argmax_{\momseq} L(\momseq;\hat{\momseq}_N) \pi(\momseq) = \projection{\mathcal{K}}(\hat{\momseq}_N).
\label{eq:MAP}
\end{align}
It is also the constrained maximum-likelihood estimator for this Gaussian moment model.
Note that this equivalence does not identify causal projection with posterior averaging.
A notable feature of projection mappings---and any inference problems derived from them---is that the number of states emerges dynamically.
Its value is related to the rank of the Hankel matrix at the solution point and will, in general, fluctuate across different realizations of the random variable $\hat{\momseq}_N$.

The MAP estimator does not occupy a fundamental position within Bayesian inference, where instead marginalization over the posterior plays the central role:
\begin{align}
    \avg{\momseq}
    &= \frac{1}{Z} \int d\momseq\, \momseq L(\momseq;\hat{\momseq}_N) \pi(\momseq)\\
    &= \frac{1}{Z} \int_{\mathcal{K}} d\momseq\, \momseq L(\momseq;\hat{\momseq}_N) \in \interior \mathcal{K},
    \label{eq:Bayesian_estimator_mu}\\
    Z &= \int d\momseq\, L(\momseq;\hat{\momseq}_N) \pi(\momseq).
\end{align}
The integrand has support on the causal cone and decays exponentially away from $\hat{\momseq}_N$.
The expectation lies in the interior since the posterior has positive density throughout the convex interior.
In particular, the posterior assigns zero probability to the boundary ($\partial\mathcal{K}$), even when the MAP lies there.

The expression for the posterior distribution in \cref{eq:posterior} reveals a certain conceptual tension in statistical analyses of lattice QCD data.
Since most correlation functions involve arbitrarily many states, the true result is expected to lie in the \emph{interior} of the causal cone.
Although the Monte Carlo estimator $\hat{\bm{\momseq}}_\Nsample$ is statistically unbiased in the sense that $\expectation[\hat{\bm{\momseq}}_\Nsample] =\momseq$, the finite-sample estimator $\hat{\momseq}_N$
essentially always falls
in the \emph{exterior} of the causal cone in practice for realistic sample sizes.
The following subsection makes this intuition precise by estimating the critical sample size necessary to avoid acausal fluctuations.
On the other hand, as mentioned above, $\avg{\momseq} \in \interior \mathcal{K}$.
These technical observations play a crucial role in observables in lattice field theory that require a spectral reconstruction.
For these observables, a fundamental uncertainty exists even for exact data.
This uncertainty arises because the inverse problem lacks a \emph{unique} solution.
Although the solution is not unique, the ambiguity can be characterized completely and computed numerically in many cases of practical interest~\cite{Bergamaschi:2023xzx,Abbott:2025snz,Abbott:2026wdw}.

For such quantities, the key question is how to estimate the integral over the interior of $\mathcal{K}$ in \cref{eq:Bayesian_estimator_mu}.
Ref.~\cite{Fields:2025glg} has used interpolation between boundary points, obtained via a certain gradient ascent procedure applied to the eigenvalues of the Pick matrix, to explore the interior of the causal cone near $\hat{\momseq}_N$.
In this way, interior points can be sampled and used to compute expectation values or Wertevorr{\"a}te as discussed in Ref.~\cite{Fields:2025glg}.
When combined with resampling methods, conic projection gives an alternative method to sample boundary points near $\hat{\momseq}_N$ and, one hopes, near the exact $\momseq$.
A detailed investigation of algorithms for navigation within the causal cone exceeds the scope of the present work.

Whether a projection step---or something like it, perhaps a parametric fit---is necessary depends on the typical size of statistical fluctuations.
The question is whether typical fluctuations violate causality in the sense that $H(\hat{\momseq}_N) \not\succeq 0$.
The next section estimates the size of typical acausal fluctuations.

\subsection{Critical sample size for acausal fluctuations \label{sec:critical_fluctuations}}

Consider the sample mean $\hat{\momseq}_N$ as in \cref{eq:CLT}, written as the sum of the exact result and a Gaussian fluctuation
\begin{align}
    \hat{\momseq}_\Nsample &= \momseq + \delta\hat{\momseq}_\Nsample,\\
    \delta \hat{\momseq}_\Nsample &\sim \normal(0, \Sigma_\Nsample).
\end{align}
Beckermann's bound in \cref{eq:Beckermann} suggests that, for realistically achievable sample sizes $\Nsample$, essentially all fluctuations will in practice
violate \cref{eq:Hamburger_positivity} for moment sequences of even modest length.
This intuition can be quantified by asking what value of $\Nsample$ would be necessary to make violations of \cref{eq:Hamburger_positivity} rare.
An estimate follows from a brief calculation and a simple dimensional argument.
Because the Hankel map in \cref{eq:hankel_linear_map} is linear, familiar perturbative methods can be used to estimate shifts in the eigenvalues that would violate this condition.

Fluctuations of the sample mean that violate \cref{eq:Hamburger_positivity} are related to fluctuations of the smallest eigenvalue of the Hankel matrix $H(\momseq)$, where $\momseq$ is the exact moment sequence.
Let $\lambda_{\rm min}>0$ denote the smallest eigenvalue of
$H(\momseq)$ with a unit eigenvector $\ket{\lambda_{\rm min}}$, which satisfies
\begin{align}
H(\momseq) \ket{\lambda_{\rm min}} = \lambda_{\rm min} \ket{\lambda_{\rm min}}.
\end{align}
First-order perturbation theory shows that a fluctuation $\delta \hat{\momseq}_N = \hat{\momseq}_N - \momseq$
gives rise to a linear shift in the eigenvalue
\begin{align}
    \label{eq:delta_lambda_min}
    \delta \lambda_{\rm min}
    &= \bra{\lambda_{\rm min}} H(\delta\hat{\momseq}_N) \ket{\lambda_{\rm min}}\\
    &= \bm{h}(\Lambda_{\rm min})^\dagger \delta \hat{\momseq}_N,\\
    \Lambda_{\rm min} &\equiv \ket{\lambda_{\rm min}} \bra{\lambda_{\rm min}}.
\end{align}
By linearity, the first-order shift of the eigenvalue is itself a normal random variable:
\begin{align}
    \delta \lambda_{\rm min} &\sim \normal(0, \sigma_{\rm eff}^2/N) \label{eq:delta-lambda-distribution},\\
    \label{eq:sigma_eff_def}
    \sigma_{\rm eff}^2 &\equiv \bm{h}(\Lambda_{\rm min})^\dagger \Sigma\bm{h}(\Lambda_{\rm min}).
\end{align}
Since $\lambda_{\rm min}$ is exponentially small, small fluctuations can make $\lambda_\mathrm{min}+\delta\lambda_\mathrm{min}$ negative.
This quantity is an exact variational upper bound on the smallest eigenvalue of $H(\hat\momseq_N)$, guaranteeing the validity of the linear analysis as a sufficient criterion for loss of positivity.

Violations of positivity become typical when the smallest eigenvalue and its fluctuation are of the same order, $\lambda_{\rm min} \simeq |\delta \lambda_{\rm min}|$.
Since the overall scale of the fluctuations is set by the effective variance $\sigma_{\rm eff}^2/N$, the critical scale can be defined as the dimensionless number
\begin{align}
    \Ncrit & \equiv \frac{\sigma_{\rm eff}^2}{\lambda_{\rm min}^{2}}
    = \left(\frac{\sigma_{\rm eff}^{2}}{\lambda_{\rm max}^{2}}\right) \kappa(H(\momseq))^{2}
    \label{eq:Ncrit}
    \\
    & \geq \left(\frac{\sigma_{\rm eff}}{\lambda_{\rm max}}\right)^{2} \frac{\gamma^{2(d-1)}}{256 d^2},
    \label{eq:Ncrit_bound}
\end{align}
which follows from the definition of the condition number $\kappa(H(\momseq)) \equiv \lambda_{\rm max}/\lambda_{\rm min}$ and from \cref{eq:Beckermann}.
When $N \gg N_{\rm crit}$, acausal fluctuations become rare.
Absent an exponential decrease in $\sigma_{\rm eff}/\lambda_{\rm max}$, the critical sample size will increase exponentially in $d$.
In practice, the largest eigenvalue $\lambda_{\rm max}$ often varies slowly with $d$ compared to the exponential behavior of $\lambda_{\rm min}$.

This bound on $\Ncrit$ provides an interesting perspective on the success of variational methods for spectroscopy, which employ a large set of operators $\{\hat{O}_1, \hat{O}_2, \dots, \hat{O}_{\Nop}\}$ to constrain the spectrum (cf.~\cref{eq:two_point_function}).
In the present context, the main idea is that, for a fixed total size $(\Nop d={\rm const})$, the condition number of the block Hankel matrix $\kappa(H)$ may plausibly be reduced by increasing the number of operators $\Nop$ and using fewer Euclidean times to decrease $d$.

This line of reasoning agrees with practical experience in lattice QCD spectroscopy calculations phrased in terms of a GEVP (see \cref{sec:GEVP} above).
State-of-the-art calculations utilize many dozens of discrete finite-volume energy levels.
Consistent with the conventional wisdom in the literature on lattice field theory, the preceding discussion argues strongly that such results would be practically impossible using scalar ($\Nop=1$) correlation functions, requiring exponentially large Monte Carlo statistics, possibly further worsened by signal-to-noise degradation.
On the other hand, a large variational set of operators does not necessarily imply exponential scaling of $\kappa(H)$ with $\Nop$, particularly if the operators in the set are carefully chosen.

\subsection{Stein's unbiased risk estimate \label{sec:SURE}}

In this subsection, $\Sigma$ denotes the covariance $\Sigma_N$ of the sample mean, treated as known.
Since the Moreau decomposition is invariant under an overall rescaling of the covariance matrix, the overall geometry is unaffected by the particular value $N$.
For simplicity, consider a scalar ($p=1$) moment sequence.
For a given sample mean $\hat{\momseq}_N$, the expected distance of the projection from the true mean $\momseq$ is given by
\begin{align}
    \expectation[&\norm{\projection{\mathcal{K}}(\hat{\momseq}_N)-\momseq}^2_{\Sigma^{-1}}]
    =
    \expectation[\norm{\hat{\momseq}_N - \projection{\mathcal{K^\circ}}(\hat{\momseq}_N)-\momseq}^2_{\Sigma^{-1}}]\\
    \begin{split}
    =&\expectation[
    \norm{\projection{\mathcal{K^\circ}}(\hat{\momseq}_N)}^2_{\Sigma^{-1}}
    ]
    - 2\,\expectation[\projection{\mathcal{K^\circ}}(\hat{\momseq}_N)^T \Sigma^{-1} (\hat{\momseq}_N-\momseq)]\\
    &\phantom{\expectation[
    \norm{\projection{\mathcal{K^\circ}}(\hat{\momseq}_N)}^2_{\Sigma^{-1}}
    ]}
    + \expectation[\norm{\hat{\momseq}_N-\momseq}^2_{\Sigma^{-1}}]
    \end{split}\\
    =& \expectation[ \norm{\projection{\mathcal{K^\circ}}(\hat{\momseq}_N)}^2_{\Sigma^{-1}}]
    - 2\,\expectation[\nabla \cdot \projection{\mathcal{K^\circ}}(\hat{\momseq}_N)]
    + m.
    \label{eq:stein_sure}
\end{align}
The first equality has used the Moreau decomposition.
The second equality has expanded the norm.
The third equality has used integration by parts for the middle term and evaluated the last term using the probability density function for $\hat{\momseq}_N \sim \normal(\momseq, \Sigma)$.
In the statistics literature, \cref{eq:stein_sure} is known as Stein's unbiased risk estimate (SURE)~\cite{Stein:1981}.
The final result should be compared to the expected distance of the sample mean itself to the true mean, $\expectation[\norm{\hat{\momseq}_N-\momseq}^2_{\Sigma^{-1}}]=m$.
A SURE value below $m$ estimates a reduction in mean squared error relative to $\hat{\momseq}_N$; the projected value is \emph{closer} to the true result.
A well-known but remarkable feature of SURE is that it can be computed without knowledge of the exact $\momseq$.
As shown in \cref{sec:divergence}, the divergence in \cref{eq:stein_sure} can be computed numerically in the setup described in \cref{sec:primal-dual}.

Limiting cases of \cref{eq:stein_sure} are enlightening.
In the large-statistics limit where $N\gg \Ncrit$ and $\hat{\momseq}_N \in \mathcal{K}$, the polar component vanishes identically, as does the middle term, which counts the dimension of the polar projection.
One thus recovers the expected result:
\begin{align}
    \expectation[&\norm{\projection{\mathcal{K}}(\hat{\momseq}_N)-\momseq}^2_{\Sigma^{-1}}]
    = m, \text{ for } N \gg \Ncrit.
\end{align}
For the noisy case $N < \Ncrit$, it is useful to imagine the counterfactual case of \emph{linear} projection operators $\Pi_{\mathcal{K}}, \Pi_{\polar{\mathcal{K}}}$ projecting onto linear subspaces instead of cones.
Since the true central values are assumed causal, $\momseq\in\mathcal{K}$, the expectation values in \cref{eq:stein_sure} would both evaluate to $\rank \Pi_{\mathcal{K^\circ}}$, giving
\begin{align}
\begin{split}
\expectation[\norm{\Pi_{\mathcal{K}}(\hat{\momseq}_N)-\momseq}^2_{\Sigma^{-1}}]
\xrightarrow{\substack{\text{fictitious linear}\\\text{projection}}}
 m &- \rank \Pi_{\mathcal{K^\circ}}\\
 &= \rank \Pi_{\mathcal{K}},
\end{split}
\end{align}
which shows that projection removes the noisy directions, bringing the causal part closer on average to the true mean.
Moreover, $\rank \Pi_{\mathcal{K}}$ gives a natural count for the degrees of freedom.
In the actual nonlinear case at hand, \cref{eq:stein_sure} represents the natural generalization relevant for the Moreau decomposition.
In particular, gradients of the projection mappings define effective dimensions satisfying $m = \nabla \cdot \projection{\mathcal{K}}(\hat{\momseq}_N) + \nabla \cdot \projection{\polar{\mathcal{K}}}(\hat{\momseq}_N)$ at any sample size $N$.
\section{Connections to the causal bootstrap}\label{sec:causal_bootstrap}

Conic projection identifies the causal moment sequence closest to the measured data.
In many applications, the final quantity of interest is neither the moment sequence nor the underlying spectral function but rather a smeared spectral observable, $\rho[\mathscr{K}]$, as in \cref{eq:rho_kappa} for a given smearing kernel $\mathscr{K}(\omega)$.
In the spirit of Refs.~\cite{Abbott:2026wdw,Lawrence:2024hjm,Lawrence:2025nmh,Mutzel:2026vyw}, it is possible to use the same conic geometry to construct bounds on $\rho[\mathscr{K}]$, optimizing over causal moment sequences compatible with data with a given statistical tolerance.
To explain this connection, \cref{sec:finite-dim} begins by discussing the close similarity between the finite-dimensional reduction in Ref.~\cite{Abbott:2026wdw} and the optimization over moment variables in the present work.
Next, \cref{sec:calibration} shows how the conic projection can be used to calibrate the statistical bounds in the causal bootstrap to construct observable-specific confidence sets with approximately calibrated coverage.

\subsection{Finite-dimensional reduction
\label{sec:finite-dim}}

A set of optimization problems similar to those discussed so far for conic projection also enters the causal bootstrap described in Ref.~\cite{Abbott:2026wdw}.
In that context, one performs a finite-dimensional reduction by first dualizing \cref{eq:opt-problem-def-inexact} and then converting the dual problem to an SDP using the fact that a polynomial (in a single variable) is non-negative if and only if it can be written as a sum of squares.
The equivalence between causal projection and the causal bootstrap at $\chi_\mathrm{cut}^2=\chi^2_{\rm min}$ can be seen as a consequence of the fact that the moment cone and the cone of positive polynomials are dual to each other~\cite{Nesterov2000}.

Consider for simplicity a scalar Hamburger moment problem ($I = \R$) with exact data.
The causal bootstrap solves problems of the form
\begin{equation}
\label{eq:direct-primal-opt-rho}
\begin{aligned}
\minimize_{\tilde{\rho}} & \int_\R d\kappavar\, \mathscr{K}(\kappavar) \tilde\rho(\kappavar) \\
  \st& ~ \mu_t^{(\rho)} = \hat{\mu}_{t} \quad \forall t \in \Tmeas \\     & ~ \tilde\rho(\kappavar) \geq 0 \quad \forall \kappavar \in \R,
\end{aligned}
\end{equation}
where $\Tmeas$ denotes the set of time slices on which data $\hat\mu_t$ have been measured and $\momseq^{(\rho)}$ is the correlator expressed as a moment sequence associated with the spectral function, see \cref{eq:C_lambda_t}, with components $\mu_t^{(\rho)}$.
For notational brevity, as done in \cref{eq:direct-primal-opt-rho}, the smearing kernel $\mathscr{K}(\omega) \equiv \mathscr{K}(\kappavar(\omega))$ will be written simply as $\mathscr{K}(\lambda)$.

Further suppose that the kernel is polynomial in the variable $\kappavar=\kappavar(\omega)$, i.e., $\mathscr{K}(\kappavar) = \mathscr{K}_0 + \mathscr{K}_1 \kappavar + \dots + \mathscr{K}_n \kappavar^n$.
In this case the objective of \cref{eq:direct-primal-opt-rho} can be rewritten as
\begin{equation}
    \label{eq:polynomial-kernel-expand-corr}
    \int_\R d\lambda\, \mathscr{K}(\kappavar) \tilde\rho(\kappavar) = \sum_{t = 0}^n \mathscr{K}_t \mu_t^{(\rho)}.
\end{equation}
Applying \cref{eq:polynomial-kernel-expand-corr} to \cref{eq:direct-primal-opt-rho}, the optimization problem depends on $\rho$ only through the finite set of moments $\mu_t^{(\rho)}$.
Writing the $\mathscr{K}_t$ as the components of a vector then gives the lower bound:
\begin{equation}\label{eq:CB_moments}
    \begin{aligned}
    \minimize_{\momseq }
    &~ \bm{\mathscr{K}}^\dagger \momseq \\
      \st&~\mu_t = \hat\mu_t \quad \forall t \in \Tmeas \\
        &H(\momseq) \succeq 0,
    \end{aligned}
\end{equation}
which is essentially \cref{eq:primal_problem}, except with a linear objective and $\chi_{\rm cut}^2=0$.
As usual, an upper bound can be obtained by maximization.
Note that the optimization vector includes the moments required by the data, the polynomial kernel, and the Hankel constraints; only the measured components are fixed.
If the kernel involves unmeasured moments, the bounds may retain nonzero width even for exact data.
The point is that infinite-dimensional optimization over the space of positive spectral functions is reduced to finite-dimensional optimization over moment sequences subject to a linear constraint~\cite{Lawrence:2024hjm}.
In the language of Ref.~\cite{Abbott:2026wdw}, this provides a direct translation of the original infinite-dimensional primal problem into the finite-dimensional primal SDP.
As in Ref.~\cite{Abbott:2026wdw}, the finite-dimensional reduction here can also be straightforwardly generalized to piecewise rational $\mathscr{K}$.

For noisy data, the interpolating condition in \cref{eq:CB_moments} can be relaxed as in \cref{eq:opt-problem-def-inexact}, effectively enlarging the feasible set.
The next section builds on these ideas, showing how enlarging the feasible set by considering a change in the chi-squared around $\chi^2_{\rm min}$ can be used to define causal-bootstrap bounds with a particular frequentist statistical interpretation.

\subsection{Statistical calibration\label{sec:calibration}}

The preceding section has discussed the connection between the finite-dimensional reduction in the causal bootstrap and moment sequences.
In this spirit, and given the discussion in \cref{sec:positivity}, the causal cone defined in \cref{eq:def_causal_cone} can be equivalently understood (up to singular points, as explained in \cref{app:spectral_decomp}) as the space of moment sequences arising from an underlying positive spectral function
\begin{align}\label{eq:cone_free_q}
    \mathcal{K} = \{ \momseq^{(\rho)} \,:\, \rho \succeq 0 \},
\end{align}
where the support of $\rho$ for particular moment problems is left implicit.

To make the connection with the causal bootstrap, consider a smearing kernel $\mathscr{K}$ and impose, for some proposed value $q$, a further constraint that $\rho[\mathscr{K}]=q$.
This constraint restricts the feasible set, giving rise to
\begin{equation}\label{eq:cone_fixed_q}
    \mathcal{K}_q = \left\{ \momseq^{(\rho)}: \rho \succeq 0,\ \rho[\mathscr{K}] = q \right\} .
\end{equation}
These two convex sets lead to two natural projections: the ordinary causal projection onto $\mathcal{K}$, and the projection onto the restricted set $\mathcal{K}_q$.
Define the squared distances from the data as
\begin{align}\label{eq:chi2_min_cb}
    \chi_\mathrm{min}^2 &= \min_{\momseq\in\mathcal{K}} \Vert \hat\momseq_N - \momseq \Vert_{\Sigma_N^{-1}}^2 \,, \\
    \label{eq:chi_q}
    \chi_q^2 &= \min_{\momseq\in\mathcal{K}_q} \Vert \hat\momseq_N - \momseq \Vert_{\Sigma_N^{-1}}^2
\end{align}
and assume that both minima exist.
The difference
\begin{equation}\label{eq:Delta_chi2q}
    \Delta\chi^2_q \equiv \chi_q^2 - \chi_\mathrm{min}^2 \geq 0
\end{equation}
measures the additional $\chi^2$ penalty required to enforce that the causal sequence reproduce $\rho[\mathscr{K}]=q$, relative to the ordinary projection that removes the noisy component of the data incompatible with positivity.
Note that $\Delta\chi^2_q$ is a random variable because it depends on $\hat{\momseq}_N$.
Using the Gaussian likelihood defined in \cref{eq:bayesian_formulation}, \cref{eq:Delta_chi2q} becomes
\begin{equation}
    \Delta\chi^2_q = -2 \log \frac{\max_{\momseq \in \mathcal{K}_q} L(\momseq; \hat\momseq_N)}{\max_{\momseq \in \mathcal{K}} L(\momseq; \hat\momseq_N)} \,.
\end{equation}
Thus, $\Delta\chi^2_q$ is a profile-likelihood ratio statistic.

For any fixed tolerance $c\geq0$, the preceding definitions give the following equivalent formulation of
the causal-bootstrap bounds~\cite{Lawrence:2024hjm,Lawrence:2025nmh,Abbott:2026wdw,Mutzel:2026vyw} in \cref{eq:opt-problem-def-inexact}:
\begin{equation}\label{eq:CB_equivalence}
\begin{aligned}
    \mathcal{I}(c) &\equiv
    \left\{ q: \, \Delta\chi^2_q \leq c \right\} \\
    &= \left\{
        \rho[\mathscr{K}] : \rho \succeq 0 \,, \, \Vert \hat\momseq_N - \momseq^{(\rho)} \Vert_{\Sigma_N^{-1}}^2 \leq \chi_\mathrm{min}^2 + c
    \right\}.
\end{aligned}
\end{equation}
The notation $\mathcal{I}(c)$ emphasizes that the bounds form an interval that depends on the tolerance $c$.
Indeed, a value $q$ lies in this set precisely when at least one causal spectral function with $\rho[\mathscr{K}]=q$ satisfies the stated bound.
The set in the second line is manifestly an interval, being the image of a convex set under a linear map, and its endpoints can be computed via causal-bootstrap optimizations.
Note that \cref{eq:CB_equivalence} is merely an identity between optimization problems; it does not yet attach a confidence level to a given choice of $c$.

A confidence level arises through the choice of $c$ in the first line of \cref{eq:CB_equivalence}.
The statistic $\Delta\chi^2_q$ measures how much the description of the data worsens when the observable is fixed to $q$, with large values serving as evidence against the hypothesis $\rho[\mathscr{K}]=q$.
The condition $\Delta\chi^2_q \leq c$ is therefore a  hypothesis test, and the interval $\mathcal{I}(c)$ collects the values of $q$ that survive it.
A set constructed in this way, by inverting a family of hypothesis tests, has coverage at least $1-\alpha$ provided each test rejects a true hypothesis with probability at most $\alpha$, for every positive true moment sequence.
In other words, $\Delta\chi^2_q \leq c$ must hold with probability at least $1-\alpha$ whenever $\rho[\mathscr{K}]=q$ is true (see, e.g., Chapter 22 of Ref.~\cite{Stuart1999} or Chapter 40 of Ref.~\cite{ParticleDataGroup:2020ssz}; for recent developments in constrained inverse problems, see Refs.~\cite{Batlle2024,Stanley2026}).
Choosing $c$ according to this goal can be accomplished in several ways.

A conservative choice, already discussed in Refs.~\cite{Lawrence:2024hjm,Abbott:2026wdw}, retains $q$ whenever $\chi^2_q \leq \chi^2_{m,1-\alpha}$, where $m$ is the number of measured real components (time slices in the scalar case) and $\chi^2_{m,1-\alpha}$ is the $(1-\alpha)$ quantile of the chi-squared distribution with $m$ degrees of freedom.
In the notation of \cref{eq:CB_equivalence}, this corresponds to the data-dependent but observable-independent cutoff $c = \chi^2_{m,1-\alpha} - \chi^2_\mathrm{min}$.
For Gaussian data with known covariance, this choice guarantees coverage of at least $1-\alpha$ by constructing the confidence region for all $m$ measurements.
When targeting a single scalar observable, this procedure can give an unnecessarily large interval.
For this reason, confidence intervals resulting from this choice of cutoff will be called \emph{uncalibrated confidence intervals}.

Here we instead propose to determine the distribution of $\Delta\chi_q^2$ under the hypothesis that the true spectrum satisfies $\rho[\mathscr{K}]=q$, and to take a $q$-dependent cutoff $c_{1-\alpha}(q)$ to be its $(1-\alpha)$ quantile.
Values of $q$ satisfying $\Delta\chi_q^2 \leq c_{1-\alpha}(q)$ are retained.
We distinguish two scenarios.

\emph{(a) Smearing kernel in the finite measured span.}
Suppose that, for a fixed nonzero coefficient vector $\bm{g}$,
\begin{equation}\label{eq:kernel_span_N}
    \mathscr{K}(\omega) = \sum_{t=0}^{m-1} g_t b_t(\lambda(\omega)) \,, \quad
    \mu_t = \int_I d\lambda\, b_t(\lambda) \tilde\rho(\lambda) .
\end{equation}
Here, $b_t(\lambda(\omega))$ are the functions that define the spectral representation of the correlator, for instance $b_t(\lambda(\omega)) = \lambda(\omega)^t = e^{-\omega t}$ in the zero-temperature Wilson case; see \cref{eq:Laplace_transform}.\footnote{
    When $\rho$ and $\momseq$ are matrix-valued, so is $\mathscr{K}$, and trace pairing in operator space is implicit in the equations above; see Ref.~\cite{Abbott:2026wdw}. In this case, the coefficients $g_t \in \mathbb{H}_p$ are $p\times p$ Hermitian matrices.
}
This hypothesis is more restrictive than that discussed around \cref{eq:polynomial-kernel-expand-corr}, limiting the span to the measured basis functions only.
Integrating both sides of \cref{eq:kernel_span_N} against $\tilde\rho(\lambda)$ yields the exact representation $\rho[\mathscr{K}] = \sum_{t=0}^{m-1} g_t \mu_t = \bm{g}^T \momseq$.
Hence,
\begin{align}
    \mathcal{K}_q = \mathcal{K} \,\cap\, \{\momseq: \bm{g}^T \momseq = q\}.
\end{align}
That is, fixing $q$ imposes one linear constraint and thus locally removes one free moment direction at an interior point of $\mathcal{K}$.
Minimizing the distance to $\hat{\momseq}_N$ subject only to the linear constraint $\bm{g}^T \momseq = q$ gives
\begin{equation}
    \momseq_q = \hat\momseq_N + \frac{\Sigma_N \bm{g}}{\bm{g}^T \Sigma_N \bm{g}} \left( q-\bm{g}^T \hat\momseq_N \right) \,.
\end{equation}
This vector solves the constrained causal projection in \cref{eq:chi_q} whenever it belongs to $\mathcal{K}$.
If both $\momseq_q$ and $\hat\momseq_N$ belong to $\mathcal{K}$, then $\chi_\mathrm{min}^2=0$ (i.e., the causal projection acts trivially as the identity) and
\begin{equation}\label{eq:Delta_chi2_q_solution}
    \Delta\chi^2_q = \frac{(\bm{g}^T \hat\momseq_N - q)^2}{\bm{g}^T \Sigma_N \bm{g}} \,.
\end{equation}
For a fixed number of moments and a true moment sequence in the interior of $\mathcal{K}$,
sufficiently small changes of its moments remain causal; cf.~\cref{sec:critical_fluctuations}.
At the true value $q=\rho_\mathrm{true}[\mathscr{K}]=\bm{g}^T \momseq_\mathrm{true}$, both $\hat{\momseq}_N$ and $\momseq_q$ then lie in $\mathcal{K}$ with probability tending to one as the number of samples $N$ increases.
Moreover, $\bm{g}^T (\hat\momseq_N - \momseq_\mathrm{true}) / \sqrt{\bm{g}^T \Sigma_N \bm{g}}$ is asymptotically a standard normal random variable, so $\Delta\chi^2_q$ in \cref{eq:Delta_chi2_q_solution} is asymptotically chi-squared with one degree of freedom, $\Delta\chi^2_q \sim \chi_1^2$.
This is a simple instance of Wilks' theorem~\cite{Wilks:1938dza}; for the use of such profile-likelihood tests in high-energy physics, see Refs.~\cite{Cranmer:2003vt,Cranmer:2005hi,Cowan:2010js}.
The corresponding asymptotic cutoff is the $(1-\alpha)$ quantile of $\chi_1^2$.
Confidence intervals resulting from this choice of cutoff will be called \emph{asymptotic confidence intervals}.
The associated observable-independent calibration, $\Delta\chi^2_q \leq \chi_{1,1-\alpha}^2$, will be called \emph{asymptotic calibration}.

In practice, the regime $N \gg \Ncrit$ of \cref{sec:critical_fluctuations} is rarely reached, so that $\hat{\momseq}_N$ is typically acausal and the $\chi_1^2$ law is only approximate even for kernels in the measured span.
\Cref{ssec:numerics_cb_calibration} investigates calibration in this noisy regime for a kernel outside the measured span.

\emph{(b) General smearing kernels.}
If the kernel is not in the measured span, fixing the target $\rho[\mathscr{K}]=q$ need not remove a moment direction, since the measured moments need not determine $\rho[\mathscr{K}]$ uniquely.
This is just a hint of the more familiar fact that a finite set of moments generally admits a nonzero interval of values for a smeared spectral observable, enclosed within the bounds of a Wertevorrat~\cite{Bergamaschi:2023xzx,Abbott:2025snz,Abbott:2026wdw}.
In the general situation where \cref{eq:kernel_span_N} does not hold exactly, the distribution of $\Delta\chi_q^2$ can differ from $\chi_1^2$.

In either scenario, therefore, the distribution of $\Delta\chi_q^2$ under the hypothesis $\rho[\mathscr{K}]=q$ is not available analytically.
For general kernels, the constraint $\rho[\mathscr{K}]=q$ is no longer a single linear constraint on the measured moments.
Therefore, fixing $q$ need not remove a measured moment direction, and the shape of the constrained convex set $\mathcal{K}_q$ depends on $q$.
Even for kernels in the measured span, positivity constraints usually remain active at finite sample size (due to acausal fluctuations), so the approximation $\Delta\chi_q^2 \approx \chi_1^2$ need not be accurate.
We therefore estimate the distribution numerically.

A practical calibration, similar to the profile construction used in Ref.~\cite{NOvA:2022wnj}, is obtained separately for each proposed $q$ using a constrained parametric bootstrap.
One begins by computing $\momseq_q \equiv \projection{\mathcal{K}_q}(\hat\momseq_N)$, the causal moment sequence closest to the data among those satisfying $\rho[\mathscr{K}]=q$, and generating Gaussian replicas $\hat\momseq_N^{(b)} \sim \mathcal{N}(\momseq_q,\Sigma_N)$; the bootstrap is \emph{constrained} in the sense that the replicas are centered on $\momseq_q$, which satisfies the hypothesis under test, rather than on the data $\hat\momseq_N$.
Then, for each replica $b$ and at fixed $q$, one repeats the constrained and unconstrained projections to compute the difference $(\Delta\chi^2_q)^{(b)} = \chi_q^2(\hat\momseq_N^{(b)}) - \chi_\mathrm{min}^2(\hat\momseq_N^{(b)})$, with the arguments indicating the input data for each projection.
From the distribution of $(\Delta\chi_q^2)^{(b)}$ computed using the replica values, one determines the empirical $(1-\alpha)$ quantile $\hat c_{1-\alpha}(q)$.
This cutoff is then compared with the value $\Delta\chi_q^2$ computed from the original measured data: the proposed $q$ is retained if $\Delta\chi_q^2\leq\hat c_{1-\alpha}(q)$, and excluded otherwise.
Repeating this procedure for different values of $q$ yields a nominal $(1-\alpha)$ confidence set for $\rho[\mathscr{K}]$.
For this reason, confidence intervals resulting from this observable-dependent choice of cutoff will be called \emph{calibrated confidence intervals}.
The associated observable-dependent calibration will be called \emph{conic calibration}.
The procedure is summarized in \cref{alg:conic-calibration}.
A concrete example of the procedure is illustrated in \cref{ssec:numerics_cb_calibration} below.

\begin{algorithm}[t]
\caption{Conic calibration}
\label{alg:conic-calibration}
\begin{algorithmic}[1]
    \Require Measured moments $\hat\momseq_N$, covariance $\Sigma_N$, smearing kernel $\mathscr K$, confidence level $1-\alpha$, and number of replicas $B$.
    \Ensure Nominal $(1-\alpha)$ confidence set for $\rho[\mathscr K]$.

    \Statex All projections use the metric induced by $\Sigma_N^{-1}$. The sets $\mathcal K$ and $\mathcal K_q$ are defined in \cref{eq:cone_free_q,eq:cone_fixed_q}.

    \State Compute $\chi_{\min}^2$ for the observed data using \cref{eq:chi2_min_cb}.

    \For{each trial value $q$ with $\mathcal K_q\neq\varnothing$}
        \StateAligned Compute the constrained projection
        $\momseq_q=\mathcal P_{\mathcal K_q}(\hat\momseq_N)$ by solving \cref{eq:chi_q}.
        \State Evaluate the observed statistic $\Delta\chi_q^2$ using \cref{eq:Delta_chi2q}.

        \For{$b=1,\ldots,B$}
            \State Draw an independent replica $\hat\momseq_N^{(b)}\sim\mathcal N(\momseq_q,\Sigma_N)$.
            \StateAligned Recompute $\chi_{\min}^2(\hat\momseq_N^{(b)})$ and $\chi_q^2(\hat\momseq_N^{(b)})$ using \cref{eq:chi2_min_cb,eq:chi_q}, keeping $q$ fixed.
            \State Set $(\Delta\chi_q^2)^{(b)} = \chi_q^2(\hat\momseq_N^{(b)}) - \chi_{\min}^2(\hat\momseq_N^{(b)})$.
        \EndFor

        \StateAligned Determine $\hat c_{1-\alpha}(q)$ as the empirical $(1-\alpha)$ quantile of $\{(\Delta\chi_q^2)^{(b)}\}_{b=1}^{B}$.

        \If{$\Delta\chi_q^2 \leq \hat c_{1-\alpha}(q)$}
            \State Retain $q$.
        \Else
            \State Exclude $q$.
        \EndIf
    \EndFor
    \State \Return the accepted set $\{ q: \Delta\chi_q^2 \leq \hat c_{1-\alpha}(q) \}$.
\end{algorithmic}
\end{algorithm}

The distribution estimated in this way is the one that would hold if the true moment sequence were $\momseq_q$.
Replacing the unknown true mean by $\momseq_q$ remains an approximation, which may raise or lower the actual attained coverage, a possibility explored in the numerical examples in \cref{ssec:numerics_cb_calibration}.\footnote{
    Refinements are possible. For instance, Ref.~\cite{Stanley2026} \emph{guarantees} rigorous coverage of at least $1-\alpha$ by replacing the quantile computed at the single point $\momseq_q$ with the largest $1-\alpha+\eta$ quantile of $\Delta\chi_q^2$ over the so-called Berger--Boos set~\cite{Berger1994} of admissible moments at fixed $q$ compatible with measured moments at a fixed $\chi^2$ cutoff within a $1-\eta$ confidence region, where $0<\eta<\alpha$.
}

\section{Comparison to related work\label{sec:comparison}}

This section provides a detailed comparison of the present work to the ideas and methods of recent and closely related proposals in the literature, including the Truncated Hankel Correlator method from Ostmeyer and Urbach~\cite{Ostmeyer:2025igc} in \cref{ssec:THC}, the Hankel denoising projection by Gull and collaborators~\cite{Yu_2024} in \cref{ssec:Gull}, Krylov-based methods in \cref{ssec:Krylov}, and Nevanlinna--Pick interpolation by Fields and Christ~\cite{Fields:2025glg} in \cref{ssec:noisyNP}.

\subsection{Truncated Hankel Correlator method}\label{ssec:THC}

The Truncated Hankel Correlator (THC) method was introduced in Ref.~\cite{Ostmeyer:2025igc} and provides a method for extracting an approximate spectral decomposition by constructing a low-rank approximation of the Hankel matrix $H(\hat{\momseq})$.
Given a weight matrix $\Omega$ and a cutoff $k \in \N$, the THC method assembles the noisy correlator data $\hat{C}$ into a noisy Hankel matrix $\hat{H}$, and then via a linear algebraic procedure computes the best rank-$k$ approximation of $\hat{H}$ defined by
\begin{equation}
\label{eq:thc-hankel-norm}
    \hat{H}_k = \argmin_{H:\mathrm{rank}(H) \leq k}
    \norm{\Omega ( \hat{H} - H ) \Omega^\dag}_F,
\end{equation}
where $\norm{\cdot}_F$ denotes the Frobenius norm.
The truncated matrix $\hat{H}_k$ is not in general a Hankel matrix, but still can be used to determine $k$ generalized eigenvalues via a least-squares generalization of a GEVP. Once the eigenvalues are determined, their associated overlaps can be determined via a least-squares problem, resulting in an approximate spectral decomposition for the original correlation function containing $k$ states.
Provided that the weights $\Omega$ are chosen appropriately, the norm in \cref{eq:thc-hankel-norm} can be chosen to mimic the behavior of the $\chi^2$ function, and hence the reconstructed spectral decomposition can be considered an approximation to the $\chi^2$-minimizing correlator defined by
\begin{equation}
C^{(k)} = \argmin_{\{Z_i\}, \{E_i\}} \chi^2 \left(
\hat{C}_{t,ab} - \sum_{i = 1}^k Z_{ia} Z_{ib}^* e^{-E_i t} \right).
\end{equation}
Ref.~\cite{Ostmeyer:2025igc} argues that, under certain assumptions, computing $C^{(k)}$ directly requires exponential computational effort, and hence solving \cref{eq:thc-hankel-norm} provides an appropriate proxy for the in general intractable optimization problem.

The conic projection proposed in the present work differs in several important respects.
First, the optimization operates directly on the moment sequence (i.e., the correlator values).
Second, the conic projection is defined in terms of the Moreau decomposition using the projection mappings in \cref{eq:projection_K,eq:projection_Kpolar} above, which by default use the full covariance matrix, although other choices are possible.
Third, by convexity, the Moreau decomposition identifies a unique partition of a generic moment sequence (with $H(\momseq)\not\succeq 0$) into causal and acausal pieces.
Unlike the THC method, where the truncation size is a free parameter, the causal component of the projection automatically identifies a definite number of states.
Fourth, the conic projection makes explicit use of the positivity conditions relevant for a given spectroscopy problem.
Requiring $H\succeq 0$ is correct for the Hamburger moment problem with $\lambda \in \R$.
Constraints on the spectral support (e.g., $\lambda \in \R^+$ for the Stieltjes moment problem) follow by imposing \cref{eq:Stieltjes_positivity} or \cref{eq:Hausdorff_positivity} at the level of the conic geometry.

A strength of the THC method is its computational complexity, which enjoys polynomial runtime.
Ref.~\cite{Ostmeyer:2025igc} has argued that direct parametric fits to the truncated spectral decomposition are generally exponentially hard and thus cannot be determined in polynomial runtime.
An algorithmic insight of the present work is that constrained optimization over the correlator itself remains convex and thus can be solved with primal-dual interior-point algorithms.
These algorithms typically have polynomial complexity.
In practice, the cost is usually dominated by linear solves whose size is determined by the number of input correlator values.

\subsection{Hankel denoising projection}\label{ssec:Gull}

The Hankel denoising projection proposed by Gull and collaborators in Ref.~\cite{Yu_2024} has several important similarities and differences with respect to the present work.

To the best of the authors' knowledge, Ref.~\cite{Yu_2024} was the first work to recognize the possibility of leveraging the convexity of the causal space to construct a projection operation, mapping a generic $\momseq$ with $H(\momseq) \not\succeq 0$ onto a causal moment sequence $\momseq'$ with $H(\momseq')\succeq 0$.
Ref.~\cite{Yu_2024} elegantly leverages Dykstra's algorithm~\cite{BoyleDykstra1986} of iterative projection to locate the intersection of two separate convex sets: the set of positive semidefinite (PSD) matrices and the set of Hankel matrices.
Projection onto PSD matrices is accomplished by setting negative eigenvalues equal to zero.
Projection of a generic PSD matrix to a Hankel matrix is accomplished by averaging over the antidiagonal.
The individual projection steps are known to yield the closest matrix in Frobenius norm to the starting matrix~\cite{Higham1988,EberleMaciel2003}.
The full covariance structure of realistic Monte Carlo data does not feature prominently in Ref.~\cite{Yu_2024}, although the authors discuss a method to enforce consistency with a particular point (say, $t=1$) if it happens to be known with greater precision.

The algorithm of Ref.~\cite{Yu_2024}
imposes positivity of both the Hankel matrix and its one-step shifted counterpart (also called ``small Hankel matrix'' in that reference), corresponding to the Stieltjes conditions in the present notation. As we already noted, different constraints are needed to restrict to suitable moment problems.
Imposing convex constraints based on domain knowledge is a natural extension of the procedure of Ref.~\cite{Yu_2024}.
For instance, since the Hausdorff conic projection consists of an optimization over a smaller convex set, it is expected to yield tighter bounds, where appropriate, than Stieltjes conic projection.

On a conceptual level, the algorithm of Ref.~\cite{Yu_2024} utilizes convexity in an essential way.
In contrast, the present work further exploits the conic structure of the problem, which allows for a purely geometric description of the noise as the closest dual-feasible obstruction to positivity.
At an operational level, the algorithms are different.
The present work utilizes interior-point methods in which all constraints are simultaneously satisfied at each iteration, while the algorithm of Ref.~\cite{Yu_2024} proceeds from the exterior and terminates when all constraints are satisfied after a series of iterative projection steps.

\subsection{Krylov-based methods}\label{ssec:Krylov}

Many existing large-scale lattice QCD spectroscopy studies are based on variants of the GEVP described in \cref{sec:GEVP}.
The GEVP operates with a single pair of Euclidean times to obtain estimates for the lowest $p$ effective states.
It can be generalized to include data from more than two Euclidean times to constrain the spectrum further, building up a finite-dimensional Krylov approximation of the transfer matrix and the associated physical Hilbert space.
The upshot is that the GEVP can be viewed as a single iteration of the block Lanczos algorithm or phrased equivalently~\cite{Wagman:2024rid,Ostmeyer:2024qgu,Abbott:2025yhm} in terms of the Prony Generalized Eigenvalue method~\cite{Fischer:2020bgv} or the Generalized Pencil of Function method~\cite{Aubin:2010jc,Aubin:2011zz}.

Roughly speaking, regardless of the precise algorithmic approach, each subsequent iteration provides information about $p$ additional states until the spectral content of a given correlator has been exhausted.
More precisely, $N_{\rm iter}$ iterations provide an interpolating representation for the correlator of the form~\cite{Hackett:2024nbe}
\begin{align}
\begin{split}
    C_{ab}(t) &= \sum_{n=1}^{n_{\rm max}} Z_{na}^{R(N_{\rm iter})*} Z_{nb}^{L(N_{\rm iter})} \left(\lambda_n^{(N_{\rm iter})}\right)^t,\\
    t &\leq 2 N_{\rm iter} -1,
\end{split}
\label{eq:block_Lanczos}
\end{align}
where
$Z_{na}^{R(N_{\rm iter})*}$ and $Z_{nb}^{L(N_{\rm iter})}$ are Ritz overlap factors and $\lambda_n^{(N_{\rm iter})} \in \C$ are the Ritz values.
Remarkably, this representation holds even in the presence of noise.
For noisy data, the right and left overlap factors generically differ, and the Ritz values are generically complex.
The appearance of complex Ritz values signals the fact that a given set of correlator data cannot be described solely in terms of a Hermitian transfer matrix.
Meanwhile, for exact data associated with a positive semidefinite spectral function, the right and left overlap factors agree and the Ritz values are real.
Throughout this work, the phrase \emph{spectral decomposition} refers exclusively to the case when a Hermitian transfer matrix exists with $\lambda_n \in \R$.
For the present discussion, the important point is that the conic decomposition in \cref{eq:conic_decomp} differs from that in \cref{eq:block_Lanczos}, although both interpolate the given input data.

\subsection{Nevanlinna--Pick interpolation from noisy data}\label{ssec:noisyNP}

Building on work in Ref.~\cite{PhysRevLett.126.056402}, Nevanlinna--Pick (NP) interpolation was applied in Ref.~\cite{Bergamaschi:2023xzx} to obtain bounds implied by causality on the energy-dependent two-point Green's function $G(E)$ evaluated at energies in the upper half of the complex plane when $G(E)$ is constrained to take known values $\{G(E_k)\}_{1 \le k \le N}$ at a finite number of energies $E_k$ on the imaginary axis.  In Ref.~\cite{Fields:2025glg}, Fields and Christ extended the method to the case of noisy data.  Just as for the case of the Hankel matrix discussed above, noisy values for the quantities $G(E_k)$ will typically be inconsistent with causality.  For the energy-dependent Green's function this inconsistency is equivalent to the $N\times N$ Hermitian Pick matrix $\mathcal{P}$ failing to be positive semidefinite, i.e., failing the Pick condition $\mathcal{P} \succeq 0$.

In the method of Ref.~\cite{Fields:2025glg}, one begins with an ensemble of statistically distributed noisy data. Starting with each sample, one follows a constrained path of steepest ascent maximizing the most negative eigenvalue, $\lambda_\mathrm{min}(\{G(E_k)\})$, of the Pick matrix as a function of the $N$ values of the Green's function. Causal data on the boundary of the Pick-consistent region are obtained at the end of each path when $\lambda_\mathrm{min}(\{G(E_k)\}) \geq0$. Causal boundary points that are statistically inconsistent with the original data are discarded.  In order to increase the number of statistically consistent endpoints, these paths were constrained to hold fixed particular Green's function values on which $\lambda_\mathrm{min}(\{G_k\}) \geq 0$ was least sensitive.

The result of this procedure is a large collection of points on the  boundary of the causal region.  The convexity of the causal region is then exploited to draw chords between pairs of boundary points and then to choose, as causal data, points that are equally spaced along these chords.  Each such point determines causal complex values for $G(E_k)$ evaluated for the $N$ imaginary energies $E_k$.  These points were then NP interpolated and their resulting statistical distribution studied.

 We note that attempting the strategy of following a path along which the most negative eigenvalue of the Hankel matrix increases does not lead to points on the boundary of the causal space but instead to degenerate negative eigenvalues, stalling the optimization.
 Since the Hankel matrix is real while the Pick matrix is complex, one might expect degeneracy to be more likely in the former case.
 For Hankel matrices, this situation motivates the need for an alternative scheme, for instance the Moreau projection described in \cref{sec:moreau} above.

\section{Numerical results \label{sec:numerics}}

This section presents numerical examples of the conic projection and various analytic predictions, on both real and mock data.
Examples using real data are given in \cref{ssec:numerics_real_scalar} for a high-precision kaon two-point correlation function and in \cref{ssec:numerics_real_matrix} for a noisy $2\times 2$ matrix of vector-vector correlation functions.
\Cref{ssec:numerics_mock_scalar} uses realistic mock data as input, offering a controlled demonstration of the method and various analytic predictions.
Finally, \cref{ssec:numerics_cb_calibration} illustrates the calibration of confidence intervals for the causal bootstrap using mock data.

The chosen examples all involve staggered-quark correlation functions, which require solving the most generic spectroscopy problem, in the sense that other cases follow by imposing additional constraints to restrict spectral support; see the discussion below \cref{eq:C_lambda_t} and \cref{eq:Hamburger_positivity,eq:Stieltjes_positivity,eq:Hausdorff_positivity}.

\subsection{High-statistics scalar $(p=1)$ Monte Carlo data \label{ssec:numerics_real_scalar}}

\begin{figure*}[t!]
\centering
\includegraphics[width=0.49\textwidth]{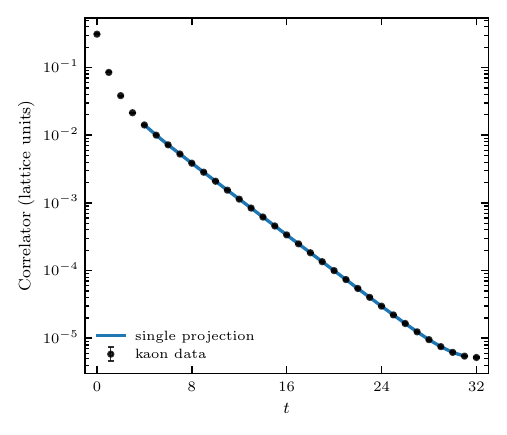}
\includegraphics[width=0.49\textwidth]{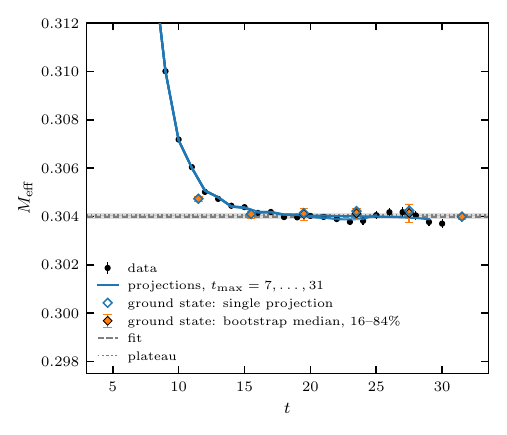}
\caption{
    Projection of the kaon two-point correlation function onto the signed cone using times $t\in\{4,\dots,31\}$ (shaded).
    \textbf{Left:} Binned data with errors (black) and the projection of the sample mean (blue).
    \textbf{Right:}
    Effective mass ($\Delta t=2$) in lattice units of the kaon two-point correlation function (black) and of its projections onto the signed cone for several different values of $t_{\rm max}$.
    Diamonds, offset just to the right of each $t_{\rm max}$, indicate the ground-state energy extracted from the projected sequence.
    Open diamonds are the result of a single projection, while filled diamonds show bootstrap results with $68\%$ uncertainties.
    The dashed and dotted horizontal lines show the ground-state energy extracted from a standard multi-exponential fit and from averaging over the plateau region of the data, respectively.
    \label{fig:kaon_projection}
    }
\end{figure*}

\Cref{fig:kaon_projection} shows data for a two-point kaon correlation function,
\begin{align}
C(t) =\sum_{\bm{x},\bm{y}} \langle \hat{P}(t,\bm{x}) \hat{P}^\dagger(0, \bm{y})\rangle,
\end{align}
where $\hat{P}$ is a pseudoscalar interpolating operator with the flavor quantum numbers of the kaon.
This correlation function has been analyzed in previously published work by the Fermilab Lattice and MILC collaborations~\cite{FermilabLattice:2022gku}.
In brief, the correlator was computed using highly improved staggered quarks (HISQ) near the physical point on an ensemble with $a\approx 0.12$~fm.
The data for this correlator were computed on 1352 configurations, with 32 source times per configuration.
Additional details can be found in Ref.~\cite{FermilabLattice:2022gku}.
In the left panel of \cref{fig:kaon_projection}, the blue curve shows the result of applying conic projection using the signed cone described in \cref{sec:signed_hamburger}.
The projection was repeated for 200 bootstrap iterations (resampling 135 bins of ten consecutive configurations).
Correlator data like these are routinely used to extract ground-state masses and amplitudes, so it is useful to see how such information arises in conic projections.
The right panel of \cref{fig:kaon_projection} shows the effective mass computed using
\begin{align}
    M_{\rm eff}(t) = \frac{1}{\Delta t} \mathrm{arccosh}\left[ \frac{C(t-\Delta t) + C(t+\Delta t)}{2\, C(t)} \right]
\end{align}
for the data and the conic projections, with $\Delta t = 2$ to remove the leading oscillating contribution.
The gray horizontal band shows the result for the ground-state energy coming from a standard multi-exponential fit using three decaying and two oscillating states using times $t \in \{4, \dots, 31\}$.
The orange points with error bars show bootstrap estimates of the ground-state energy, obtained by applying the Lanczos algorithm to the projected data.
The different points indicate projections using $t\in[4, t_{\rm max}]$ for different values of $t_{\rm max}$.
The results are consistent with the fit results, both in terms of central values and uncertainties.

\subsection{Matrix-valued $(p=2)$ Monte Carlo data \label{ssec:numerics_real_matrix}}

In 2019, the Fermilab Lattice, MILC, and HPQCD collaborations computed the light-quark connected contribution to hadronic vacuum polarization using lattice QCD~\cite{FermilabLattice:2019ugu}.
The results in that calculation were based on a set of vector-vector correlation functions computed using the HISQ action.
The interpolating operators included both local and spatially smeared vector currents, yielding a $2 \times 2$ matrix of correlation functions,
\begin{align}
    C_{ab}(t) &= \sum_{i=1}^3\int d^3\bm{x}\, \langle V_a^i(t, \bm{x}) V_b^{i\dagger}(0, \bm{0}) \rangle,
\end{align}
where the labels $a,b$ distinguish ``point'' and ``smeared'' operators.

\begin{figure*}[t!]
    \centering
    \includegraphics[width=0.48\textwidth]{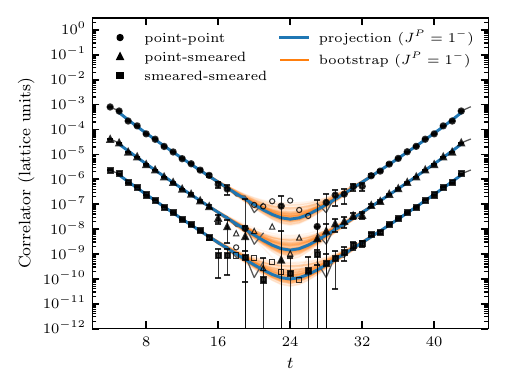}
    \includegraphics[width=0.48\textwidth]{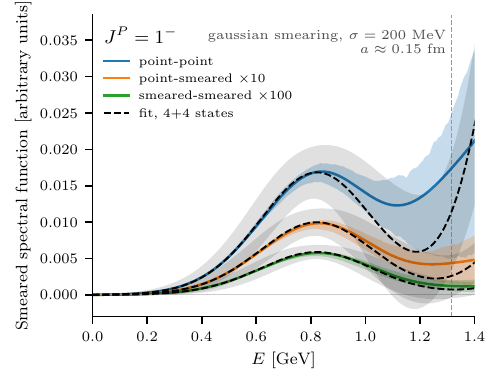}
    \caption{
    \textbf{Left:}
    Lattice QCD data from Ref.~\cite{FermilabLattice:2019ugu} for a $2\times 2$ matrix of vector-vector correlation functions, with the three independent components shown in black.
    Results are shown for the $J^P=1^-$ component of the projection of the black points (indicated in blue) as well as for projections of bootstrap replicas (indicated in orange).
    The main text describes the data in greater detail.
    \textbf{Right:}
    Gaussian-smeared ($\sigma=200$~MeV) spectral functions for the three independent components of the projected $2\times 2$ correlator in the left panel.
    Solid lines show the result from a single projection.
    Colored bands show bootstrap $68\%$ uncertainties.
    Black dashed lines with gray bands show the smeared spectral function of the multi-exponential fit described in the text.
    For visibility, the point-smeared and smeared-smeared components are multiplied by 10 and 100, respectively.
    \label{fig:hisq_vv_projection}
    }
\end{figure*}

\begin{figure}[t]
\centering
\includegraphics[width=0.48\textwidth]{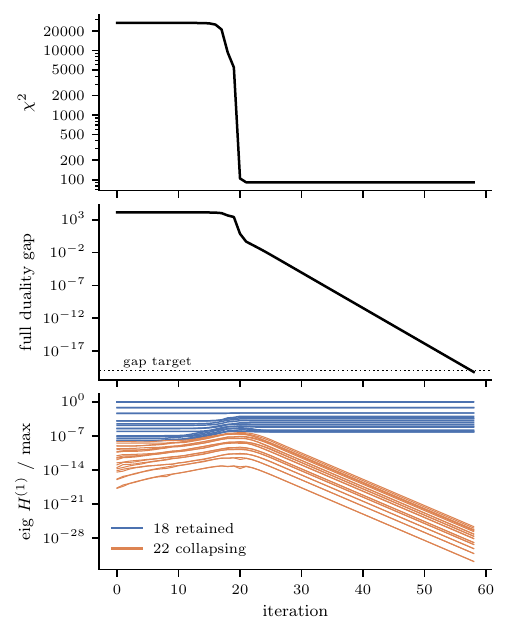}\hfill
\caption{
    Solver diagnostics for the projection shown in \cref{fig:hisq_vv_projection}.
    \textbf{Top:} Evolution of the $\chi^2$ versus iteration.
    \textbf{Middle:} Evolution of the full duality gap versus iteration; the dotted line marks the gap target of $10^{-20}$.
    \textbf{Bottom:} Eigenvalues of the Hankel matrix $H^{(1)}$ versus iteration, normalized by the largest eigenvalue.
    The blue eigenvalues remain finite, while those in orange collapse with the duality gap.
    }
    \label{fig:matrix_solver_diagnostics}
\end{figure}

The numerical example here is based on data from Ref.~\cite{FermilabLattice:2019ugu} taken from a single ensemble with lattice spacing $a\approx 0.15$~fm and near-physical pion masses.
The correlation functions were computed with degenerate light-quark masses of $a m_u=a m_d = 0.002426$ on $9362$ gauge-field configurations, with $48$ loose solves per configuration using the truncated solver method.
The left panel of \cref{fig:hisq_vv_projection} shows the Monte Carlo estimates for the $2\times 2$ matrix of correlation functions.
Note that the data exhibit a severe degradation in the signal-to-noise ratio at large Euclidean times near the lattice midpoint.
Since Ref.~\cite{FermilabLattice:2019ugu} was published, much effort in the community has been applied to reduce such statistical uncertainties.
The main point of applying conic projection to these data is to illustrate that such noise does not present any particular obstruction to the methods at hand.
Since large Euclidean times correspond to low energies, precise determination of the low-lying finite-volume energies will remain difficult with such data.

\Cref{fig:matrix_solver_diagnostics} provides additional information regarding the numerical solution using primal-dual interior-point methods (see \cref{sec:primal-dual} above).
The conic projection converged with a target duality gap of $10^{-20}$ after roughly 60 iterations.
The top panel of \cref{fig:matrix_solver_diagnostics} shows that the primal objective function (i.e., the $\chi^2$ function) nearly reaches its minimum after roughly 25 iterations, after which the duality gap plotted in the middle panel decreases rapidly.
The bottom panel shows how the expected singular structure of the solution arises using interior-point methods, plotting the normalized eigenvalue spectrum of the Hankel matrix versus the iteration number.
At early times, the spectrum is relatively dense, with normalized eigenvalues between unity and roughly $10^{-17}$.
This behavior persists during early iterations, during which the spectrum initially compresses.
However, once the primal objective function is nearly minimized, a group of eigenvalues separates and collapses exponentially to zero.
By adjusting the target duality gap, one can choose the tolerance for these zero modes.
The same qualitative behavior was observed in all projections studied in the present work.

Determining the ground-state energy from the data in the left panel of \cref{fig:hisq_vv_projection} remains a difficult challenge given the noise level at large Euclidean times.
Instead, one can investigate the approximation of the Gaussian-smeared spectral function, which is related to the $R$-ratio~\cite{Cabibbo:1970mh} and which can be extracted from the finite number of resolved states.
The right panel of \cref{fig:hisq_vv_projection} shows the Gaussian-smeared spectral function with a smearing width of $200$~MeV.
To obtain these results, the Lanczos algorithm was applied to the conic projections to extract an estimate for the finite-volume spectral function, which is then integrated against the Gaussian smearing kernel.
The solid blue, orange, and green lines show the results from applying the conic projection to the data central values, while the bands show bootstrap $68\%$ uncertainties.
Meanwhile, the black dashed curves (with gray bands) show the result of smearing the spectral function obtained from a multi-exponential fit.
In all cases, one observes a broad spectral peak around the known location of the rho resonance, as expected.

It bears emphasizing that these smeared spectral functions are simply the result of taking the spectral function obtained from conic projection at face value as a statistical estimator.
In particular, these curves do not enjoy any particular bounding properties.
Statistical calibration of causal-bootstrap bounds is discussed separately in \cref{ssec:numerics_cb_calibration} below.

\subsection{Scalar mock data \label{ssec:numerics_mock_scalar}
}

\begin{figure*}[t!]
\centering
\includegraphics[width=0.48\textwidth]{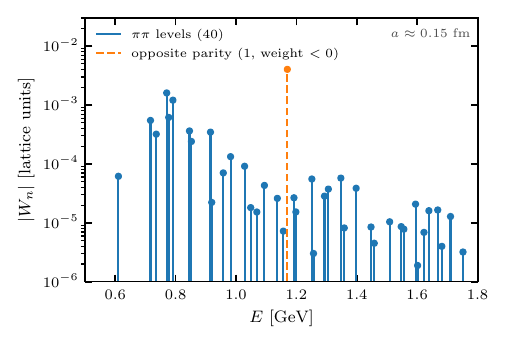}
\includegraphics[width=0.45\textwidth]{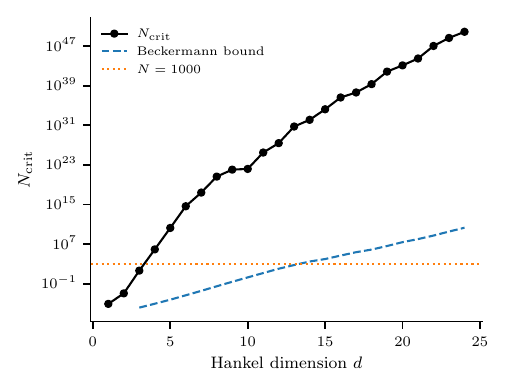}
\caption{
    \textbf{Left:}
    Exact finite-volume spectral function of the MLLGS mock model.
    The blue vertical lines show the locations of the $\pi\pi$ energy levels.
    The orange vertical line shows the location of the opposite-parity state.
    In both cases, the heights are equal to the spectral weights.
    \textbf{Right:}
    Critical sample size $N_{\rm crit}$ for the exact values of the mock data, computed using \cref{eq:Ncrit}, versus the Hankel dimension $d$.
    The dashed blue line shows the lower bound from \cref{eq:Ncrit_bound}.
    }
\label{fig:mock_data}
\end{figure*}

The example in this section uses realistic mock data to explore analytic properties discussed in the preceding sections.
To mirror the point-point vector-vector two-point functions in \cref{ssec:numerics_real_matrix}, following the original construction of Ref.~\cite{Borsanyi:2020mff} realized in Ref.~\cite{MILC:2024ryz}, we consider mock data with exact central values given by the Meyer--Lellouch--L{\"u}scher--Gounaris--Sakurai (MLLGS) model.
This model utilizes a tower of finite-volume $\pi\pi$ energy levels derived from the L{\"u}scher quantization condition and a Gounaris--Sakurai phase shift parametrization, with the corresponding amplitudes extracted via Lellouch--L{\"u}scher relations.
In order to mirror closely the HISQ data at $a\approx 0.15$~fm in \cref{ssec:numerics_real_matrix} above, for the mock data a similar lattice spacing of $a^{-1}=1.2617$~GeV is used;
taste-breaking is included following Ref.~\cite{MILC:2024ryz}, and a single opposite-parity state is included.
For $48$ total Euclidean times on a periodic lattice, the exact moment sequence is strictly causal, lying in the interior of the causal cone.

Samples of mock data are drawn according to
\begin{align}
    \hat{\momseq}_N \sim \mathcal{N}(\momseq, \Sigma/N)
    \label{eq:mock_data_model},
\end{align}
where the covariance matrix $\Sigma$ is taken from the observed covariance matrix of the point-point correlator in \cref{ssec:numerics_real_matrix} via $\Sigma_{N_\mathrm{data}} = \Sigma / N_\mathrm{data}$.
This model allows mock data to be sampled for \emph{any} desired value of $N$, including exponentially large values, which enables numerical illustration of many analytic results described in the preceding sections.

\subsubsection{Critical sample size}
The left panel of \cref{fig:mock_data} shows the exact finite-volume spectral function for the MLLGS model, where the height of each line is proportional to the weight of the discrete state.

The present model is useful for exploring the $N$-dependence of the conic projection.
The right panel of \cref{fig:mock_data} shows the critical sample size $N_{\rm crit}$, computed with the exact model using \cref{eq:Ncrit}, as a function of the dimension $d$ of the scalar Hankel matrix.
As expected, $N_{\rm crit}$ grows exponentially and lies well above the lower bound in \cref{eq:Ncrit_bound}.
The horizontal line indicates $N=1000$, which is often typical of practical Monte Carlo calculations.
This line falls exponentially below $N_{\rm crit}$ for the modest values of $d$ on the horizontal axis.
Modern lattice QCD calculations routinely employ lattices with more than 100 Euclidean times.

\begin{figure*}[t]
\centering
\includegraphics[width=\textwidth]{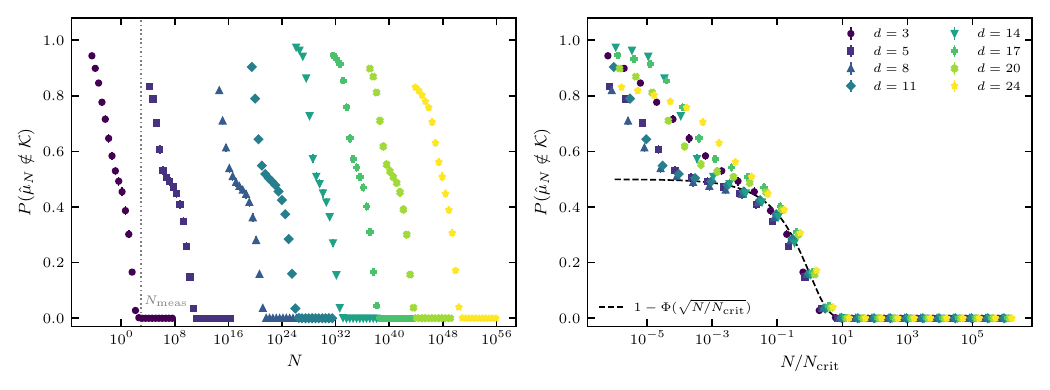}
\caption{
    The probability of acausal sampling for different values of $d$ and $N$, computed using 1000 independent samples drawn from \cref{eq:mock_data_model} at each point.
    \textbf{Left:} Fraction of samples $\hat\momseq_N$ with $H^{(1)}(\hat\momseq_N)\not\succeq0$ versus $N$.
    \textbf{Right:} The same results plotted versus $N/N_{\rm crit}(d)$.
    The dashed black line shows the predicted leading-order bound, \cref{eq:first-order-bound}.
    The data in the right panel are offset slightly horizontally for visibility.
    }
\label{fig:mock_data_causal_fraction}
\end{figure*}

\Cref{fig:mock_data_causal_fraction} investigates the frequency of acausal fluctuations.
To obtain this figure, 1000 samples were drawn from \cref{eq:mock_data_model} for various values of $d$ and various sample sizes $N$ above and below $N_{\rm crit}$.
The left panel shows the probability $P(\hat{\momseq}_N \not\in \mathcal{K})$ for samples to be acausal as a function of $N$.
As expected, the curves fall off sharply above $N_{\rm crit}$, which according to the right panel of \cref{fig:mock_data} rises exponentially with $d$.
The right panel shows the same results plotted instead versus the ratio $N/N_{\rm crit}$, exhibiting a striking universal behavior as $N$ approaches and exceeds $N_{\rm crit}$.
The behavior is easily understood quantitatively in terms of the Gaussian cumulative distribution function (CDF), which appears because the leading fluctuations (in the $N \to \infty$ limit) of the smallest eigenvalue $\delta \lambda_{\rm min}$ are Gaussian.
Since the leading-order shift in the eigenvalue gives a well-known variational upper bound, the prediction for the probability also gives a lower bound for acausal sampling:
\begin{align}
\begin{split}
    P(\hat{\momseq}_N \not\in \mathcal{K}) & \geq 1 - \Phi(\sqrt{N/N_{\rm crit}}), \\
    \Phi(x) &= \frac{1}{2}\left( 1 + \erf(x/\sqrt{2})\right).
\end{split}
\label{eq:first-order-bound}
\end{align}
As $N$ decreases below $N_{\rm crit}$, higher-order corrections become important and increase the probability of acausal data compared to the leading-order result.

\subsubsection{SURE and effective dimension}
\begin{figure*}[t!]
    \centering
    \includegraphics[width=0.48\textwidth]{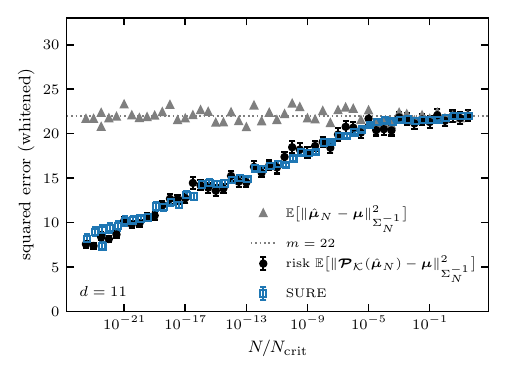}
    \centering
    \includegraphics[width=0.48\textwidth]{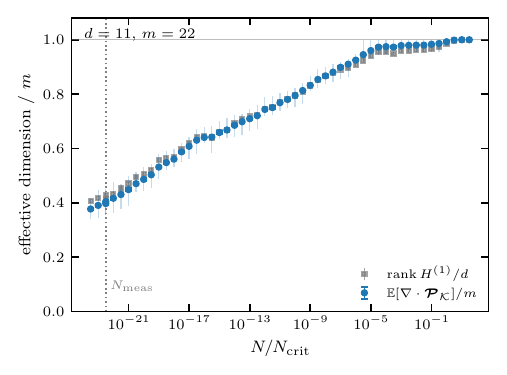}
    \caption{
        \textbf{Left:}
        The $L_2$ risk of the projection and Stein's unbiased risk estimate (SURE) for mock data, for fixed $d=11$ as a function of $N/N_{\rm crit}$.
        The gray points show the raw risk of the sample data.
        The black points show the risk after projection, computed using the known values for the exact data.
        The blue points are computed using SURE, \cref{eq:stein_sure}, without using the known exact data.
        The black and blue points differ by sampling errors only; each point was computed using 100 samples.
        As $N \to N_{\rm crit}$, the reduction in risk disappears because the raw samples are increasingly likely to be causal.
        \textbf{Right:}
        Effective dimension of the projection as a function of $N/N_{\rm crit}$ for a fixed size of $d=11$.
        Each point was estimated from 100 independent samples from \cref{eq:mock_data_model}.
        The blue points show direct estimation of  $\expectation[\nabla\cdot \projection{\mathcal{K}}]/m$.
        The faint gray squares show an alternative proxy for the same quantity, the rank of the projected Hankel matrix normalized by $d$.
        }
    \label{fig:sure_test}
\end{figure*}

The mock-data model in \cref{eq:mock_data_model} gives a controlled environment to understand the quantitative behavior of SURE discussed in \cref{sec:SURE} above.
The left panel of \cref{fig:sure_test} displays the $L_2$ risk before and after projection, $\expectation[ \norm{\hat{\momseq}_N -\momseq}^2_{\Sigma_N^{-1}} ]$ and $\expectation[ \norm{\projection{\mathcal{K}}(\hat{\momseq}_N) -\momseq}^2_{\Sigma_N^{-1}} ]$, respectively. To obtain this figure, the size of the Hankel matrix was fixed to $d=11$ and $N/N_{\rm crit}$ was varied between roughly $10^{-23}$ and $10^2$, with 100 independent samples at each $N$.
As expected, conic projection dramatically reduces the $L_2$ risk compared to the sample mean before projection when $N \ll N_{\rm crit}$.
\Cref{eq:stein_sure} above showed that SURE can also be used to compute the $L_2$ risk without reference to the exact data, which is usually unknown in practical problems.
The left panel of \cref{fig:sure_test} shows this equality borne out in practice, where the blue and black points track each other closely.
Note that \cref{eq:stein_sure} holds at the level of expectation values, so agreement is only expected at the level of statistical agreement.

Building on the discussion in \cref{sec:SURE}, the divergence of the projection mapping appearing in \cref{eq:stein_sure} can be interpreted as the effective dimension of the projection, which can range between one and the total number of moments $m$,
\begin{align}
\left( \text{effective dimension}\right) \equiv \expectation[ \nabla \cdot \projection{\mathcal{K}}(\hat{\momseq}_N) ].
\end{align}
The right panel of \cref{fig:sure_test}
shows the effective dimension normalized by $m$ in blue for $d=11$ and $N/N_{\rm crit}$ ranging between roughly $10^{-25}$ and $10^2$, with 100 independent samples at each $N$.
The vertical dotted line shows $N_{\rm meas}=1000$ as a proxy for typical practical Monte Carlo statistics.
As an alternative proxy for the dimension, the light gray squares in the background show the rank of the Hankel matrix normalized by $d$.
The curves exhibit matching behavior within the statistical uncertainties.
Evidently the effective dimension can be well approximated for many decades in $N$ by a linear function of $\ln(N/\Ncrit)$.

\subsection{Causal bootstrap calibration \label{ssec:numerics_cb_calibration}
}

\begin{figure*}[t]
    \centering
    \includegraphics[width=0.9\textwidth]{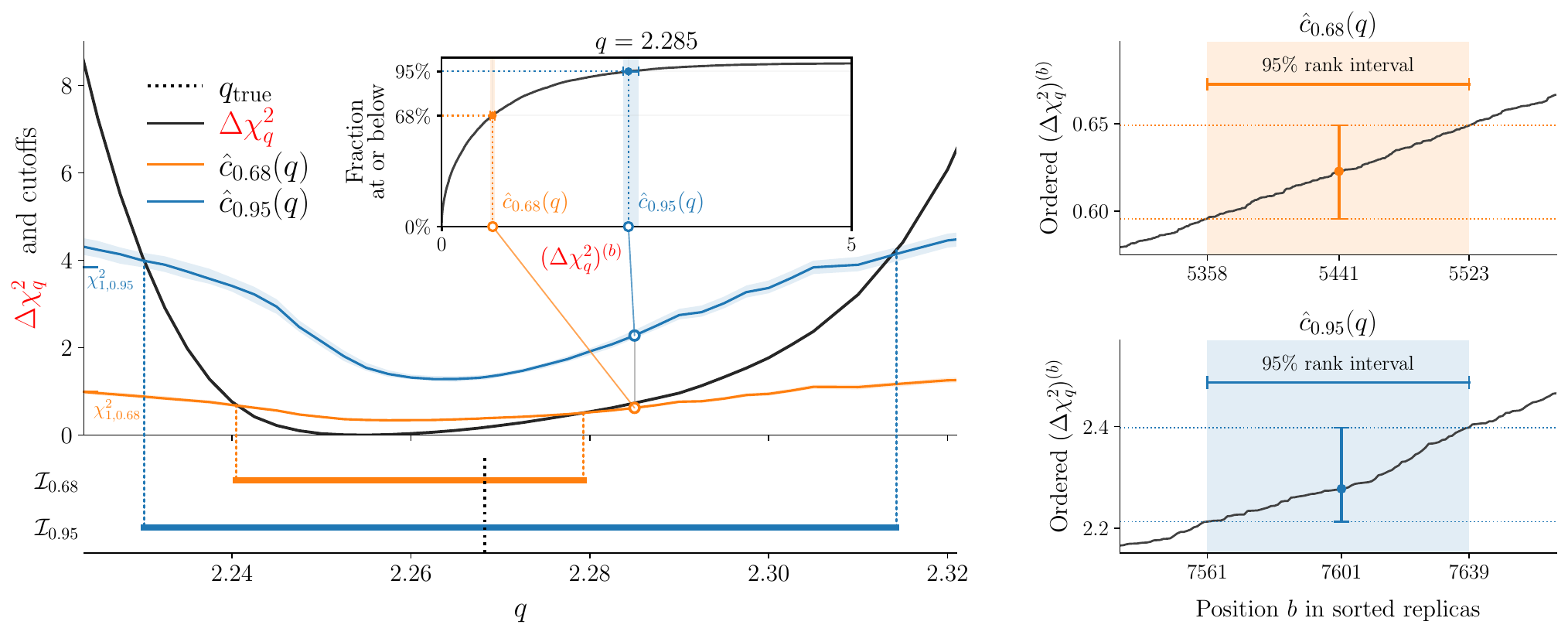}
    \caption{Illustration of the causal-bootstrap statistical calibration.
    \textbf{Main figure:}
    The solid black curve shows the profile-likelihood statistic, $\Delta\chi^2_q$.
    The calibrated cutoffs $\hat{c}_{1-\alpha}(q)$ are shown by the blue and orange curves for two confidence levels, $(1-\alpha)\in\{68\%, 95\%\}$;
    their shading indicates pointwise $95\%$ binomial uncertainty intervals.
    The intersections of these curves with $\Delta \chi^2_q$ define the boundary of the confidence intervals, as indicated by the colored vertical dashed lines.
    Short colored ticks on the left mark the $\chi_1^2$ reference thresholds at the familiar values of $\chi_{1,0.68}^2=1$ and $\chi_{1,0.95}^2 \approx 3.84$.
    \textbf{Inset and right plots:}
    The inset illustrates the construction of the calibrated cutoffs $\hat{c}_{1-\alpha}(q)$ at a fixed $q$.
    The black curve shows the empirical CDF for $\Delta\chi^2_q$, computed from bootstrap resampling.
    The vertical bands are defined by the intersection of the CDF with the target coverage, $(1-\alpha)$; the locations give the calibrated cutoff $\hat{c}_{1-\alpha}(q)$.
    The right panels illustrate the calculation of the uncertainty on the calibrated cutoffs $\hat{c}_{1-\alpha}(q)$.
    }
    \label{fig:calibration}

    \vspace{0.1cm}

    \centering
    \subfloat[Bounds and coverage at nominal $68\%$ confidence level.\label{fig:calibration_comparison68}]{%
        \includegraphics[width=0.48\textwidth]{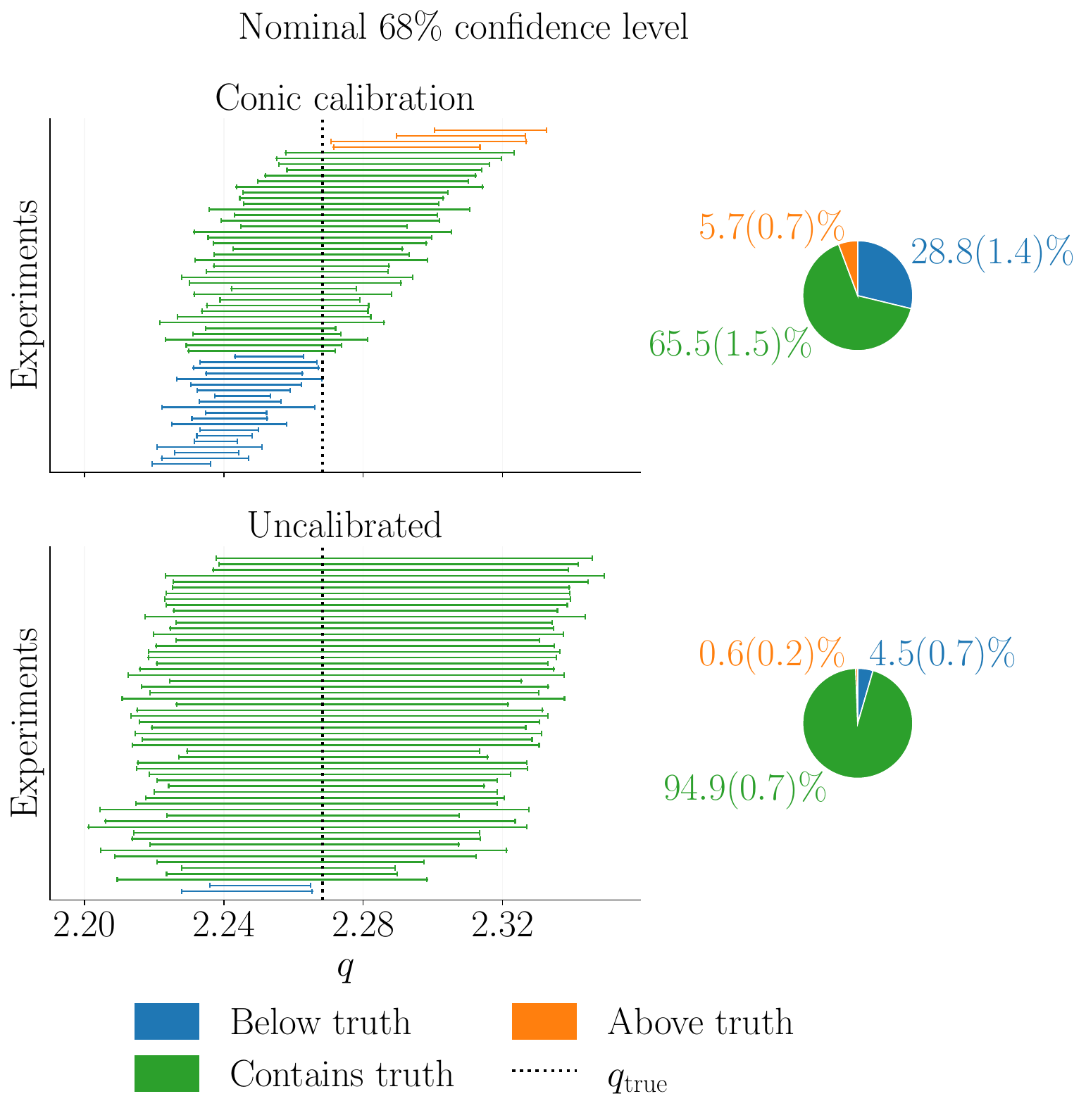}
    }%
    \hfill
    \subfloat[Bounds and coverage at nominal $95\%$ confidence level.\label{fig:calibration_comparison95}]{%
        \includegraphics[width=0.48\textwidth]{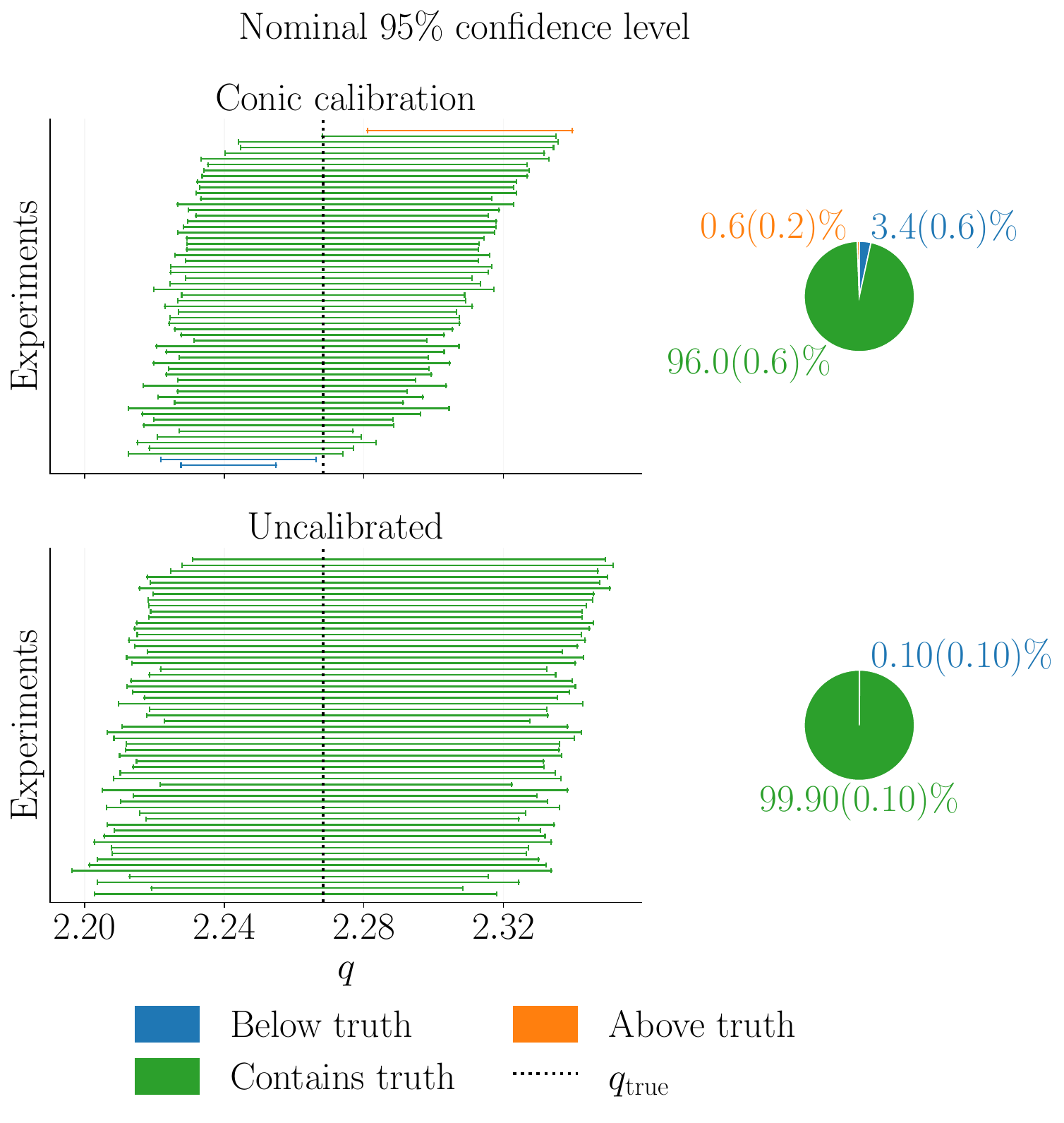}
    }%
    \caption{Comparison of bounds and attained coverage for calibrated and uncalibrated causal bootstrap at two nominal confidence levels, $68\%$ for \cref{fig:calibration_comparison68} and $95\%$ for \cref{fig:calibration_comparison95}, computed from $N=10^4$ samples. Each panel features $60$ selected intervals for different experiments, ordered and colored according to their position relative to the mock truth, marked by the black dotted line. Pie charts report the attained coverage for $1000$ experiments, with estimated binomial standard errors in parentheses. The interval panels therefore illustrate widths and failure modes, while the pie charts quantify coverage. The proposed calibration strategy yields less conservative bounds with the approximate desired statistical interpretation.
    }
    \label{fig:calibration_comparison}
\end{figure*}

\begin{figure*}[t]
    \centering
    \includegraphics[width=0.9\textwidth]{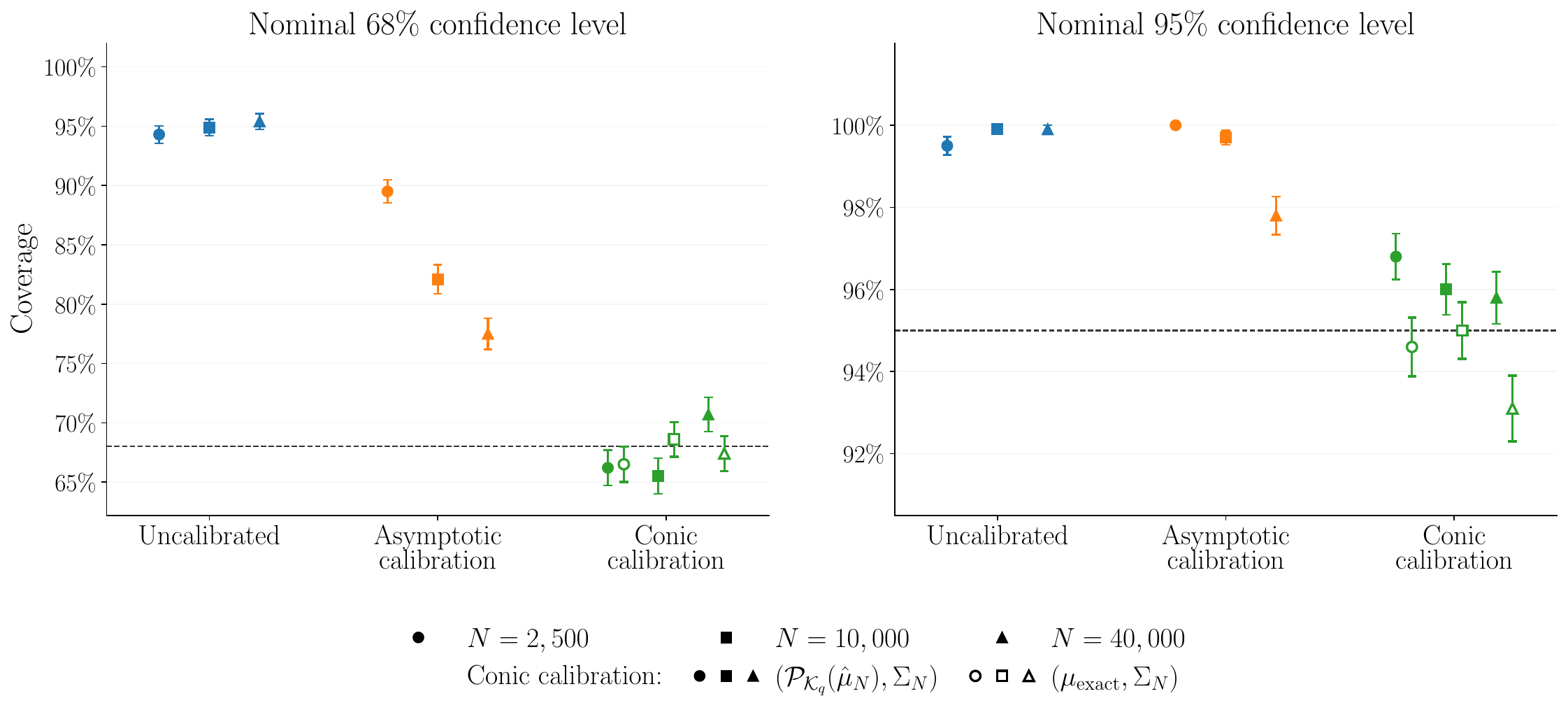}
    \caption{Coverage at nominal $68\%$ and $95\%$ for different methods to tune the $\chi^2$ cutoff, for three values $N$ of the sample size.
    The uncalibrated strategy is consistently conservative. The $\chi_1^2$ prescription reduces overcoverage substantially at nominal $68\%$, but much less at nominal $95\%$. Observable-specific calibration brings coverage substantially closer to nominal at both levels. Calibrating by sampling around the true correlator removes part of this approximation, providing a benchmark for the practically attainable coverage.
    }
    \label{fig:coverage}
\end{figure*}

This section demonstrates the calibration of causal-bootstrap bounds described in \cref{sec:calibration}, applied to the scalar mock correlator of \cref{ssec:numerics_mock_scalar}.
The input data consist of $m=22$ time slices, $t=3,\dots,24$, with the same model as above.
The target observable is $\rho[\mathscr{K}]$ with kernel
\begin{equation}
    \mathscr{K}(\omega) = 12\pi^2 \frac{5}{9} \frac{1}{\omega^2} \frac{e^{-\frac{(\omega-E)^2}{2\sigma^2}}}{\sqrt{2\pi\sigma^2}}
\end{equation}
at $E=750$~MeV and $\sigma=200$~MeV, which corresponds to the light-connected component of the $R$-ratio, smeared with a Gaussian kernel centered at energy $E$ and with smearing width $\sigma$.
The prefactors include the sum of charges and compensate the quadratic ultraviolet growth of the vector spectral density.
The Gaussian kernel is handled using semi-infinite programming, as in Ref.~\cite{Jay:2026qoh}.

\Cref{fig:calibration} shows the construction of the profile statistic and conic calibration of cutoffs $\hat{c}_{1-\alpha}(q)$ for two confidence levels, $1-\alpha \in \{68\%,95\%\}$.
For each fixed value of $q$, the inset shows the procedure outlined in \cref{sec:calibration} to compute the cutoffs $\hat{c}_{1-\alpha}(q)$ from the quantiles of the profile statistic $\Delta\chi^2_q$.
As \cref{fig:calibration} shows graphically, the boundaries of the confidence intervals are defined by the intersections of the cutoffs $\hat{c}_{1-\alpha}(q)$ and $\Delta\chi^2_q$, which are computed from $\momseq_q$ and the measured data $\hat\momseq_N$, respectively.

The inset and the right panels provide additional visual details about the empirical estimation of the cutoffs $\hat{c}_{1-\alpha}(q)$.
These are computed from bootstrap samples $\hat\momseq_N^{(b)}$, $b=1,\dots,B$, normally distributed around the projected causal moment sequence $\momseq_q$ at fixed $q$.
Conic projection gives access to bootstrap samples $(\Delta\chi^2_q)^{(b)}$.
Ordering them by their observed value and counting the fraction of observations that fall below a given percentage allows one to  estimate empirically the CDF of $\Delta\chi^2_q$, represented by the black curve in the inset.
The horizontal lines are given by $68\%$ and $95\%$.
Their intersection with the empirical CDF defines the cutoffs $\hat{c}_{1-\alpha}(q)$.
The right panels illustrate instead the calculation of their uncertainties, estimated using the fact that the number of bootstrap observations $n_B$ falling below the population quantile $c_{1-\alpha}(q)$ follows a binomial distribution, $n_B \sim {\rm Binomial}(B, 1-\alpha)$.
The right panels then show the intervals $[(\Delta\chi_q^2)^{(b_\ell)},(\Delta\chi_q^{2})^{(b_u)}]$ satisfying $\mathrm{Pr}(b_\ell \leq n_B < b_u) \geq 95\%$,
where $b_\ell$ and $b_u$ are chosen from the $2.5\%$ and $97.5\%$ quantiles of ${\rm Binomial}(B, 1-\alpha)$.

The present construction is designed to give nominal $(1-\alpha)$ confidence intervals.
Repeated experiments (at fixed statistics) provide a way to check the coverage level obtained in practice.
\Cref{fig:calibration_comparison} shows the coverage achieved in repeated experiments before and after conic calibration at fixed $N=10^4$ for two choices of nominal confidence level, namely $68\%$ (left panel) and $95\%$ (right panel).
The uncalibrated procedure fixes the conservative $\chi_{m,1-\alpha}^2$ cutoff independently of the observable.
The pie charts summarize the results of all $1000$ independent mock-data experiments, while the horizontal bands show $60$ selected intervals.
Intervals that contain, fall below, and lie above the true result appear in green, blue, and orange, respectively.
Empty sets arising due to $\chi^2_{m,1-\alpha} < \chi^2_\mathrm{min}$ are counted as failures and assigned to ``below truth'' or ``above truth'' according to their best-fit estimate.
For both confidence levels, the observed coverage of the calibrated confidence interval agrees statistically with the desired $(1-\alpha)$ target.\footnote{
    We assign binomial errors $\sqrt{C(1-C)/M}$ across the $M=1000$ experiments associated with coverage $C$ measuring the fraction of experiments where the true value of $\rho[\mathscr{K}]$ lies within the reconstructed interval $\mathcal{I}$.
}
Meanwhile, the uncalibrated results are manifestly more conservative:
the intervals are larger and contain the true result much more often than the nominal $(1-\alpha)$ would suggest.
This conservatism is expected: tuning the $\chi^2$ cutoff solely based on compatibility with the measured data yields a simpler but deliberately more conservative observable-agnostic procedure to compute causal bounds.
Overall, \cref{fig:calibration_comparison} supports nominal coverage of the calibrated $(1-\alpha)$ confidence intervals in this example.

\Cref{fig:coverage} explicitly tests the coverage for the three calibrations described in \cref{sec:calibration}
and for three different sample sizes, in each case using $1000$ independently repeated experiments.
Results with $N=10^4$ samples correspond to those shown in \cref{fig:calibration_comparison}.
The figure contains three important comparisons worth highlighting.
The first comparison is between the coverage in the uncalibrated and conically calibrated results. \Cref{fig:coverage} confirms that the uncalibrated results are consistently more conservative than the calibrated ones: while their coverage is formally at least nominal, it is practically always much higher.
Across experiments with conic calibration, coverage is not uniformly nominal, as expected from an approximate procedure.
However, compared to uncalibrated results, conic calibration proves useful in achieving coverage levels which are much closer to the targeted nominal ones.
No statistically significant undercoverage is observed in these tests.
The second useful point of comparison is to the asymptotic calibration based on quantiles of the $\chi^2_1$ distribution.
The Gaussian smearing kernel lies outside the measured span, so the asymptotic $\chi_1^2$ limit derived in \cref{sec:calibration} does not, strictly speaking, apply.
Nevertheless, this prescription reduces overcoverage substantially at nominal $68\%$, but much less at nominal $95\%$. Observable-specific conic calibration brings coverage substantially closer to nominal at both levels.
In the third and final comparison, the open symbols indicate conic-calibrated results where the true correlator is used as the mean for generating calibration replicas ($\hat\momseq_N^{(b)}$), while retaining the same noisy observations and covariance.
This replacement removes the approximation of the fitted reference ($\momseq_q$) and isolates its effect on coverage.
The conic calibration from samples around exact data is closer to the desired nominal coverage in most of the examples shown in \cref{fig:coverage}.
These observations remain consistent when the number of experiments is increased, which only determines the precision with which coverage is estimated.

\section{Conclusions \label{sec:conclusions}}

The present work has presented a unique decomposition for noisy matrix-valued Euclidean-time correlation functions, viewed as moment sequences, into a manifestly causal piece and acausal noise.
This decomposition arises from the interplay of two facts, namely, that positivity constraints from causality lead to a convex conic geometry and that statistical uncertainties give rise, via the covariance matrix, to a natural measure of distance.
With help from recent results in the mathematical literature (see \cref{app:flat} and especially Ref.~\cite{Kimsey:2022moment}), the causal piece can be shown to possess a unique, minimal spectral representation whenever the decomposition is nontrivial.
Meanwhile, the acausal piece has a clean geometric interpretation as the closest dual-feasible obstruction to positivity.
From a practical perspective, an important  upshot of the present work is that this decomposition can be computed efficiently using well-established primal-dual interior-point methods from the literature on convex optimization. 
Two applications for the methods studied are readily envisioned.

First, these methods naturally apply to lattice QCD spectroscopy studies, especially in noisy systems. Common spectroscopic analyses use noisy data directly as inputs to the GEVP or other Krylov-based methods. One then studies the extent to which various analytic properties (e.g., variational bounds) are satisfied by the resulting transfer-matrix eigenvalues.
The present proposal turns the pipeline on its head: the conic projection first identifies the causal and acausal components.
The minimal spectral representation can then be extracted algebraically from the projected data.
A detailed comparison of spectroscopy results using state-of-the-art data would be an interesting direction for future work. 

Second, the conic projection provides the starting point for calibrated confidence intervals for smeared spectral observables computed using the causal bootstrap or similar techniques.
After calibration, the confidence intervals have a clear frequentist interpretation. 
Our preferred method, conic calibration, uses an observable-specific constrained parametric bootstrap to reduce overcoverage compared to the uncalibrated bounds in Ref.~\cite{Abbott:2026wdw} or the asymptotic calibration described in \cref{sec:causal_bootstrap}, which nevertheless already provides a valuable improvement over the original proposal in Ref.~\cite{Abbott:2026wdw}. 
As a related benefit, the calibrated confidence intervals are often numerically tighter than those from other methods.

Besides these applications, we also presented a dimensional argument for the critical sample size below which acausal fluctuations become unavoidable. 
The scale is expected to be exponentially large in the length of the moment sequence, as confirmed explicitly with realistic mock data.
The effective dimension of the projection quantifies the accessible spectral content for a given noise level, acting as a sort of statistical phase boundary separating fully causal data from noisy samples.
In the scalar mock-data example, the effective dimension tracks the normalized Hankel rank and grows logarithmically with the number of samples. 
These observations provide a new quantitative perspective on the success of variational methods in lattice field theory.

\section*{Acknowledgments}

We thank
Norman Christ,
Dan Hackett,
Shaun Lahert,
Christoph Lehner,
and Mike Wagman
for useful conversations.

We thank our collaborators in the Fermilab Lattice and MILC collaborations for allowing us to use data from the published work in Refs.~\cite{FermilabLattice:2019ugu,FermilabLattice:2022gku}.
The research reported in this work made use of computing and long-term storage facilities of the USQCD Collaboration, which are funded by the Office of Science of the U.S. Department of Energy.

RA is an Ernest Kempton Adams Postdoctoral Fellow supported in part by the Ernest Kempton Adams Fund for Physical Research of Columbia University.
RA and SF are supported in part by U.S. DOE Grant No. DE-SC0011941.
SF is also supported in part by the National Science Foundation Graduate Research Fellowship Program under Grant No. DGE-2036197.
Any opinions, findings, conclusions or recommendations expressed in this material are those of the author(s) and do not necessarily reflect the views of the National Science Foundation.
PO is supported in part by the U.S. Department of Energy, Office of Science, Office of Nuclear Physics under grant Contract Number DE-SC0012704 (BNL).

Numerical calculations in this work made use of \textsc{NumPy}~\cite{vanderWalt:2011bqk,Harris:2020xlr}, \textsc{mpmath}~\cite{mpmath}, and \textsc{cvxpy}~\cite{diamond2016cvxpy,agrawal2018rewriting}.
Figures were generated using \textsc{matplotlib}~\cite{Hunter:2007} and \textsc{seaborn}~\cite{Waskom2021}.

We wrote the original implementation of the ideas described in \cref{app:Newton}.
This hand-written baseline served as a validation benchmark for performance optimization.
We used Claude Code and Codex by OpenAI to assist with implementation and comparison of optimization strategies.
In preparation of the manuscript, the same tools were used to assist in final copyediting.

\appendix
\crefalias{section}{appendix}
\crefalias{subsection}{appendix}
\section{Hankel matrices and the spectral decomposition \label{app:spectral_decomp}}

This appendix collects additional technical details regarding the spectral decomposition and its interplay with positive semidefinite Hankel matrices.
\Cref{app:degeneracies} provides additional comments about the spectral decomposition in the presence of degenerate energy levels.
\Cref{app:definitions} collects the necessary definitions and notation.
\Cref{app:flat} states the flat-extension theorem of the truncated moment problem and derives from it the two facts used in the main text: the first $m-1$ moments of a causal sequence determine a unique minimal spectral decomposition, and the last moment may carry an endpoint excess, given by a Schur complement, that controls whether further moments can be appended consistently with causality.
\Cref{app:examples} illustrates the possibilities with concrete examples.

\subsection{Degeneracies \label{app:degeneracies}}

The spectral decomposition in \cref{eq:spectral_function_FV} contains the possibility for degenerate energy levels.
If the $n^\mathrm{th}$ level contains $\nu$ such states, $E_n$ is said to have a $\nu$-fold degeneracy:
\begin{align}
    E_{n'} &= E_n \text{ for } n' \in \{1, 2, \dots, \nu\},
\end{align}
where $n'$ labels the distinct physical states.
The matrix of overlap factors then satisfies
\begin{align}
    W_n = \sum_{n'=1}^\nu \bm{Z}_{n'} \bm{Z}_{n'}^\dagger, && \rank W_n = \nu,
\end{align}
where $\nu \leq p$ and assuming independent overlap factors.
Of course, each level $E_n$ will in principle have a different degeneracy $\nu$.

In infinite volume, such degeneracy arises as a matter of principle whenever multiple inelastic partial-wave channels are present, e.g., isoscalar $\pi\pi$--$K\bar{K}$ scattering.
In a finite volume, degeneracy is expected to be rare for operators transforming irreducibly (with a given fixed irrep and row), since the von Neumann--Wigner theorem governing level crossings for real-symmetric mixing matrices requires the tuning of two parameters, like the lattice volume $V=L^3$ and a quark mass.
On the other hand, near-degeneracies are an important concern in numerical work.
For example, one expects near-degeneracies for continuum spins (say, $J=0,4$) that transform in the same finite-volume irrep but with small energy splittings.
As a rule of thumb, resolving a $\nu$-fold degeneracy or near-degeneracy typically requires at least $\nu$ operators.

\subsection{Definitions and notation \label{app:definitions}}

Consider the truncated Hamburger problem.
Moment sequences are written as $\momseq=(\moment{0},\dots,\moment{m-1})\in(\H_\Nop)^m$ with $m=2d-1$, $d \in \N$.
Nested Hankel matrices built from the first $2k-1$ moments are
\begin{align}
    \Hk{k}(\momseq)=\big[\moment{i+j}\big]_{i,j=0}^{k-1}\in\H_{k\Nop},
    \qquad 1\leq k\leq d .
\end{align}
It will always be assumed that a given moment sequence $\momseq$ is causal, so that $\Hk{d}(\momseq)\succeq0$.
Since $\Hk{k}$ is a principal submatrix of $\Hk{d}$, every $\Hk{k}$ is then positive semidefinite.
An (effective) spectral decomposition of $\momseq$ is a finite sum
\begin{align}
    \moment{t}=\sum_{n=1}^{\Nstates}W_n\lambda_n^{\,t},\qquad t=0,\dots,m-1,
    \label{eq:spectral_decomposition}
\end{align}
with distinct real $\lambda_n$ and nonzero $W_n\succeq0$, as in \cref{eq:spectral_function_FV}.
\Cref{eq:spectral_decomposition} is effective in the sense that it need not agree with the spectral decomposition of the full theory.
As the discussion in \cref{ssec:Krylov} reveals, it is often useful to think of an effective spectral decomposition as the result of applying Krylov methods to a causal moment sequence.
The same notion applies to a sequence of any length, in particular to the first $m-1$ moments $(\moment{0},\dots,\moment{m-2})$.
A spectral decomposition is \emph{minimal} if $\sum_n\rank W_n$ is as small as possible; in the scalar case $\Nop=1$ this is simply the number of states.
A spectral decomposition of $\momseq$ exists whenever $\Hk{d}(\momseq)\succ0$, but a boundary sequence, i.e., one with singular $\Hk{d}(\momseq)$, need not admit one, as the examples of \cref{app:examples} show.

An \emph{extension} of $\Hk{k}$ is a Hankel matrix $\Hk{k+1}$ obtained by appending two further moments $\moment{2k-1}$ and $\moment{2k}$.
A positive semidefinite extension is called \emph{flat} if $\rank\Hk{k+1}=\rank\Hk{k}$~\cite{CurtoFialkow1998,Kimsey:2022moment}.
Physically, a positive semidefinite extension specifies two further time slices consistently with causality, and a flat one does so without revealing any new state.

To analyze the role of the \emph{last} moment, it is useful to partition $\Hk{d}(\momseq)$ into its leading block and its last block row and column,
\begin{align}
\begin{split}
    \Hk{d}(\momseq) &=
    \begin{pmatrix}
        \Hk{d-1}                & \tilde{\momseq} \\
        \tilde{\momseq}^\dagger & \mu_{2d-2}
    \end{pmatrix}
    \equiv
        \begin{pmatrix}
        A         & B \\
        B^\dagger & C
        \end{pmatrix},\\
    \tilde{\momseq} &\equiv B \equiv \big(\moment{d-1}\ \ \moment{d}\ \ \cdots\ \ \moment{2d-3}\big)^\dagger .
\end{split}
    \label{eq:block_form}
\end{align}
A standard result on Schur complements (known variously as the fundamental lemma~\cite{Kovalishina1984} or Smuljan's lemma in the literature on moment problems; see Lemma~2.1 of Ref.~\cite{Kimsey:2022moment} and Lemma~2.3 of Ref.~\cite{CurtoFialkow1991}) says that, given $\Hk{d-1}\succeq0$, one has $\Hk{d}\succeq0$ if and only if
\begin{align}
    \ker A\subseteq\ker B^\dagger
    \qquad\text{and}\qquad
    C\succeq B^\dagger A^{+}B,
    \label{eq:smuljan}
\end{align}
with $A^{+}$ the Moore--Penrose pseudo-inverse.
In this case, $\Hk{d}$ is congruent to $\diag(A,\,C-B^\dagger A^{+}B)$, so that
\begin{align}
    \rank\Hk{d}=\rank\Hk{d-1}+\rank\big(C-B^\dagger A^{+}B\big).
    \label{eq:smuljan_rank}
\end{align}
The first condition in \cref{eq:smuljan} says that the new block column $\tilde{\momseq}$ lies in the range of $\Hk{d-1}$: any linear combination of the first $d-1$ time slices that has zero norm also has zero overlap with the new time slice.
The second condition bounds the last moment from below.
These considerations motivate the definition of the flat completion $\lastflat$ and the endpoint excess $\Delta$ via
\begin{align}
\begin{split}
    \lastflat &\equiv  B^\dagger A^{+}B = \tilde{\momseq}^\dagger (\Hk{d-1})^+ \tilde{\momseq}, \\
    \Delta &\equiv \moment{2d-2}-\lastflat \succeq 0.
\end{split}
    \label{eq:schur_min}
\end{align}
The moment sequence can be written in terms of the flat completion and excess via
\begin{align}
\begin{split}
    \momseq&=\muflat+\momseq_\Delta,
    \qquad
    \momseq_\Delta\equiv(0,\dots,0,\Delta),
    \\
    \muflat&\equiv(\moment{0},\dots,\moment{2d-3},\lastflat).
\end{split}
    \label{eq:endpoint_split}
\end{align}
By the rank equation in \cref{eq:smuljan_rank}, $\Hflat\equiv\Hk{d}(\muflat)$ is a flat extension of $\Hk{d-1}$, and
\begin{align}
    \rank\Delta=\rank\Hk{d}-\rank\Hk{d-1}.
    \label{eq:rank_Delta}
\end{align}
In words, $\muflat$ is the causal sequence that agrees with $\momseq$ in all but the last moment and has the smallest possible last moment.
The endpoint excess $\Delta$ measures by how much, and in how many operator directions, the measured last moment exceeds this minimum.

\subsection{The flat-extension theorem \label{app:flat}}

The existence and uniqueness of spectral decompositions for truncated moment sequences is governed by a flat-extension theorem due to Curto and Fialkow for scalar moments~\cite{CurtoFialkow1991,CurtoFialkow1998} and extended to matrix-valued moments by Kimsey and Trachana~\cite{Kimsey:2022moment}.
To keep the present work self-contained, we state the one-variable version in the present notation and refer to the literature for the proofs.

\begin{theorem}[Flat-extension theorem]
\label{thm:flat}
Let $\bm{\nu}=(\nu_0,\dots,\nu_{2k-2})\in(\H_\Nop)^{2k-1}$ be a truncated moment sequence with $\Hk{k}(\bm{\nu})\succeq0$.
\begin{enumerate}[label=(\alph*),leftmargin=2em]
\item Every spectral decomposition of $\bm{\nu}$ satisfies $\sum_n\rank W_n\geq\rank\Hk{k}(\bm{\nu})$.
\item $\bm{\nu}$ admits a spectral decomposition with $\sum_n\rank W_n=\rank\Hk{k}(\bm{\nu})$ if and only if $\Hk{k}(\bm{\nu})$ admits a flat extension $\Hk{k+1}$.
    For each flat extension, the extended sequence $(\nu_0,\dots,\nu_{2k})$ has exactly one spectral decomposition, and it is minimal.
\item $\bm{\nu}$ admits a spectral decomposition if and only if $\Hk{k}(\bm{\nu})$ admits a positive semidefinite extension $\Hk{k+1}$, i.e., if and only if two further moments can be appended consistently with causality.
    If $\Hk{k}(\bm{\nu})\succ0$, this is possible for every choice of $\nu_{2k-1}$, and the resulting spectral decompositions are distinct.
\end{enumerate}
\end{theorem}
\noindent
For $\Nop=1$, (a) is Lemma~3.2, while (b) and (c) are Theorems~3.9 and 3.10 of Ref.~\cite{CurtoFialkow1991}, restated in the language of flat extensions as Theorems~1.2 and 1.3 of Ref.~\cite{CurtoFialkow1998}.
For general $\Nop$, (a) is Lemma~5.57 and (b) is Theorem~6.2 together with Corollary~5.66 of Ref.~\cite{Kimsey:2022moment}; (c) follows from (b), because by \cref{eq:smuljan_rank} the last moment of a positive semidefinite extension can always be lowered to its Schur-complement value, which makes the extension flat.

The idea behind the theorem is the construction of a finite-dimensional transfer matrix.
A positive semidefinite Hankel matrix defines an inner product for vectors with ``time-slice'' indices.
Together, this vector space and the inner product give rise to a Hilbert space whose dimension is the rank of the matrix.
A flat extension says that the next time slice adds no spectral content.
Translation by one unit in Euclidean time acts as a Hermitian operator on this space, an effective transfer matrix.
Its eigenvalues are the $\lambda_n$, and its eigenprojectors give rise to $W_n$.

Applied to the first $m-1$ moments and to the full sequence, the theorem yields the two facts used in the main text.

\begin{corollary}
\label{cor:main}
Let $\Hk{d}(\momseq)\succeq0$, with $\muflat$ and $\Delta$ as in \cref{eq:endpoint_split}.
\begin{enumerate}[label=(\alph*),leftmargin=2em]
\item The first $m-1$ moments $(\moment{0},\dots,\moment{2d-3})$ admit a unique minimal spectral decomposition.
    It satisfies
    \begin{align}
        \sum_{n=1}^{\Nstates}\rank W_n=\rank\Hk{d-1}\leq\Nop(d-1),
        \label{eq:rank_counts_flat}
    \end{align}
    and it is the unique spectral decomposition of $\muflat$.
    In particular, its prediction for the last moment is $\lastflat$.
    The endpoint excess $\Delta=\moment{2d-2}-\lastflat\succeq0$ is the discrepancy between this prediction of the minimal spectral decomposition and the data.
    $\Delta$ is nonzero in $\rank\Delta=\rank\Hk{d}-\rank\Hk{d-1}$ operator directions.
\item The full sequence $\momseq$ admits a spectral decomposition if and only if further moments can be appended to it consistently with causality, i.e., if and only if $\Hk{d}(\momseq)$ admits a positive semidefinite extension.
    This is the case if $\Delta=0$, where the decomposition is unique and coincides with that of (a), and if $\Hk{d}(\momseq)\succ0$, where it is not unique.
    If $\Delta\neq0$ and $\Hk{d}(\momseq)$ is singular, both outcomes occur.
\end{enumerate}
\end{corollary}
\noindent
Indeed, $\Hflat$ is a flat extension of $\Hk{d-1}$, so \cref{thm:flat}(b) with $k=d-1$ gives the unique spectral decomposition of $\muflat$ and \cref{eq:rank_counts_flat}.
By \cref{thm:flat}(a), every spectral decomposition of the first $m-1$ moments has $\sum_n\rank W_n\geq\rank\Hk{d-1}$; if equality holds, the moments of the decomposition form a positive semidefinite $\Hk{d}$ of rank at most $\rank\Hk{d-1}$ that contains $\Hk{d-1}$, hence a flat extension of it, whose last moment is $\lastflat$ by \cref{eq:smuljan_rank}.
Such a decomposition is therefore a decomposition of $\muflat$, and (a) follows.
Part (b) is \cref{thm:flat}(c) with $k=d$; for $\Delta=0$ one has $\momseq=\muflat$, and the examples of \cref{app:examples} supply the last statement.

\subsection{Example of an endpoint excess \label{app:examples}}

Let $\Nop=2$, write $P_1=\bm{e}_1\bm{e}_1^T$ and $P_2=\bm{e}_2\bm{e}_2^T$ for projectors onto the two operator directions, and consider the following moments
\begin{align}
\begin{split}
    \moment{0} &= \identity,\\
    \moment{1} &= \moment{3}=0,\\
    \moment{2} &= P_1,\\
    \moment{4} &= P_1+\bm{u}\bm{u}^T.
\end{split}
    \label{eq:block_example}
\end{align}
Different values of $\bm{u}\in\R^2$ will be chosen below for the purpose of illustration.
Using the block notation defined above, one finds $A=\Hk{2}=\diag(\identity,P_1)$ and $B^\dagger=(P_1\ \ 0)$, so $\lastflat=P_1$ and $\Delta=\bm{u}\bm{u}^T$.
That is, the value of $\bm{u}$ determines the endpoint excess.
The sequence $\muflat$ is reproduced by the three-state spectral function
\begin{align}
    \tilde{\rho}(\lambda)=P_2\,\delta(\lambda)+\tfrac12P_1\,\delta(\lambda-1)+\tfrac12P_1\,\delta(\lambda+1),
    \label{eq:rho_flat}
\end{align}
whose weight matrices have total rank $3=\rank\Hk{2}$, in agreement with \cref{cor:main}(a), and
\begin{align}
    \Hk{3}
    =\underbrace{\begin{pmatrix}\identity&0&P_1\\0&P_1&0\\P_1&0&P_1\end{pmatrix}}_{\equiv\,\Hflat}
    +\begin{pmatrix}0&0&0\\0&0&0\\0&0&\Delta\end{pmatrix},
\end{align}
with $\rank\Hflat=3=\rank\Hk{2}$ and $\rank\Hk{3}=4$ for any $\bm{u}\neq0$.
Three cases are possible depending on the value of the endpoint excess $\Delta$.
\begin{enumerate}
    \item For $\bm{u}=0$, $\Delta=0$ and \cref{eq:rho_flat} is the unique spectral decomposition of $\momseq$.
    \item For $\bm{u}=\bm{e}_1+\bm{e}_2$, no further moments can be appended consistently with causality, and hence no spectral decomposition exists.
    To see this, note that the block vector $\bm{v}=(0,\bm{e}_2,0)$ lies in $\ker\Hk{3}$, but for any $\moment{5}$ the new block column of $\Hk{4}$ has nonzero overlap $(\moment{3}\ \moment{4}\ \moment{5})\bm{v}=\moment{4}\bm{e}_2=\Delta\bm{e}_2=\bm{u}\neq0$ with it, violating the first condition in \cref{eq:smuljan}.
    \item For $\bm{u}=\bm{e}_1$, the same computation gives vanishing overlap $\Delta\bm{e}_2=0$, and indeed $\momseq$ is reproduced by
\begin{align}
\begin{split}
    \tilde{\rho}'(\lambda)
    =&\left(P_2+\frac{1}{2}P_1 \right)\delta(\lambda)\\
    &+\frac{1}{4}P_1\left[\delta(\lambda-\sqrt{2})+\delta(\lambda+\sqrt{2})\right].
\end{split}
\end{align}
The weight matrices have total rank $4=\rank\Hk{3}$.
This decomposition is not unique: the last moment fixes the endpoint excess in direction $\bm{e}_1$, but not the individual spectral weights and energies.
\end{enumerate}
Note that the two choices $\bm{u}=\bm{e}_1$ and $\bm{u}=\bm{e}_1+\bm{e}_2$ give the same value for $\rank\Delta$, the same $\muflat$, and the same ranks of $\Hk{2}$ and $\Hk{3}$;
what distinguishes them is whether $\Delta$ is supported on the direction $\bm{e}_1$ that is still ``open'' after two time slices or leaks into the direction $\bm{e}_2$ that $\Hk{2}$ has already exhausted.
This is the situation described in the last statement of \cref{cor:main}(b).

\section{Direct numerical solution of convex quadratic problems \label{app:Newton}}

This appendix contains technical details relevant for the direct numerical solution of the convex quadratic program in \cref{eq:primal_problem} using primal-dual interior-point methods.

\subsection{Hamburger moment problem: Newton system}

In the primal-dual interior-point algorithm approach, the solution point is defined by the perturbed complementarity condition (cf.~\cref{eq:perturbed complementarity}).
One seeks the roots of
\begin{align}
\begin{split}
    \Sigma^{-1}(\momseq-\hat{\momseq}) - \bm{h}(\dualmomseq) &= 0,\\
    H(\momseq) \dualmomseq - \centralpath \mathds{1} &= 0,
\end{split}
\label{eq:central_path_zeros}
\end{align}
where $\mathds{1}$ is a $(pd) \times (pd)$ unit matrix.
The first-order variations are
\begin{align}
    \delta \left[ \Sigma^{-1}(\momseq-\hat{\momseq}) - \bm{h}(\dualmomseq)\right]
     &= \Sigma^{-1}\delta\momseq - \bm{h}(\delta \dualmomseq),  \\
    \delta \left[  H(\momseq) \dualmomseq - \centralpath \mathds{1}\right] &= H(\delta\momseq) \dualmomseq + H(\momseq)\delta \dualmomseq.
\end{align}
Define residuals at the point $(\momseq_n, \dualmomseq_n)$ by
\begin{align}
    \bm{r}_n &\equiv  \Sigma^{-1}(\momseq_n-\hat{\momseq}) - \bm{h}(\dualmomseq_n),  \label{eq:residual_stationarity}\\
    R_n &\equiv H(\momseq_n) \dualmomseq_n - \centralpath \mathds{1} \label{eq:residual_perturb_comp}.
\end{align}
The Newton system for the roots of \cref{eq:central_path_zeros} is therefore
\begin{align}
    \begin{bmatrix}
        \Sigma^{-1} & -\bm{h}(\cdot ) \\
        H(\cdot) \dualmomseq  & H(\momseq)
    \end{bmatrix}
    \begin{bmatrix}
        \delta\momseq \\ \delta \dualmomseq
    \end{bmatrix}
    = -\begin{bmatrix}
        \bm{r}_n \\
        R_n
    \end{bmatrix},
    \label{eq:Newton_system_Hamburger}
\end{align}
where $H(\cdot)$ and $\bm{h}(\cdot)$ are the linear maps in \cref{eq:hankel_linear_map,eq:hankel_transpose_linear_map}.
A vectorized form of \cref{eq:Newton_system_Hamburger} is discussed in \cref{sec:Newton_vectorization}.
At each iteration, a line search is used to maintain feasibility along a given Newton search direction $(\delta \momseq, \delta \dualmomseq)$.

To keep the discussion self-contained, it is useful to describe how these considerations can be used in a simple implementation of a primal-dual interior-point algorithm.
The algorithm begins by choosing an initial strictly feasible point $(\momseq_0, \dualmomseq_0)$, a target tolerance $\epsilon>0$ for the duality gap, a centering parameter $\sigma \in (0,1)$ that controls the speed at which $\centralpath \to 0^+$ along the central path, and a line-search parameter $\alpha_{\rm max}$.
While the duality gap remains above the desired tolerance, the iterations of the algorithm proceed as follows:
\begin{enumerate}
    \item Compute the normalized $\gap_n = \gap(\momseq_n, \dualmomseq_n)/(dp)$.
    \item Set $\centralpath_n = \gap_n \times \sigma$.
    \item Compute the Newton directions $(\delta \momseq_n, \delta \dualmomseq_n)$.
    \item Symmetrize the Newton directions $\delta \momseq_n \to \frac{1}{2}(\delta \momseq_n + \delta \momseq_n^\dagger)$ and $\delta M \to \frac{1}{2}(\delta M + \delta M^\dagger)$.
    \item Determine the largest feasible update in the Newton direction via a line search on $\alpha\in(0, \alpha_{\rm max}]$.
    \item Update $(\momseq_{n+1}, \dualmomseq_{n+1}) = (\momseq_n, \dualmomseq_n) + \alpha (\delta \momseq_n, \delta \dualmomseq_n)$.
\end{enumerate}
Values for $\sigma$ and $\alpha$ are chosen to improve convergence; values of $\sigma \approx 0.25\text{--}0.4$ and $\alpha \approx 0.95 \alpha_\mathrm{max}$ are not atypical.
The target for the duality gap depends on the details of the system.

The only substantive difference for moment problems besides the Hamburger moment problem is the form of \cref{eq:central_path_zeros} and the associated Newton system, \cref{eq:Newton_system_Hamburger}.
Additional details are given below for vectorization of the Hamburger moment problem, the Stieltjes moment problem, and the signed Hamburger moment problem.
The latter arises in practical lattice QCD calculations involving staggered quarks (see, e.g., Appendix G of Ref.~\cite{Follana:2006rc}).

\subsection{Hamburger moment problem: vectorization \label{sec:Newton_vectorization}}

\Cref{eq:Newton_system_Hamburger} gives the Newton system for the Hamburger moment problem in terms of matrix-valued variables.
It is useful to cast the system in a standard form using vectorization.
Using the vectorization formula
$\vectorization(AB) = (I \otimes A) \vectorization(B)$,
the Newton system can be rewritten in flattened form
\begin{align*}
    \begin{bmatrix}
        (\Sigma^{-1})_{ij} & -(\bm{h}(\cdot ))_{i\alpha} \\
        (H(\cdot) \dualmomseq)_{\alpha j}  & (I \otimes H(\momseq))_{\alpha\beta}
    \end{bmatrix}
    \begin{bmatrix}
        (\delta\momseq)_j \\
        \vectorization(\delta \dualmomseq)_{\beta}
    \end{bmatrix}
    = -\begin{bmatrix}
        (\bm{r})_i \\
        \vectorization(R)_\alpha
    \end{bmatrix},
\end{align*}
where $i,j$ index the coordinates of $\R^m$ and $\alpha,\beta$ those of a flattened $\Sym^d$.
This form is convenient for numerical solution.
Note that the solution to a block linear system of the form
\begin{align}
    \begin{bmatrix}
        A & B \\ C & D
    \end{bmatrix}
    \begin{bmatrix}
        \delta \momseq \\ \delta \dualmomseq
    \end{bmatrix}
    =
    -
    \begin{bmatrix}
        \bm{r} \\ R
    \end{bmatrix}
\end{align}
can be written
\begin{align}
    \begin{split}
    \delta \momseq
        &= -\left[A - B D^{-1} C\right]^{-1} \left[ \bm{r} - B D^{-1} R\right], \\
    \delta \dualmomseq
        &= -\left[D - C A^{-1} B\right]^{-1} \left[ R - C A^{-1} \bm{r}\right],
    \end{split}
    \label{eq:block_solve}
\end{align}
where the vectorization on $R$ and $\delta M$ has been left implicit.
This form is manifestly symmetric under the simultaneous exchange
$(\momseq, A, C, \bm{r}) \longleftrightarrow (\dualmomseq, D, B, R)$
and conveniently avoids computation of the inverse of the full $2\times 2$ block matrix.
Analogous vectorization can be used to solve the Newton systems for other moment problems.
Direct usage of the second line in \cref{eq:block_solve} involves an expensive linear solve.
Assuming that $\delta\momseq$ has already been obtained from the first equation, this linear solve can be avoided, since the Newton system in \cref{eq:Newton_system_Hamburger} immediately gives
\begin{align}
    \delta M = - H(\momseq)^{-1}
    \left(
    R_n + H(\delta \momseq) M
    \right).
\end{align}

\subsection{Divergence of the projection mappings \label{sec:divergence}}

The numerical solution described above also gives access to the divergence of the projection mapping appearing in SURE, \cref{eq:stein_sure}.
Since $\mathds{1} = \projection{\mathcal{K}} + \projection{\polar{\mathcal{K}}}$, it follows that
$m = \nabla\cdot \projection{\mathcal{K}}(\hat{\momseq}) + \nabla\cdot \projection{\polar{\mathcal{K}}}$, so it suffices to consider $\nabla\cdot \projection{\mathcal{K}}(\hat{\momseq})$.
At the solution point ($\varepsilon \to 0$), the residuals in \cref{eq:residual_stationarity,eq:residual_perturb_comp} both vanish.
Differentiating these two conditions with respect to the data
$\hat\momseq$ at fixed $\varepsilon$
and eliminating $\delta M$ as in the preceding section gives
\begin{equation}
\left[\Sigma^{-1}+\bm{h}\left(H(\momseq)^{-1}H(\cdot)M\right)\right]\,\delta\momseq = \Sigma^{-1}\delta\hat{\momseq}.
\label{eq:kktjac}
\end{equation}
In other words, the Jacobian of the projection at the solution point ($\varepsilon\to 0$) is given by
\begin{align}
    \frac{\partial \projection{\mathcal{K}}}{\partial \hat{\momseq}} = \left[\Sigma^{-1}+\bm{h}\left(H(\momseq)^{-1}H(\cdot)M\right)\right]^{-1} \Sigma^{-1},
\end{align}
so that the divergence of the projection mapping is
\begin{align}
    \nabla \cdot \projection{\mathcal{K}} = \Tr\left[\frac{\partial \projection{\mathcal{K}}}{\partial \hat{\momseq}}\right].
\end{align}
Entirely analogous steps can be used to compute this quantity for projection onto the Stieltjes, Hausdorff, or signed Hamburger cones.

\subsection{Stieltjes moment problem: Newton system}

The Stieltjes moment problem requires both the Hankel matrix and shifted Hankel matrix to be PSD.
The primal problem is
\begin{align}
\begin{split}
    \minimize_{\momseq\in \R^m} &\quad p(\momseq) \equiv\frac{1}{2} \chi^2(\hat{\momseq} - \momseq)\\
    \st &\quad H(\momseq) \succeq 0,\\
        &\quad H^{(1)}(\momseq) \succeq 0.
\end{split}
\end{align}
The associated Lagrange function is
\begin{align}
    \mathcal{L}(\momseq, \dualmomseq, Z) = \frac{1}{2} \chi^2(\momseq- \hat{\momseq}) - \langle \dualmomseq, H(\momseq)\rangle - \langle Z, H^{(1)}(\momseq)\rangle,
\end{align}
where $Z\in \Sym^{d-1}$.
The stationarity condition reads
\begin{align}
    \nabla_{\momseq} \mathcal{L} = \Sigma^{-1}(\momseq-\hat{\momseq}) - \bm{h}(\dualmomseq) - \bm{h}^{(1)}(Z) = 0,
\end{align}
where $\bm{h}^{(1)}(Z) \in \R^m$ is the dual map associated with the shifted Hankel matrix, defined analogously to \cref{eq:hankel_transpose_linear_map}.
The perturbed complementarity conditions read
\begin{align}
    H(\momseq) \dualmomseq - \centralpath_\dualmomseq \identity  &= 0,\\
    H^{(1)}(\momseq) Z - \centralpath_Z \identity &=0.
\end{align}
First-order variations give the associated Newton system
\begin{align}
    \Sigma^{-1} \delta \momseq - \bm{h}(\delta \dualmomseq) - \bm{h}^{(1)}(\delta Z) &= -\bm{r},\\
    H(\delta\momseq) \dualmomseq + H(\momseq)\delta \dualmomseq &= - R_M,\\
    H^{(1)}(\delta\momseq) Z + H^{(1)}(\momseq) \delta Z &= - R_Z,
\end{align}
where the residuals are defined according to
\begin{align}
    \bm{r} &\equiv \Sigma^{-1}(\momseq - \hat{\momseq}) - \bm{h}(\dualmomseq) - \bm{h}^{(1)}(Z),\\
    R_\dualmomseq &\equiv H(\momseq) \dualmomseq - \centralpath_\dualmomseq \identity, \\
    R_Z &\equiv H^{(1)}(\momseq) Z - \centralpath_Z \identity.
\end{align}
The definitions of the residuals $R_\dualmomseq, R_Z$ have used the conventional choice to take $\centralpath_\dualmomseq = \centralpath_Z \equiv \centralpath$.
In matrix-vector notation, the Newton system takes the form
\begin{align}
    \begin{bmatrix}
        \Sigma^{-1} & -\bm{h}(\cdot ) & -\bm{h}^{(1)}(\cdot)\\
        H(\cdot) \dualmomseq  & H(\momseq) & 0 \\
        H^{(1)}(\cdot)Z & 0 & H^{(1)}(\momseq)
    \end{bmatrix}
    \begin{bmatrix}
        \delta\momseq \\ \delta \dualmomseq \\ \delta Z
    \end{bmatrix}
    = -\begin{bmatrix}
        \bm{r} \\ R_\dualmomseq \\ R_Z
    \end{bmatrix}.
    \label{eq:Newton_system_Stieltjes}
\end{align}
As above, large linear solves for $\delta M$ and $\delta Z$ can be avoided by using the Newton system after first solving for $\delta\momseq$.

\subsection{Signed Hamburger moment problem: Newton system \label{sec:signed_hamburger}}

Practical staggered-quark spectroscopy problems in lattice QCD may involve
a signed spectral function.
For example, see Appendix G of Ref.~\cite{Follana:2006rc}.
Consider the convention
\begin{equation}
 \widetilde\rho(\lambda)=
 \begin{cases}
  +\rho_+(\lambda),  & \lambda \geq 0,\\
  -\rho_-(-\lambda), & \lambda < 0,
 \end{cases}
 \quad \quad \lambda  \in \R,
 \label{eq:c4-sign}
\end{equation}
where $\rho_\pm(\lambda)\succeq0$ are defined on the positive real line.
The corresponding moments are
\begin{align}
 \mu_t&=\int_{\mathbb R}d\lambda\,\lambda^t\widetilde\rho(\lambda)
 \nonumber\\
 &=\int_0^\infty d\lambda\,\lambda^t\rho_+(\lambda)
   +(-1)^{t+1}\int_0^\infty d\lambda\,\lambda^t\rho_-(\lambda).
 \label{eq:c4-moments}
\end{align}
That is, the second term enters with a plus sign at odd $t$ and
a minus sign at even $t$.
The relevant positivity condition follows from
$\lambda\widetilde\rho(\lambda)\succeq0$.
For a vector-valued polynomial
$f(\lambda)=\sum_{i=0}^{d-1}\lambda^i\bm{v}_i$,
\begin{equation}
 \sum_{i,j=0}^{d-1}\bm{v}_i^\dagger\mu_{i+j+1}\bm{v}_j
 =\int_{\mathbb R}d\lambda\,
 f(\lambda)^\dagger\lambda\widetilde\rho(\lambda)f(\lambda)
 \geq 0,
 \label{eq:c4-positivity}
\end{equation}
which implies that the shifted Hankel matrix $H^{(1),[d]}(\momseq)$ is positive
semidefinite.
The convex cone associated with the signed spectral function is therefore
\begin{equation}
 \begin{aligned}
 \mathcal{K}_{\rm{signed}}&=
 \{\momseq:H^{(1),[d]}(\momseq)\succeq0\},\\
 H^{(1),[d]}(\momseq)&=[\mu_{i+j+1}]_{i,j=0}^{d-1}.
 \end{aligned}
 \label{eq:c4-cone}
\end{equation}
Projection onto the boundary of $ \mathcal{K}_{\rm{signed}}$ is obtained by
\begin{align}
    \begin{split}
    \projection{\mathcal{K}_{\rm signed}}(\momseq)
    =&
    \argmin_{\momseq'\in (\H_\Nop)^m} \lVert\momseq'-\momseq \rVert^2_{\Sigma_N^{-1}} \\
    &\st \momseq' \in \mathcal{K}_{\rm signed}.
    \end{split}
\end{align}
The shifted Hankel matrix $H^{(1),[d]}$ does not involve the zeroth moment, so the optimization over that variable is unconstrained.

The solution can be written compactly in block form.
Let
\begin{align}
\momseq = (\momseq_1, \momseq_2),
&&
\Sigma = \begin{pmatrix}
\Sigma_{11} & \Sigma_{12} \\
\Sigma_{21} & \Sigma_{22}
\end{pmatrix},
\end{align}
where $\momseq_1$ contains any unconstrained moment(s) and $\momseq_2$ contains the constrained moments.
A brief calculation shows that unconstrained optimization over $\momseq_1$  gives
\begin{align}
    \momseq_1' = \momseq_1 + \Sigma_{12}\Sigma_{22}^{-1}(\momseq_2' -\momseq_2),
\end{align}
where $\Sigma_{22}^{-1}$ denotes the inverse of the lower-right block of the covariance matrix (and \emph{not} the corresponding block of the full inverse covariance matrix).
As expected, $\momseq_1' = \momseq_1$ in the absence of cross correlations ($\Sigma_{12}=0$).
Substituting this value into the original objective function results in constrained optimization over $\momseq_2'$ with primal objective function given by
\begin{align}
    \primal(\momseq_2') = (\momseq_2' - \momseq_2)^\dagger \Sigma_{22}^{-1} (\momseq_2' - \momseq_2).
\end{align}
Numerical solution to the constrained optimization problem proceeds analogously to that of the Hamburger moment problem.

\subsection{Symmetric correlators}

Many important correlation functions in lattice QCD are symmetric about the lattice midpoint.
Consider a lattice of even temporal extent $T$.
Consider a mesonic two-point function with periodic boundary conditions, which satisfies
\begin{align}
 \moment{t}=\moment{T-t}, && t=0,\ldots,T.
 \label{eq:c6-reflection}
\end{align}
This symmetry is a property of the exact correlator but is usually not satisfied by Monte Carlo estimators.
Imposing the symmetry therefore amounts to an additional linear constraint alongside Hankel positivity.
One possibility consists in applying the above construction to the first half of the measured moments, $t=0,\dots,T/2$, possibly after averaging about the midpoint.
Here, instead, we focus on employing the entire, unmodified moment sequence.
As above, early times can be excluded to avoid contact terms.
The construction below applies to any window $t_0\leq t\leq T-t_0$ after relabeling the time origin.
The discussion will focus on scalar $(p=1)$ correlators, since matrix-valued correlators do not present any conceptual complications.

Let $R$ denote the operation of time reflection $(R\momseq)_t=\moment{T-t}$.
Let $S$ be the subspace of reflection-symmetric sequences (i.e., the fixed points of the reflection $R$), a linear subspace of dimension $T/2+1$.
The full sequence $\momseq = (\moment{0},\ldots,\moment{T})$ contains an odd number of moments, so that the Hankel matrices of \cref{ssec:moments} have dimension $d=T/2+1$ for $H^{[d]}$ and $T/2$ for $H^{(1),[d]}$;
note that the endpoint $\moment{T}=\moment{0}$ is inferred by symmetry rather than by explicit measurement.
Causal, symmetric sequences form the set
\begin{equation}
 \mathcal{K}_S\equiv \mathcal{K}\cap S,
 \label{eq:c6-cone}
\end{equation}
which is again a closed convex cone.
The conic geometry of \cref{sec:cones} therefore applies without modification after replacing $\mathcal{K}$ with $\mathcal{K}_S$.
The symmetric projection $\projection{\mathcal{K}_S}(\hat\momseq)$ minimizes $\chi^2(\hat{\momseq}-\momseq)$ over $\momseq \in \mathcal{K}_S$, where $\chi^2$ is evaluated on the measured times with their
covariance $\Sigma$, as in \cref{eq:chi2_ab}.
Symmetric projection can be solved immediately using the methods developed above, thanks to a pair of useful facts.

The first fact is that the symmetric minimization follows from a simple change of variables.
Let $\hat{\momseq}_S\in S$ be the symmetric sequence closest to the data in
the inverse-covariance metric, i.e., the best fit to $\hat{\momseq}$ subject to the linear constraint in \cref{eq:c6-reflection} but \emph{ignoring} positivity.
Since the residual $\hat{\momseq}-\hat{\momseq}_S$ is orthogonal (with respect to the inverse covariance matrix) to $S$, one has
\begin{align}
\begin{split}
    \chi^2(\hat{\momseq}-\momseq)&=\chi^2_S+\chi^2(\hat{\momseq}_S-\momseq),\\
    \chi^2_S &\equiv \chi^2(\hat{\momseq}-\hat{\momseq}_S)
\end{split}
\label{eq:c6-split}
\end{align}
for every $\momseq \in S$.

The constant $\chi^2_S$ is the minimum cost of imposing the symmetry
alone and is manifestly independent of $\momseq$.
Consequently,
$\projection{\mathcal{K}_S}(\hat{\momseq})=\projection{\mathcal{K}_S}(\hat{\momseq}_S)$.
In other words, the symmetric conic projection can be obtained by first projecting onto $S$ and then projecting onto the cone within $S$.
In terms of the $T/2+1$ independent moments that parameterize $S$, the second step reads
\begin{equation}
\minimize_{\momseq\in \mathcal{K}_S} ~
 (\momseq-\hat{\momseq}_S)^T \Sigma_S^{-1} (\momseq-\hat{\momseq}_S),
 \label{eq:c6-reduced}
\end{equation}
where $\Sigma_S$ is the covariance matrix of $\hat{\momseq}_S$.
\Cref{eq:c6-reduced} has precisely the form of the original primal problem in \cref{eq:primal_problem} after the substitution $(\hat{\momseq},\Sigma)\to(\hat{\momseq}_S,\Sigma_S)$, with the Hankel maps restricted to symmetric sequences.
Existing numerical methods thus carry over almost immediately to this minimization.
In the language of \cref{sec:conic_decomposition}, the acausal component of the data now consists of two pieces that are orthogonal with respect to $\Sigma^{-1}$: a symmetry-violating piece with cost $\chi^2_S$ and a positivity-violating piece within the symmetric subspace $S$.

A superficial obstruction to using existing solvers, nominally defined for the full moment sequence $\momseq$, seems to arise from the fact that $\Sigma_S$ will be singular on the full space, since points related by symmetry are completely correlated.
The following elegant group-invariance argument shows how to sidestep this challenge in practice.

The second fact is that reflection acts on Hankel matrices by a permutation of the time indices and thus preserves the causal cone.
That is, one can write
\begin{align}
\begin{split}
    H(R\momseq) &= JH(\momseq)J, \\
    H^{(1)}(R\momseq) &= JH^{(1)}(\momseq)J,
\end{split}
\label{eq:hankel-congruence}
\end{align}
where $J$ is a permutation matrix, since $\moment{T-i-j}=\moment{(d-1-i)+(d-1-j)}$ with $d-1=T/2$, and similarly for the shifted matrix.
Since $J$ is orthogonal, one has $H(R\momseq)\succeq0$ if and only if $H(\momseq)\succeq0$.
Thus, the Hamburger, Stieltjes, and signed Hamburger cones are invariant under $R$. In terms of the spectrum, reflection exchanges the contributions $W\lambda^t$ and $(\lambda^TW)\lambda^{-t}$, i.e., $\lambda \in(0,1) \leftrightarrow 1/\lambda \in(1,\infty)$
with a positive weight for even $T$, which preserves the sign of $\lambda$.
The Hausdorff cone is clearly not invariant, since $1/\lambda$ lies
outside $[0,1]$, reflecting the obvious physical fact that a zero-temperature correlation function is not periodic.

Whenever the data and the metric are themselves reflection invariant,
the unique minimizer over $\mathcal{K}$ is automatically symmetric.
To see this, note that if $\momseq^\star$ were a minimizer, $R\momseq^\star$ would be too, and uniqueness forces $R\momseq^\star=\momseq^\star$.
Essentially the same group-averaging argument was given in Ref.~\cite{Gatermann:2004} for invariant semidefinite programs;
the core idea is that averaging a feasible point with its reflection preserves feasibility and cannot increase a convex invariant objective.
This fact immediately allows reuse of existing numerical solvers.
To use existing solvers (nominally defined on the full moment sequence $\momseq$), one extends $\Sigma_S^{-1}$ to any reflection-invariant positive-definite quadratic form acting on the full space of sequences that happens to agree with $\Sigma_S^{-1}$ on $S$.
Minimization over $\mathcal{K}$ is then symmetric and gives $\projection{\mathcal{K}_S}(\hat{\momseq}_S)$, and is independent of the extension chosen.
The same invariance guarantees that the Newton directions of the primal-dual iteration remain symmetric when the iterates are, so that positivity and symmetry are maintained simultaneously at every step~\cite{Kanno:2001,Toh:2008}.

\bibliography{refs.bib}

\end{document}